\documentclass[12pt]{article}
\pdfoutput=1
\usepackage{mathtools,amssymb,mathrsfs,microtype,ytableau,tikz,array}
\usepackage[
 linktoc=all,
 colorlinks=true,
 linkcolor=blue,
 urlcolor=blue,
 citecolor=red,
 ]{hyperref}
\usepackage[backend=bibtex,style=numeric-comp,sorting=none,maxbibnames=99, minbibnames=99]{biblatex}
\ytableausetup{centertableaux,boxsize=.5em}

\usepackage[framemethod=TikZ]{mdframed}
\mdfsetup{%
skipabove=3pt,
skipbelow=2pt,
linecolor=black!0,
backgroundcolor=black!15,
roundcorner=3pt}

\makeatletter\@addtoreset{equation}{section}\makeatother

\newtheorem{lemma}{Lemma}

\DeclareMathOperator{\diag}{diag}

\DeclareMathOperator{\tr}{tr}
\DeclareMathOperator{\re}{re}
\DeclareMathOperator{\im}{im}

\renewcommand{\title}[1]{\vbox{\center\LARGE{#1}}\vspace{5mm}}
\renewcommand{\author}[1]{\vbox{\center\large#1}\vspace{5mm}}
\newcommand{\address}[1]{\vbox{\center\em#1}}

\newcommand{\normord}[1]{\vcentcolon\mathrel{#1}\vcentcolon}
\providecommand{\vcentcolon}{\mathrel{\mathop{:}}}
\begin{document}

\begin{titlepage}
\begin{center}
\vspace{5mm}
\hfill {\tt }\\
\vspace{8mm}

\title{\makebox[\textwidth]{\Huge{Infrared phases of 2d QCD}}}
\vspace{10mm}
Diego Delmastro,${}^{ab}$ \footnote{\href{mailto:ddelmastro@perimeterinstitute.ca}{\tt ddelmastro@perimeterinstitute.ca}}
Jaume Gomis,${}^{a}$ \footnote{\href{mailto:jgomis@perimeterinstitute.ca}{\tt jgomis@perimeterinstitute.ca}}
Matthew Yu${}^{a}$ \footnote{\href{mailto:myu@perimeterinstitute.ca}{\tt myu@perimeterinstitute.ca}}
\vskip 7mm
\address{
${}^a$Perimeter Institute for Theoretical Physics,\\
Waterloo, Ontario, N2L 2Y5, Canada}
\address{
${}^b$Department of Physics, University of Waterloo,\\ Waterloo, ON N2L 3G1, Canada}
\end{center}

\vspace{5mm}
\abstract{
We derive the necessary and sufficient conditions for a $2d$ QCD theory of massless gluons and left and right chiral quarks in arbitrary representations of a gauge group $G$ to develop a mass gap. These results are obtained from spectral properties of the lightcone and temporal QCD Hamiltonians. The conditions can be explicitly solved, and we provide the complete list of all $2d$ QCD theories that have a quantum mechanical gap in the spectrum, while any other theory not in the list is gapless. The list of gapped theories includes QCD models with quarks in vector-like as well as chiral representations. The gapped theories consist of several infinite families of classical gauge groups with quarks in rank 1 and 2 representations, plus a finite number of isolated cases. We also put forward and analyze the effective infrared description of QCD --- TQFTs for gapped theories and CFTs for gapless theories --- and exhibit several interesting features in the infrared, such as the existence of non-trivial global 't Hooft anomalies and emergent supersymmetry. We identify $2d$ QCD theories that flow in the infrared to celebrated CFTs such as minimal models, bosonic and supersymmetric, and Wess-Zumino-Witten and Kazama-Suzuki models.
}
 
\vfill\eject


{\hypersetup{linkcolor=black}
\tableofcontents
\thispagestyle{empty}
}

\end{titlepage}

\section{Introduction}
\setcounter{footnote}{0}

A central theme in physics is unraveling the low energy phenomena that emerges from a physical system described by a collection of microscopic degrees of freedom and interactions. The long distance behavior of the system crucially depends on whether the spectrum of the Hamiltonian is gapped or gapless, but determining which phase is realized is often a nonperturbative problem.

In broad terms, a gapped system is described at low energies by a topological quantum field theory (TQFT), while the asymptotic low energy dynamics of a gapless one is captured by a conformal field theory (CFT).\footnote{The CFT can be either a symmetry preserving nontrivial fixed point of the renormalization group, the extreme infrared limit of the nonlinear theory of Goldstone bosons when the vacuum spontaneously breaks a continuous symmetry, or free massless particles in a symmetric vacuum (e.g.~infrared free gauge theories).} Ascertaining whether a system flows to a TQFT or a CFT, and to which one, can be out of reach because of large quantum fluctuations, which are responsible for a wealth of low energy phenomena. 
 
QCD theories are an important class of strongly coupled systems in which it is nontrivial to postulate the infrared dynamics.
Determining whether Yang-Mills theory with gauge group $G$ coupled to massless quarks in a representation $R$ of $G$ in $d\leq 4$ spacetime dimensions is gapped or gapless remains an open problem. We henceforth refer to such theories of massless quarks and gluons as \emph{QCD theories}. The following qualitative picture is expected:

\begin{itemize}
\item QCD theories without quarks, that is, pure Yang-Mills theory, are believed to be gapped. For simply connected gauge group $G$, the infrared is described by the trivial TQFT.\footnote{In $4d$ the theory has a unique vacuum for $\theta\neq \pi$ while for $\theta= \pi$ the time-reversal symmetry is spontaneously broken and there are two trivially gapped vacua~\cite{Gaiotto:2017yup}. Yang-Mills theory in $3d$ can be enriched by a Chern-Simons term and then the theory in the infrared is gapped and described by a nontrivial TQFT.} 
\item QCD theories with a large number of quarks -- more precisely, with a large Dynkin index\footnote{$N_F$ fermions in a representation $R$ of $G$ has Dynkin index $N_F\times I(R)$, where $\tr(t_R^at_R^b)=I(R) \delta^{ab}$.} -- are gapless. In $4d$, this is by virtue of the beta function~\cite{Gross:1973ju,Politzer:1973fx} being positive for a sufficiently large number of quarks, which implies that the infrared is described by a CFT of free massless particles. In $3d$, the fact that QCD theories flow to a weakly coupled CFT can be established in the limit of large Dynkin index~\cite{Appelquist:1988sr,Appelquist:1989tc}.
\end{itemize}
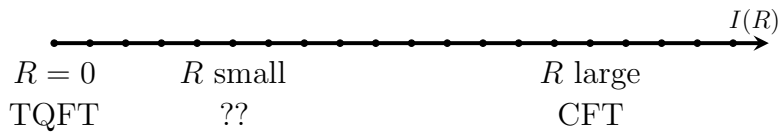
\begin{figure}[!h]
\centering
\begin{tikzpicture} 
[scale=0.95]
\draw[->,>=stealth, line width = 1.5pt] (0,.1) -- (10,.1);
\foreach \x in {0,...,19} {\filldraw (.5*\x,.1) circle (1.3pt);}
\node[text width=1.2cm,align=center,anchor=north] at (0,0) {$R=0$ TQFT};
\node[text width=1.5cm,align=center,anchor=north] at (2.5,0) {$R$ small ??};
\node[text width=1.5cm,align=center,anchor=north] at (7.5,0) {$R$ large CFT};
\node[anchor=south,scale=.8] at (9.8,.1) {$I(R)$};
\end{tikzpicture}
\caption{Diagram describing infrared dynamics of QCD with massless quarks in a representation $R$ of the gauge group $G$. The theory without quarks is expected to be gapped, and is gapless for large enough $R$. The intermediate regime where the representation $R$ is small remains an open problem.}
\label{fig:IR_picture}
\end{figure}
While gapped QCD theories in $4d$ have been known for some time~\cite{WITTEN1982253,AFFLECK1985557}, it is only recently that examples of gapped QCD theories in $3d$, together with their infrared TQFTs, have been put forward~\cite{Gomis:2017ixy} (see also~\cite{Cordova:2017vab,Bashmakov:2018wts,Benini:2018umh,Choi:2018ohn,Choi:2018tuh,Choi:2019eyl}). Little is otherwise known about whether a given QCD theory is gapped or not, and which TQFT/CFT describes its infrared limit (see~fig.~\ref{fig:IR_picture}).

In this paper we determine all the QCD theories in $2d$ that are gapped, and therefore those that are gapless. The full classification of gapped QCD theories is summarized in tables~\ref{tab:classification_gapped_algebra} and~\ref{tab:classification_gapped_chiral}. 

In $2d$ QCD, the quark content is specified by a pair of representations $(R_\ell,R_r)$ of the gauge group $G$ acting on the left and right chiral quarks. We denote such a QCD theory by $(G;R_\ell,R_r)$.\footnote{See section~\ref{sec:QCD2} for the role of the topology of the gauge group $G$ in defining topological sectors, discrete theta angles, gauge anomalies, etc.} We derive the necessary and sufficient conditions for a QCD theory $(G;R_\ell,R_r)$ to be gapped by analyzing the explicit lighcone and temporal Hamiltonians of QCD. Lightcone quantization, where $x^+$ \emph{and} $x^-$ are time, and the canonical Hamiltonian formalism, where $x^0$ is time, yield exactly the same conditions.

From our Hamiltonian analysis, the following criterion is derived: a QCD theory is gapless if and only if there exists a canonical, chiral, dimension $2$ primary operator of the quark current algebra constructed from either the left chiral quarks or right chiral quarks. A QCD theory is gapped if and only if both these left and right chiral operators vanish identically. These operator equations, derived by studying the Hamiltonian(s) in the ultraviolet, can be completely solved, yielding the classification of gapped theories.\footnote{The operator equations that are the necessary and sufficient conditions for a QCD theory to be gapped 
\begin{equation*}
\begin{aligned}
T{}_{\mathfrak{so}(\dim R_\ell)_1}-T{}_{G_{I(R_\ell)}}=0 \\[+2pt]
\overline T_{\mathfrak{so}(\dim R_r)_1}-\overline T_{G_{I(R_r)}}=0
\end{aligned}
\end{equation*}
correspond to \emph{all} the conformal embeddings into the $\mathfrak{so}(\dim R_\ell)_1$ \emph{and} $\mathfrak{so}(\dim R_r)_1$ current algebras, and are in one-to-one correspondence with Cartan's classification of symmetric spaces. $T{}_{\mathfrak{so}(\dim R_\ell)_1}/\overline T_{\mathfrak{so}(\dim R_r)_1}$ is the canonical energy-momentum tensor of the left/right chiral quarks in the ultraviolet, and $T{}_{G_{I(R_\ell)}}/\overline T_{G_{I(R_r)}}$ is the left/right moving Sugawara energy-momentum tensor of the current algebra $G_{I(R)}$ at level $I(R)$. See section~\ref{sec:DLCQ} for details.} The exhaustive list of gauge groups $G$ and quark contents $(R_\ell,R_r)$ of all the QCD theories that are gapped appears in tables~\ref{tab:classification_gapped_algebra} and \ref{tab:classification_gapped_chiral}, corresponding to vector-like and chiral QCD theories respectively. Any other QCD theory not in the tables is gapless.

\begin{table}[!h]
\centering
\setlength{\tabcolsep}{1em} 
{\renewcommand{\arraystretch}{1.1}
\begin{tabular}{>{$}c<{$}|>{$}c<{$} c >{$}c<{$}|>{$}c<{$}}
\multicolumn{2}{c}{} && \multicolumn{2}{c}{} \\
\mathfrak g & R && \mathfrak g & R \\
\cline{1-2} \cline{4-5}
\forall \mathfrak g&\text{adjoint} && \mathfrak{su}(2)&\boldsymbol5 \\
\mathfrak{so}(N)&\ydiagram1 && \mathfrak{so}(9)&\boldsymbol{16} \\
\mathfrak{u}(N)&\hphantom{_q}\ydiagram1_q && F_4&\boldsymbol{26} \\
\mathfrak{so}(N)&\ydiagram2 && \mathfrak{sp}(4)&\boldsymbol{42} \\
\mathfrak{sp}(N)&\ydiagram{1,1} && \mathfrak{su}(8)&\boldsymbol{70} \\
\mathfrak u(N)&\hphantom{_q}\ydiagram{1,1}_q && \mathfrak{so}(16)&\boldsymbol{128} \\
\mathfrak u(N)&\hphantom{_q}\ydiagram2_q && \mathfrak{so}(10)+ \mathfrak u(1)&\hphantom{_q}\boldsymbol{16}_q \\
\mathfrak{su}(M)+ \mathfrak{su}(N)+ \mathfrak u(1)&\hphantom{_q}(\ydiagram1,\ydiagram1)_q && E_6+ \mathfrak u(1)&\hphantom{_q}\boldsymbol{27}_q \\
\mathfrak{so}(M)+ \mathfrak{so}(N)&(\ydiagram1,\ydiagram1) && \mathfrak{su}(2)+ \mathfrak{su}(2)&(\boldsymbol{2},\boldsymbol{4}) \\
\mathfrak{sp}(M)+ \mathfrak{sp}(N)&(\ydiagram1,\ydiagram1) && \mathfrak{su}(2)+ \mathfrak{sp}(3)&(\boldsymbol{2},\boldsymbol{14}) \\
 & && \mathfrak{su}(2)+ \mathfrak{su}(6)&(\boldsymbol{2}, \boldsymbol{20}) \\
 & && \mathfrak{su}(2)+\mathfrak{so}(12)&(\boldsymbol{2},\boldsymbol{32}) \\
 & && \mathfrak{su}(2)+ E_7&(\boldsymbol{2},\boldsymbol{56})\\\hline\hline
\bigoplus_i \mathfrak g_i & \bigoplus_i(\boldsymbol1,...,R_i,...,\boldsymbol1)_{\vec q_i}
\end{tabular}}
\caption{Classification of vector-like gapped QCD theories $(G;R,R)$. $\mathfrak{g}$ denotes the Lie algebra of the gauge group and $R$ the representation of the quarks (given in terms of a Young diagram or dimension of the representation). The global form of the gauge group $G$ is arbitrary, as long as it admits $R$ as a representation. $q\in \mathbb Z$ is the charge under the $\mathfrak u(1)$ gauge group factor. The left columns include adjoint QCD for arbitrary gauge group and families of theories for the classical groups, while the right columns contain isolated theories. Gapped QCD theories with classical gauge groups must have quarks transforming in rank-one or rank-two representations, any other representation leading to a gapless theory. The bottom entry indicates an arbitrary tensor product of gapped theories $(\mathfrak g_i,R_i)$ constructed from the entries in the table that have a $U(1)$ gauge group factor. These theories are coupled together via the $\mathfrak u(1)$ matrix of charges $\{\vec q_i\}$, which must be non-singular (see section~\ref{sec:class_gap_QCD} for details).}
\label{tab:classification_gapped_algebra}
\end{table}

Remarkably, there exist chiral QCD theories that are gapped. The complete classification of chiral gapped QCD theories is given in table \ref{tab:classification_gapped_chiral}. From the classification of vector-like gapped theories in table \ref{tab:classification_gapped_algebra}, we can construct chiral gapped theories in two ways:
\begin{itemize}
\item A chiral QCD theory $(G;R_\ell,R_r)$ is gapped if and only if $(R_\ell,R_r)=(\sigma_\ell\cdot R,\sigma_r\cdot R)$ and the vector-like theory $(G;R,R)$ appears in table~\ref{tab:classification_gapped_algebra}, where $\sigma_\ell,\sigma_r$ are outer automorphisms of $\mathfrak g$. Here $\sigma\cdot R$ denotes the action of $\sigma$ on $R$.\footnote{See table~\ref{tab:comarksone} for a list of the automorphisms of simple Lie algebras. If $G$ contains a $U(1)$ factor then $\sigma$ can also be chosen to act on the $U(1)$ charge by reversing its sign.} An example of such a chiral gapped theory is
\begin{equation}
(Spin(8);\boldsymbol8_v,\boldsymbol 8_c) \,,
\end{equation}
corresponding to the triality automorphism acting on the vector-like gapped theory $(Spin(8);\boldsymbol 8_v,\boldsymbol 8_v)$ that appears in table~\ref{tab:classification_gapped_algebra}.

\item A chiral gapped QCD theory can be constructed by taking arbitrary tensor products of the basic gapped theories with quarks in a complex representation (there are seven such entries in table~\ref{tab:classification_gapped_algebra}). The theories are coupled via the integral matrices $q_{\ell}$ and $q_{r}$ that specify the charges under the $\mathfrak u(1)$ gauge group factors for the left and right chiral quarks. In order for theory to be gapped these matrices must be non-singular. A concrete example of such a chiral gapped theory is
\begin{equation}
\prod_i U(n_i)~\text{with quarks}~ R=\bigoplus_i(\boldsymbol1,\!...,\ydiagram 1_i,\!...,\boldsymbol1)_{\vec q_{\ell,i},\vec q_{r,i}}\,,
\end{equation}
corresponding to the tensor product of the vector-like gapped theories $(U(n_i);\ydiagram 1_q,\ydiagram 1_q)$, coupled via their $U(1)$ gauge subgroups.

\end{itemize}

\begin{table}[t]
\centering
\setlength{\tabcolsep}{1em} 
{\renewcommand{\arraystretch}{1.3}
\begin{tabular}{>{$}c<{$}|>{$}c<{$}}
\mathfrak g & (R_\ell,R_r)\\ \hline
\mathfrak g&(\sigma_\ell\cdot R,\sigma_r\cdot R)\\
\bigoplus_i \mathfrak g_i & \bigoplus_i(\boldsymbol1,\dots,(\sigma_{\ell,i}\cdot R_i,\sigma_{r,i}\cdot R_i),\dots,\boldsymbol1)_{\vec q_{\ell,i},\vec q_{r,i}}
\end{tabular}
}
\caption{Classification of chiral gapped QCD theories. Here $\mathfrak g$ and $R$ label the vector-like gapped theories from table~\ref{tab:classification_gapped_algebra}. $\sigma_{\ell, i},\sigma_{r,i}$ denote outer automorphisms of $\mathfrak g_i$ (such as complex conjugation for simply-laced groups, or triality for $\mathfrak{so}(8)$). $\vec q_{\ell,i},\vec q_{r,i}$ are tuples of charges for $\mathfrak u(1)$ factors. These charge matrices must have trivial kernel and cancel gauge anomalies, but are otherwise arbitrary (see section~\ref{sec:class_gap_QCD} for details).}
\label{tab:classification_gapped_chiral}
\end{table}

Having established which QCD theories are gapped and which are gapless, our next goal is to put forward the explicit low energy description of all QCD theories. In the gapped case this means finding the specific topological degrees of freedom carried by the vacua, the infrared TQFT, and in the gapless case finding the specific massless degrees of freedom of the infrared CFT.
 
Determining the long distance description of a given QCD theory is nontrivial. Unlike the question of whether a QCD theory is gapped or gapless, which can be answered rigorously, the task of finding the specific infrared degrees of freedom requires some guesswork. The most natural and straightforward conjecture is that the infrared description of $(G;R_\ell,R_r)$ is given by the $g^2\rightarrow \infty$ limit of the QCD Lagrangian, since $g$ has mass dimension. This limit yields the gauged WZW description of a CFT with the following left and right chiral algebras
\begin{equation}\label{eq:intro_conjecture}
\frac{SO(\dim(R_\ell))_1}{G_{I(R_\ell)}}\times\frac{SO(\dim(R_r))_1}{G_{I(R_r)}}\,.
\end{equation}
This coset can be shown to be a TQFT -- and hence to correspond to a gapped QCD theory -- if and only if $(G;R_\ell,R_r)$ is in table~\ref{tab:classification_gapped_algebra} or~\ref{tab:classification_gapped_chiral}. In this sense, our results derived from the Hamiltonian analysis are perfectly consistent with the conjectured infrared description. One can study many interesting aspects of QCD beyond the existence of a gap using the infrared coset description~\eqref{eq:intro_conjecture}. For example, many well-known CFTs, such as minimal models, emerge in the infrared of QCD, which allows us to map non-trivial questions about the dynamics of QCD theories into questions about these CFTs, which can then be answered explicitly.

If a QCD theory has continuous chiral symmetries, then the infrared CFT necessarily contains a Wess--Zumino--Witten (WZW) factor for these symmetries; this sector carries the corresponding perturbative 't Hooft anomalies for the continuous symmetries. Even in the absence of continuous chiral symmetries, the infrared CFT is nontrivial when the QCD theory is gapless, and is based on a chiral algebra without spin one currents.\footnote{For example, the QCD theories with $G=SU(2)$ with a single quark in a spin $j\in \mathbb Z$ representation, that is $(SU(2), j,j)$, has an infrared chiral algebra given by the W-algebra $\mathcal W(2,4,\ldots, 2j)$. For $j=1,2$ the theory is gapped and flows to a TQFT, while for $j=3$ the spin $4$ and $6$ currents become null and the chiral algebra is the Virasoro algebra, and the theory flows to the fermionic tricritical Ising model.} Interestingly, 't Hooft anomaly matching predicts the existence of some hitherto unknown 't Hooft anomalies for discrete global symmetries in these CFTs, such as nonperturbative anomalies for time-reversal symmetry and discrete chiral symmetries.

As an example, the proposal implies that QCD with a classical gauge group and with fundamental quarks flows in the infrared to a WZW CFT:
\begin{equation}
\begin{aligned}
SU(N)+N_F\,\ydiagram1\quad&\xrightarrow{\text{ infrared }}\quad U(N_F)_N\ \text{WZW}\\
SO(N)+N_F\,\ydiagram1\quad&\xrightarrow{\text{ infrared }}\quad SO(N_F)_N\ \text{WZW}\\
Sp(N)+N_F\,\ydiagram1\quad&\xrightarrow{\text{ infrared }}\quad Sp(N_F)_N\ \text{WZW}\,.
\end{aligned}
\end{equation}
These theories indeed carry the 't Hooft anomalies for the continuous flavor symmetries. Moreover, using that the 't Hooft anomalies must match predicts that these WZW models are endowed with several non-trivial global anomalies, which can be exhibited by general arguments or by brute-force computation in specific examples. For instance, the renormalization group flow predicts that the $SO(N_F)_N$ WZW model has a global anomaly associated with time-reversal symmetry, with $\mathsf T^2=(-1)^F$, which takes the value $NN_F\mod2$. Many other such examples can be constructed, leading to a wealth of CFTs in the infrared and 't Hooft anomalies thereof.

The plan for the rest of the paper is as follows. In section~\ref{sec:QCD2} we set the stage by carefully analyzing the microscopic description of QCD, its free parameters, topological sectors, and gauge anomalies. In section~\ref{sec:symmanom} we begin our investigations of the mass gap problem; in particular, we exploit the symmetries and 't Hooft anomalies of $2d$ theories to constrain as much as possible the theories that can potentially be gapped. We find several simple criteria that automatically force the theory to be gapless, thus considerably reducing the landscape of gapped theories. These criteria alone are not enough to actually prove that a given theory is gapped, so in section~\ref{sec:DLCQ} we turn our attention to an explicit analysis of the Hamiltonian of QCD. We recover the necessary conditions laid out in the previous section, and also find sufficient conditions as well, culminating in a concrete list of gapped theories. In section~\ref{sec:Infrareddynamnics} we reconsider our results, this time in light of the conjecture~\eqref{eq:intro_conjecture} which proposes a concrete description of QCD at low energies. We give further evidence for the correctness of this conjecture, and subsequently apply it to many explicit examples.

We also include several appendices. Appendix~\ref{app:conven} can be used as a reference for our conventions, and it contains some technical computations that supplement the main text. Appendix~\ref{app:coset_cft} reviews some relevant facts about $2d$ CFTs and, in particular, cosets of the form~\eqref{eq:intro_conjecture} that conjecturally encapsulate the low-energy degrees of freedom of QCD. We also work out a few examples in some detail. Finally, appendix~\ref{sec:abelian_G} contains a separate discussion of QCD theories where the gauge group is abelian, i.e., where $G$ consists of factors of $U(1)$ only. While this type of theories is covered by our general discussion from other sections, when studied in isolation one can be more explicit in some of our claims. Also, they illustrate some general features of other non-abelian QCD theories that contain $U(1)$ factors, such as the breaking of some $U(1)$ flavor symmetries as the result of flavor-gauge mixed anomalies.

\section{$\boldsymbol{2d}$ QCD theories}
\label{sec:QCD2}

The field content of a $2d$ QCD theory is specified by a choice of gauge group $G$ and a pair of representations $R_\ell$ and $R_r$ of $G$ acting on left and right chiral quarks.\footnote{In $2d$, complex conjugation does not reverse the chirality of a fermion since the chirality matrix $\gamma_3=\gamma^0\gamma^1$ does not include an $i$, unlike in $4d$ where $\gamma_3=i \gamma^0 \gamma^1 \gamma^2 \gamma^3$ and conjugation does flip chirality. This implies that the most general $2d$ QCD theory cannot be written using just left chiral fermions, in contrast to $4d$.} 
We label such a QCD theory by the triple $(G;R_\ell, R_r)$. $G$ is an arbitrary compact, connected Lie group with Lie algebra $\mathfrak{g}=\oplus_I \mathfrak g_I \oplus_m \mathfrak u(1)_m$, a direct sum of simple Lie algebras $\mathfrak g_I$ and abelian Lie algebras ${\mathfrak u}(1)_m$. The Lagrangian of a QCD theory with massless quarks is
\begin{equation}
{\mathcal L}_{\text{QCD}}=
-\frac{1}{2}\tr(g^{-2}F_{\mu\nu}F^{\mu\nu})+i \psi^\dagger_\ell D_-\psi_\ell+i \psi^\dagger_rD_+\psi_r\,,
\label{eq:QCD2}
\end{equation}
where 
\begin{equation}
D_-\psi_\ell=(\partial_--i A^a_- t^a_\ell)\psi_\ell\,,\qquad\qquad 
D_+\psi_{r}=(\partial_+-i A^a_+ t^a_{r})\psi_{r}\,,
\end{equation}
$t^a_\ell$ ($t^a_r$) are the generators of the Lie algebra $\mathfrak{g}$ in the representation $R_\ell$ ($R_{r}$) and we have introduced lightcone coordinates $x^\pm=\frac{1}{\sqrt2}(x^0\pm x^1)$, and $A^a_{\pm} = \frac{1}{2}(A^a_0 \pm A^a_1)$. Each gauge group factor has a gauge coupling, which is captured by $g^{-2}$ inside the trace. See appendix~\ref{app:conven} for details and conventions. See also table~\ref{tab:comarksone} for a summary of simple Lie groups and relevant properties.

\begin{table}[h!]
\begin{equation*}
\hspace{-6pt}\begin{array}{|c|c|c|c|c|c|c|c|c|c|c|}\hline
G_\text{sc}&SU(N)&Sp(N)&Spin(2N+1)&Spin(4N)&Spin(4N+2)&E_6&E_7&E_8&F_4&G_2\\\hline
h&N&N+1&2N-1&4N-2&4N&12&18&30&9&4\\\hline
\operatorname{Out}(\mathfrak g)&\mathbb Z_2&\cdot &\cdot & \mathbb Z_2&\mathbb Z_2&\mathbb Z_2&\cdot &\cdot &\cdot &\cdot \\\hline
Z(G_\text{sc})&\mathbb Z_N&\mathbb Z_2&\mathbb Z_2&\mathbb Z_2\times \mathbb Z_2&\mathbb Z_4&\mathbb Z_3& \mathbb Z_2&\cdot &\cdot &\cdot
\\ \hline
\end{array}
\end{equation*}
\caption{Lie data for the simply-connected simple Lie groups $G_\text{sc}$. Here $h$ denotes the dual Coxeter number (defined as the Dynkin index of the adjoint representation). $\operatorname{Out}(\mathfrak g)$ is the group of outer automorphisms of $\mathfrak g$, which corresponds to charge conjugation symmetry of QCD. $Z(G_\text{sc})$ is the center of the gauge group, which contains the one-form center symmetry of QCD. For $SU(2)$, $\operatorname{Out}(\mathfrak g)$ is trivial, and for $Spin(8)$, it is enhanced to $\operatorname{Out}(\mathfrak g)=\mathbb S_3$ (triality).}
\label{tab:comarksone}
\end{table}

{\begin{center}\underline{\textbf{Global issues, flux tubes and theta terms}.} \end{center}}

A QCD theory requires specifying a global choice of a gauge group $G$ with Lie algebra $\mathfrak{g}$. We consider first QCD with the simply-connected form of the gauge group $G_{\text{sc}}$, which we denote by $(G_\text{sc}; R_\ell, R_r)$. Such a QCD theory may have a one-form symmetry $\Gamma$~\cite{Gaiotto:2014kfa}, where $\Gamma\subseteq Z(G_\text{sc})$ is a subgroup of the center (cf.~table~\ref{tab:comarksone}). 

$2d$ QFTs with a one-form symmetry $\Gamma$ have topological sectors labeled by a representation $\rho\in \Gamma^\vee$ of $\Gamma$, where $\Gamma^\vee$ is the Pontryagin dual group. Physically, a topological sector labeled by $\rho\in \Gamma^\vee$ describes the theory in the presence of a flux tube created by a quark-antiquark pair of charge $\rho$ at $\pm$-infinity~\cite{Witten:1978ka,Coleman:1976uz}, a background that preserves Poincar\'e invariance in $2d$. 

We now consider the theory with gauge group $G=G_{\text{sc}}/\Gamma$.\footnote{The discussion can be easily extended to the case $G_{\text{sc}}/K$, where $K\subset \Gamma$.} Since $G$-bundles are classified by $ H^2(M, \pi_1(G))\cong \Gamma$, the sum over gauge fields in the functional integral can be weighted by a discrete theta term labeled by $\rho\in \Gamma^\vee$, which takes the form of a generalized Stiefel-Whitney  class
 \begin{equation}
 i\int_M w_\rho(G)\,.
 \label{thetaterm}
\end{equation}
We label such a QCD theory by $(G; R_\ell, R_r)_\rho$. 

We proceed to prove that:
 \begin{mdframed}
$\bullet$ $(G_\text{sc}; R_\ell, R_r)$ with a $\rho$-flux tube is the same as $(G; R_\ell, R_r)_\rho$
\end{mdframed}

The one-form global symmetry $\Gamma$ of $(G_\text{sc}; R_\ell, R_r)$ implies that there is a topological local operator $U_g$, with $g\in \Gamma$, which acts on line operators.\footnote{The charge of a line operator $\mathcal L$ under $\Gamma$ is measured by $U_g$ as
\begin{equation}
U_g\, \mathcal L\, U_g^{-1}=\chi_{\rho}(g)\mathcal L\,,
\label{chargeline}
\end{equation}
where $\rho\in \Gamma^\vee$ is an irreducible representation of $\Gamma$ and $\chi_\rho(g)$ is a character of $\Gamma$ in the representation $\rho$. This means that the spectrum of line operators in the theory can be organized according to their charges under $\Gamma$ as 
\begin{equation}
[\mathcal L]=\bigoplus_{\rho \in \Gamma^\vee} [ \mathcal L]_{\rho}\,,
\end{equation}
where line operators in $[ \mathcal L]_{\rho}$ carry charge $\rho$.} Diagonalizing the topological local operators $U_g$ on the Hilbert space leads to the decomposition 
\begin{equation}
\mathcal H=\bigoplus_{\rho\in \Gamma^\vee} \mathcal H_\rho\,,
\label{decomposition}
\end{equation}
where $\rho$ is an irreducible representation of $\Gamma$ and
\begin{equation}
|\psi\rangle \in {\mathcal H}_{\rho}\quad\Longleftrightarrow \quad U_g |\psi\rangle= \chi_\rho(g)|\psi\rangle\,.
\end{equation}
The Hilbert space $\mathcal H_\rho$ corresponds to $(G_\text{sc}; R_\ell, R_r)$ in the presence of a $\rho$-flux tube.

The QCD theory $(G; R_\ell, R_r)_\rho$ can be constructed by gauging the one-form symmetry of $(G_\text{sc}; R_\ell, R_r)$ tensored with an SPT phase for the one-form symmetry $\Gamma$; such SPTs are labeled by an element $\rho$ of the reduced cobordism group $\tilde \Omega^2_\text{Spin}(B^2\Gamma)\cong \Gamma^\vee$, where $B^2\Gamma$ denotes the second Eilenberg-MacLane space of $\Gamma$. A nontrivial SPT$_\rho$ weights the sum over $g\in \Gamma$ that defines $(G; R_\ell, R_r)_\rho$ by gauging $\Gamma$ with the phase
 \begin{equation}
 \chi^*_\rho(g)\,,
 \end{equation}
where $\rho \in \Gamma^\vee$ is a representation of $\Gamma$ and $ \chi_\rho(g)$ is a character of $\Gamma$ in the representation $\rho$. This is an alternative way to think about the discrete theta term~\eqref{thetaterm}.

Consider a theory $T$ in the presence of a fixed two-form gauge field $B_2$ for the one-form symmetry $\Gamma$, which takes values in $H^2(M,\Gamma)=\Gamma$. The partition function of the theory in such a background is given by 
 \begin{equation}
 Z_{T}[g]= \sum_{\rho\in \Gamma^\vee} Z_T(\rho) \chi_\rho(g)\,,
 \label{fixedB}
 \end{equation}
 where the sum over $\rho$ is due to the Hilbert space structure~\eqref{decomposition} and $g\in \Gamma$ labels the choice of background gauge field $B_2$. $Z_T(\rho)$ is the partition function of the theory in the presence of a $\rho$-flux tube. 
 
The theory $T/\Gamma$ obtained by gauging $\Gamma$ has a dual $(-1)$-form symmetry $\Gamma^\vee$~\cite{Yu:2020twi,Cordova:2019jnf,Sharpe:2019ddn}, and $T/\Gamma$ can be coupled to a background zero-form gauge field for this symmetry, which corresponds to an element $\hat\rho$ of $\Gamma^\vee$. The partition of the gauged theory in the presence of this background gauge field is 
\begin{equation}
 Z_{T/\Gamma}[\hat \rho]=\frac{1}{|\Gamma|}\sum_{g\in \Gamma} Z_T[g] \chi^*_{\hat\rho}(g)\,,
 \end{equation}
 where $\chi^*_{\hat\rho}(\hat g)$ encodes the coupling of the background two-form gauge field for $\Gamma$ with the zero-form gauge field for $\Gamma^\vee$. This describes the theory $T/\Gamma$ with a discrete theta term labeled by $\hat \rho \in \Gamma^\vee$. Using equation~\eqref{fixedB} we arrive at 
\begin{equation}
 Z_{T/\Gamma}[\hat \rho]=\frac{1}{|\Gamma|}\sum_{g\in \Gamma}\sum_{\rho\in \Gamma^\vee} Z_T(\rho)\, \chi_\rho(g)\chi^*_{\hat\rho}(g)=Z_T(\hat \rho)\,,
 \end{equation}
 where we have used that $\sum_{g\in \Gamma} \chi_\rho(g)\chi^*_{\hat\rho}(g)=|\Gamma|\delta_{\rho,\hat \rho}$. Therefore, the partition function of the theory $T/\Gamma$ with a theta term $\rho \in \Gamma^\vee$ is the same as the partition function of the original theory $T$ in the sector with a with a $\rho$-flux tube, thus completing the proof that $(G_\text{sc}; R_\ell, R_r)$ with a $\rho$-flux tube is the same as $(G; R_\ell, R_r)_\rho$. This implies that it is sufficient to study QCD theories with a simply connected gauge group, which we will do henceforth. 

We now turn to the next result
 \begin{mdframed}
$\bullet$ $(G_\text{sc}; R_\ell, R_r)$ is gapless if and only if it is gapless in the $\rho=0$ flux tube sector. 
\end{mdframed}
This implies that for the purposes of classifying gapped QCD theories it suffices to consider QCD theories with simply connected gauge group and in the trivial flux tube sector. 

This conclusion is a consequence of the fact that the (massless) QCD theory $(G_\text{sc}; R_\ell, R_r)$ admits topological line operators ${\mathcal L}$ that carry any charge under the one-form symmetry $\Gamma$~\cite{Komargodski:2020mxz}.\footnote{The global symmetries of the QCD Lagrangian~\eqref{eq:QCD2} together with these topological lines make it technically natural to study the theory without four-fermi terms, cf.~\cite{Cherman:2019hbq,Komargodski:2020mxz}.} Since ${\mathcal L}$ carries one-form symmetry charge $\rho$, it defines a map between the Hilbert spaces ${\mathcal H}_{\rho=0}$ and ${\mathcal H}_{\rho}$: acting with ${\mathcal L}$ on ${\mathcal H}_{\rho=0}$ creates states in ${\mathcal H}_{\rho}$. Physically, acting with a topological line ${\mathcal L}$ inserts static probe charges $\rho$ at $\pm$-infinity. Such a topological line operator ${\mathcal L}$ interpolates between the Hamiltonian of the theory in distinct flux tube sectors
\begin{equation}
\mathcal L H_{\rho=0} =H_\rho \mathcal L\,,
\label{simili}
\end{equation}
where $H_{\rho=0}$ and $H_\rho$ are the Hamiltonians of the theory in the trivial and $\rho$-flux tube sector respectively. Note that in general ${\mathcal L}$ is a non-invertible topological operator and therefore~\eqref{simili} cannot be written as a similarity transformation. Since ${\mathcal L}$ is topological it carries vanishing energy density (zero tension). Therefore it cannot lower the energy and the sector with a $\rho$-flux tube is gapless if and only it is gapless in the sector with no string (that is with $\rho=0$).\footnote{If $|\Omega\rangle$ is the ground state of $H_{\rho=0}$, the ground state of $H_\rho$ is ${\mathcal L}|\Omega\rangle$.} We note that this conclusion relies on the existence of topological line operators, and these are not present generically in the theory with massive quarks, where indeed a massless particle can appear in the theory with a flux tube (see e.g.~\cite{Kutasov:1993gq}).

In summary, for the purposes of classifying all gapped QCD theories we can, without loss of generality, consider the theory with simply connected gauge group in the trivial topological sector, without a string. Henceforth, we will use $G$ to refer to the simply connected gauge group or use instead the Lie algebra $\mathfrak{g}$.

{\begin{center}\underline{\textbf{Gauge anomaly cancellation}.} \end{center}}

In order to define a consistent QCD theory, the global symmetry $G$ acting on the free fermions in the deep ultraviolet must have no obstructions to being gauged. Therefore all anomalies for $G$ gauge transformations, perturbative and nonperturbative, must cancel. Perturbative anomalies, that is, anomalies associated to $G$ gauge transformations connected to the identity, are classified by the first summand in the free part of the spin cobordism group
\begin{equation}
\operatorname{Free}\bigl(\Omega_{\text{spin}}^4(BG)\bigr)=\mathbb Z^{|I|+\frac12|m|(|m|+1)}\oplus \mathbb Z\,,
\label{eq:anompert}
\end{equation}
where $|I|$ and $|m|$ is the number of simple and abelian factors in ${\mathfrak g}$ respectively. These anomalies are determined by a one-loop diagram and encoded in the first line of the anomaly polynomial 
\begin{equation}
\begin{aligned}
\bigl(\tr_{R_\ell} e^{F/2\pi}-\tr_{R_r} e^{F/2\pi}\bigr)\hat A(\mathcal R)\bigr|_4&= \sum_{a,b} \bigl[\tr(t^a_\ell t^b_\ell)-\tr(t^a_r t^b_r)\bigr] \frac{F^a\wedge F^b}{8\pi^2}\\[+2pt]
&\quad-\frac{p_1(\mathcal R)}{24}\bigl[\dim(R_\ell)-\dim(R_r) \bigr]\,,
\label{eq:anomP}
\end{aligned}
\end{equation}
where $F^a$ is the two-form field strength and $p_1(\mathcal R)$ the first Pontryagin class for the background metric. Gauge anomaly cancelation requires that the representations $R_\ell$ and $R_r$ of the left and right chiral quarks obey 
\begin{equation}
\tr(t^a_\ell t^b_\ell)=\tr(t^a_r t^b_r)~~\qquad \forall~ a,b\,.
\label{eq:anomalies}
\end{equation}
The nontrivial anomaly constraints in~\eqref{eq:anomalies} are:\footnote{This is to be contrasted with the $4d$ anomaly cancelation equation $\tr(t_\ell^a \{t_\ell^b,t_\ell^c\})=0$ when the theory is written using left chiral fermions, which is nontrivial for the ${\mathfrak g}_I$-${\mathfrak g}_I$-${\mathfrak g}_I$, ${\mathfrak g}_I$-${\mathfrak g}_I$-$\mathfrak u(1)_m$ and $\mathfrak u(1)_m$-$\mathfrak u(1)_n$-$\mathfrak u(1)_p$ anomalies.
 In $2d$ a chiral fermion in any irreducible representation of any ${\mathfrak g}$ contributes to the ${\mathfrak g}_I$-${\mathfrak g}_I$ anomaly, while in $4d$ only chiral fermions transforming in a complex representation of $SU(N)$ contribute to the pure ${\mathfrak g}_I$-${\mathfrak g}_I$-${\mathfrak g}_I$ anomaly, because the rest of the simple Lie algebras have no cubic Casimir. Since a chiral fermion cannot be given a mass in $2d$, unlike for a $4d$ chiral fermion in a real representation, any $2d$ chiral fermion can potentially contribute to the anomaly and, indeed, it does.}
\begin{enumerate}
\item ${\mathfrak g}_I$-${\mathfrak g}_I$ anomaly: $t^a_{\ell,r}$ are generators of the simple Lie algebra ${\mathfrak g}_I$. The anomaly cancelation condition requires that 
\begin{equation}\label{eq:UV_anomaly_simple}
I(R_\ell)-I(R_r)=0\,,
\end{equation}
where $I(R)$ is the Dynkin index of the representation $R$, defined by $\tr(t^at^b)=I(R) \delta^{ab}$. The index of a reducible representation follows from $I(R_1\oplus R_2)=I(R_1)+I(R_2)$.
\item $\mathfrak u(1)_m$-$\mathfrak u(1)_n$ anomaly: $t^a_{\ell,r}$ are generators of an abelian Lie algebra. The anomaly cancelation condition is 
\begin{equation}\label{eq:abelian_gauge_anomaly_cancel}
\sum_\ell Q_{\ell,m} Q_{\ell,n}-\sum_r Q_{r,m}Q_{r,n}=0\,,
\end{equation}
where $Q_{\ell,m}$ and $Q_{r,m}$ are the left and right $U(1)_m$ charges of the quarks. 
\end{enumerate} 

A global symmetry $G$ may have a more subtle obstruction to being gauged associated to a background $G$ gauge transformation not connected to the identity, like the celebrated $SU(2)$ global anomaly in $4d$~\cite{WITTEN1982324}. If the symmetry group $G$ is gauged, like in QCD, global anomalies for $G$ must also cancel for the gauge theory to be consistent. Topologically nontrivial gauge transformations in (compactified) $2d$ flat spacetime are classified by $\pi_2(G)$, which vanishes for any continuous Lie group $G$, and $2d$ gauge theories do not have this type of global anomalies. From the cobordism point of view of anomalies, the vanishing of the anomalies is seen through the fact that $\Omega^3_{\text{spin}}(BG)=0$ (see e.g.~\cite{Wan:2018bns}).\footnote{Global anomalies for a discrete symmetry group can be nontrivial. For example $\Omega_{\text{spin}}^3(B\mathbb Z_2)=\mathbb Z_8$.} Therefore the anomaly cancelation conditions~\eqref{eq:anomalies} are necessary and sufficient for a QCD theory to be consistent.

Since gravity couples to QCD as a nondynamical background field, it can be afflicted by gravitational anomalies without rendering the theory inconsistent. These anomalies are captured by the second $\mathbb Z$ summand in~\eqref{eq:anompert} and by the second line of the anomaly polynomial in~\eqref{eq:anomP}.\footnote{A mixed ${\mathfrak u}(1)$-$\text{gravity}$ anomaly governed by $\nabla_\mu J^\mu =\alpha R$ can be written down, where $J^\mu$ is the ${\mathfrak u}(1)$ current. But $\alpha=0$ in a unitary theory. It can be nonvanishing in a nonunitary theory, like in the string theory $bc$ ghost system. Thus there are are no mixed gauge-gravity anomalies in $2d$ QCD.} We discuss in the next section the implications that 't Hooft anomalies, including gravitational anomalies, have for the infrared dynamics of QCD theories.

Of course, vector-like theories $(G;R,R)$, with $R_\ell=R_r$, are manifestly free of gauge anomalies. But in $2d$, gauge-anomaly-free chiral QCD theories are abundant. Most of these chiral theories, however, have gravitational anomalies. There are, nonetheless, chiral gauge theories with neither gauge nor gravitational anomalies, i.e., simultaneous solutions to\footnote{A simple example of a chiral theory with no gravitational anomalies is $(Spin(5);\boldsymbol{35},\boldsymbol{5}+\boldsymbol{30})$. As a matter of fact, this theory has no continuous flavor symmetries, so it does not have any perturbative 't Hooft anomalies whatsoever.} 
\begin{equation}
\begin{aligned}
\tr(t^a_\ell t^b_\ell)&=\tr(t^a_r t^b_r)~~\qquad \forall~ a,b\\[+2pt]
\dim(R_\ell)&= \dim(R_r)\,.
\label{eq:cancelable}
\end{aligned}
\end{equation} 
Unlike in $4d$, where the beta-function for the gauge coupling constrains the quark content of $4d$ QCD theories that are strongly coupled in the infrared, any $2d$ QCD flows to strong coupling at low energies. Our first goal is to determine which $2d$ QCD theories are gapped, and which are gapless.

\section{Symmetries, 't Hooft Anomalies and Gaplessness}
\label{sec:symmanom}
 
In this section we use symmetry and 't Hooft anomaly considerations to derive necessary conditions for a $2d$ QFT theory to be gapped. We start with a discussion of symmetries and 't Hooft anomalies and then use them to constraint the phases of $2d$ QFTs.

Symmetries provide a powerful organizing principle parametrizing the most general solution of a QFT consistent with the symmetries. But without further input, either perturbative or nonperturbative, symmetries do not inform the actual dynamics of a physical system.

An 't Hooft anomaly for a global symmetry, diagnosed by violations of Ward identities in the presence of nondynamical background gauge fields for global symmetries, instead, does inform the dynamics of the system. Since 't Hooft anomalies are quantized, they are invariant under symmetric deformations, and define invariants in the space of symmetric QFTs. In particular they are invariant under renormalization group transformations. While 't Hooft anomalies alone cannot determine the dynamics of a system, they rule out any dynamical scenario that does not match the microscopic 't Hooft anomalies. As such, 't Hooft anomalies provide nonperturbative guidance about the dynamics of QFTs. 
 
A system defined at short distances with an 't Hooft anomaly for a symmetry cannot flow in the deep infrared to a trivially gapped theory, as this has vanishing 't Hooft anomalies. A system with an 't Hooft anomaly can flow either to a symmetry-preserving gapless phase or a symmetry breaking phase, which is gapless if the broken symmetry is continuous\footnote{More precisely, the anomaly is not torsion.} and a TQFT if the broken symmetry is discrete.\footnote{More precisely, the anomaly is torsion.} If the anomalous symmetry is discrete, the system may also flow to a symmetry preserving gapped phase described by a TQFT with topological order, which can saturate anomalies that are torsion classes.
 
A system with an 't Hooft anomaly for a continuous symmetry cannot flow to a TQFT because an 't Hooft anomaly for a continuous symmetry implies a nonvanishing correlation function for conserved currents at separated points, and a TQFT, being topological, does not have such correlation functions. This implies that a system with perturbative anomalies, corresponding to anomalies for continuous symmetry transformations connected to the identity, can only flow to a symmetry preserving gapless phase or a symmetry breaking gapless phase.

In $2d$, the fate of a system with an 't Hooft anomaly is further constrained by important theorems. These theorems, once combined with the discussion above, leads to the following implications:
\begin{enumerate}
\item Coleman-Mermin-Wagner theorem~\cite{PhysRevLett.17.1307,Coleman:1973ci}: a continuous global symmetry cannot be spontaneously broken in $2d$.
\begin{mdframed}
 A $2d$ system with an 't Hooft anomaly for a continuous symmetry must flow to a symmetry preserving gapless phase.
\end{mdframed}

\item A $2d$ TQFT does not have intrinsic topological order~\cite{PhysRevB.84.235128}: in $2d$ a symmetry preserving gapped phase cannot saturate 't Hooft anomalies.
\begin{mdframed}
 A $2d$ system with an 't Hooft anomaly for a discrete symmetry must flow to a symmetry preserving gapless phase or a symmetry breaking gapped phase described by a TQFT.
\end{mdframed}

\end{enumerate}
 
We are now ready to state the following far-reaching result for the dynamics of $2d$ QFTs:

\begin{mdframed}
\begin{lemma}\label{lm:ch_sym_gap}\normalfont 
A $2d$ QFT with a continuous chiral global symmetry is symmetry preserving and gapless.
\end{lemma}
\end{mdframed}

Consider a QFT with a $U(1)$ global symmetry. The one-form current for the $U(1)$ global symmetry is $J = J_\mu\,\mathrm dx^\mu\equiv J_+\mathrm dx^++J_-\mathrm dx^-$, with $J_{\pm} = \frac{1}{\sqrt{2}}(J_0 \pm J_1)$. This obeys the conservation equation (see appendix~\ref{app:conven} for conventions)
\begin{equation}
\partial_- J_++\partial_+ J_-=0\,.
\end{equation}
This is an operator equation that holds inside any correlation function as long as the location of the current $J$ does not coincide with any operator insertions (conservation may fail at coincident points). Poincar\'e's lemma implies that locally the current takes the following form 
\begin{equation}
J_\mu=\epsilon_{\mu \nu}\partial^\nu\phi \Longleftrightarrow J_\pm=\pm \partial_\pm \phi(x^+,x^-)\,,
\label{eq:currentcons}
\end{equation}
where $\phi(x^+,x^-)$ is a scalar operator in the theory. 

Consider now a chiral symmetry. A right-moving $U(1)_r$ symmetry is implemented by a conserved current that is antiselfdual
\begin{equation}
U(1)_r:\, J=-{\star} J\Longleftrightarrow J_+=0\Longleftrightarrow \partial_+J_-=0\,.
\end{equation}
By virtue of~\eqref{eq:currentcons}, a theory with a $U(1)_r$ symmetry contains a scalar operator $\phi$ that is right-moving 
\begin{equation}
 J_+=0\Longrightarrow \phi=\phi(x^-)\,.
\end{equation}
Likewise, a left-moving $U(1)_\ell$ symmetry is generated by a conserved current that is selfdual
\begin{equation}
U(1)_\ell:\, J=+{\star} J\Longleftrightarrow J_-=0 \Longleftrightarrow \partial_- J_+=0\,,
\end{equation}
and a $U(1)_\ell$ symmetry implies the existence of a left-moving scalar operator 
\begin{equation}
J_-=0\Longrightarrow \phi=\phi(x^+)\,.
\end{equation}
This implies that a QFT with either a left or a right moving $U(1)$ symmetry is necessarily gapless: the theory has a chiral scalar operator
that creates chiral massless states when acting on the vacuum. 
 
This lemma may seem at odds with our discussion above since, typically, symmetries alone cannot determine the infrared phase of a system.
The reason that it does in this case is that a chiral $U(1)$ symmetry automatically leads to an 't Hooft anomaly for that symmetry, as we show below. And as we mentioned above, an 't Hooft anomaly for a continuous global symmetry in $2d$ necessarily leads to a symmetry preserving gapless phase.

Consider the renormalization group flow out of a CFT with a $U(1)_\ell$ symmetry that is triggered by a $U(1)_\ell$-invariant relevant operator.\footnote{In QCD, the CFT in the ultraviolet is the CFT of free fermions, and the renormalization group flow is triggered by the gauge coupling.} Since the flow preserves the $U(1)_\ell$ symmetry, the most general two-point function for the $U(1)_\ell$ current consistent with dimensional analysis and Poincar\'e invariance is 
\begin{equation}
 \langle J_+(x) J_+(0)\rangle=\frac{K_\ell(x^+x^-/\mu^2)}{x^+x^+}\,, 
\end{equation}
where $\mu$ is a scale generated along the renormalization group flow. In a unitary theory $K_\ell\geq 0$, with $K_\ell=0$ if and only if $J_+=0$.
Demanding conservation law of $U(1)_\ell$ symmetry current at separated points $\partial_- J_+=0$ implies that $K_\ell$ is a renormalization group invariant
\begin{equation}
\frac{\partial}{\partial \rho^2} K_\ell =0\,, 
\end{equation}
where we have introduced Rindler coordinates $x^\pm=\rho\, e^{\pm \sigma}$, so that 
\begin{equation}
\langle J_+(x) J_+(0)\rangle=\frac{k_\ell}{x^+x^+}\,. 
\label{eq:contacttt}
\end{equation}
 In the deep ultraviolet, $K^{\text{UV}}_\ell=k_\ell\in \mathbb Z$ is the level of the $U(1)_\ell$ current algebra of the ultraviolet CFT. Therefore, $k_\ell$ is the 't Hooft anomaly coefficient for the $U(1)_\ell$ symmetry.\footnote{The contact term implied by~\eqref{eq:contacttt} leads to a violation of the conservation equation $\partial_-j_+=\frac{k_\ell}{2\pi} \partial_+ A_-$ upon coupling system to a background gauge field for $U(1)_\ell$ via $\int d^2x A_-J_+$. In QCD, $k_\ell=\sum_i q_{i,\ell}^2$, where $q_{i,\ell}$ are the $U(1)_\ell$ charges of chiral fermions.} Since $k_\ell\neq 0$ implies that correlators have support at separated points~\eqref{eq:contacttt}, and $U(1)_\ell$ cannot be spontaneously broken, the infrared of a system with a $U(1)_\ell$ symmetry must be symmetry preserving and gapless.

This argument admits an interesting generalization. Consider now the renormalization group flow of a $U(1)$-symmetric CFT triggered by a $U(1)$-invariant relevant operator. In a unitary CFT with a normalizable vacuum, the conservation law for the $U(1)$ current $J=J_+\mathrm dx^++J_-\mathrm dx^-$ implies a separate conservation law for chiral $U(1)_\ell$ and $U(1)_r$ symmetries, generated by $J_-$ and $J_+$ respectively~\cite{Affleck:1985jc}. The most general current two-point functions consistent with dimensional analysis and Poincar\'e invariance are 
\begin{equation}
\begin{aligned}
\langle J_+(x) J_+(0)\rangle&=\frac{K_\ell(x^+x^-/\mu^2)}{x^+x^+}\,, \\[+3pt]
\langle J_-(x) J_-(0)\rangle&=\frac{K_r(x^+x^-/\mu^2)}{x^-x^-}\,, \\[+3pt]
\langle J_+(x) J_-(0)\rangle&=\frac{G(x^+x^-/\mu^2)}{x^+x^-}\,.
\end{aligned}
\end{equation}
In a parity invariant system $K_\ell=K_r$. Conservation of the $U(1)$ current $\partial_-J_++ \partial_+J_-=0$ at separated points 
implies that 
\begin{equation}
\rho^2\frac{\partial}{\partial \rho^2}(K_\ell+G )=G\,, \qquad \rho^2\frac{\partial}{\partial\rho^2}(K_r+G )=G\,. 
\label{RGconst}
\end{equation}
Therefore, the quantity $K_\ell-K_r$ is a renormalization group invariant 
\begin{equation}
\rho^2\frac{\partial}{\partial \rho^2}(K_\ell-K_r)=0\,.
\end{equation}
In the deep ultraviolet, $K^{\text{UV}}_\ell=k_\ell\in\mathbb Z$ and $K^{\text{UV}}_r=k_r\in\mathbb Z$ are the levels of the $U(1)_\ell$ and $U(1)_r$ current algebras of the ultraviolet CFT. $K_\ell-K_r=k_\ell-k_r$ is the 't Hooft anomaly coefficient for the $U(1)$ symmetry and is constant everywhere in the flow.\footnote{In QCD $k_\ell-k_r= \tr_{\text{fermions}}(\gamma^3 Q^tQ)$, where $Q$ is the $U(1)$ charge of the fermions.} 
This must be reproduced by be infrared phase, and it can only be realized by a symmetry preserving gapless phase.

\begin{mdframed}
\begin{lemma}\normalfont 
A $2d$ QFT with a gravitational anomaly is gapless.
\end{lemma}
\end{mdframed}

An almost identical reasoning applies to the conservation law of the energy-momentum tensor $T_{\mu\nu}$ along a renormalization group flow out of a CFT, for which we have
\begin{equation}
\partial_-T_{+\pm}+ \partial_+T_{-\pm}=0\,.
\label{eq:consemom}
\end{equation}
In a unitary CFT with a normalizable vacuum, $T_{+-}=0$ and $T_{++}$ and $T_{--}$ are chiral, that is $\partial_-T_{++}=\partial_+T_{--}=0$, so that in the ultraviolet CFT, the correlators with support at separated points are
\begin{equation}
\text{UV CFT:}\quad \langle T_{++}(x) T_{++}(0)\rangle =\frac{c^{\text{UV}}_\ell}{2 (x^+)^4}\,,\qquad \langle T_{--}(x) T_{--}(0)\rangle =\frac{c^{\text{UV}}_r}{2 (x^-)^4}\,,
\end{equation}
where $c^{\text{UV}}_\ell$ and $c^{\text{UV}}_r$ are the central charges of the left and right-moving Virasoro algebras. In a parity invariant theory $c^{\text{UV}}_\ell=c^{\text{UV}}_r$. The quantity $c^{\text{UV}}_\ell-c^{\text{UV}}_r$ detects a gravitational 't Hooft anomaly, and must be matched by the infrared phase.\footnote{This can be derived by imposing energy-momentum conservation law~\eqref{eq:consemom} on the most general two-point functions of $T_{++}, T_{--}$ and $T_{+-}$ at separated points.} While the $c$-theorem~\cite{Zamolodchikov:1986gt} says that $c_\ell$ and $c_r$ decrease along a renormalization group flow, the difference $c_\ell-c_r$ must remain constant. In a theory with a gravitational 't Hooft anomaly the energy-momentum tensor is a nontrivial operator, with separated point correlation functions. Since such correlation functions cannot be realized by a TQFT, a $2d$ theory with a gravitational 't Hooft anomaly is necessarily gapless.

Comparing with the anomaly polynomial~\eqref{eq:anomP}, we have that the gravitational 't Hooft anomaly of the QCD theory $(G;R_\ell,R_r)$ is
\begin{equation}
c^{\text{UV}}_\ell-c^{\text{UV}}_r=\frac12\left(\dim(R_\ell)- \dim(R_r)\right)\,.
\end{equation}
$(G;R_\ell,R_r)$ with a gravitational 't Hooft anomaly in the ultraviolet is gapless.

\subsection{Towards Gapped QCD Theories}
\label{sec:towards}

In the previous section we established that a $2d$ theory with a continuous chiral symmetry is automatically gapless. Therefore, if we wish to classify QCD theories that are gapped, the first step is to determine which QCD theories have no such symmetries. These can be expressed as conditions on the 
quark content of the QCD theory as follows:\footnote{We discuss QCD with a reductive gauge group, that is with abelian gauge group factors, below.}

\begin{mdframed}
\begin{lemma}\normalfont 
A necessary condition for $(G;R_\ell,R_r)$ with semisimple $G$ to be gapped is that the representations $R_\ell$ and $R_r$ of $G$ are the direct sum of distinct, real irreducible representations of $G$. A QCD theory with a quark content that is not of this type is necessarily gapless. 
\end{lemma}
\end{mdframed}

$(G;R_\ell,R_r)$ is obtained by gauging a diagonal subgroup $G$ of the global symmetry acting on the left and right chiral quarks, and giving the gauge fields a kinetic term. For the purpose of identifying the continuous global symmetries of a QCD theory it suffices to discuss the Lie algebra of symmetries. The continuous global symmetry algebra acting on the quarks in the ultraviolet is
\begin{equation}
\mathfrak{so}(\dim(R_\ell))\oplus \mathfrak{so}(\dim(R_r))\,,
\end{equation}
where $\dim(R_{\ell/r})$ is the real dimension of the representation $R_{\ell/r}$ of $\mathfrak{g}$. Our immediate task is to answer for what choices of $R_\ell$ and $R_r$ does a QCD theory admit a continuous chiral global symmetry, and is therefore gapless (cf.~lemma~\ref{lm:ch_sym_gap}).

In order to answer this question it suffices to consider the left chiral fermions, as an identical discussion holds for the right chiral ones. 
Consider left chiral fermions transforming in an irreducible representation $R$ of a semisimple Lie algebra $\mathfrak{g}$. A QCD theory with this quark content has a left chiral flavor symmetry if and only if the embedding 
\begin{equation}
\mathfrak{so}(\dim(R)) \supset \mathfrak{g} 
\end{equation}
has a nontrivial commutant, that is, there exists an algebra $\mathfrak{h}\subset\mathfrak{so}(\dim(R))$ such that $[\mathfrak{g},\mathfrak{h}]=0$. This depends on the nature of the representation $R$, which for now we take to be irreducible:

 \medskip

\noindent $\bullet$ A chiral quark in a \underline{complex representation} has a chiral $U(1)$ global symmetry. 
 
\medskip

A complex representation $R$ of a semisimple Lie algebra $\mathfrak{g}$ is described by traceless, antihermitian $(\dim(R)/2)\times (\dim(R)/2)$ matrices $t$. Therefore the pair $(\mathfrak{g}, R)$ defines the following Lie algebra embedding and branching 
\begin{equation}
\begin{aligned}
\mathfrak{su}(\dim(R)/2 )&\supset \mathfrak{g}\\
\text{fundamental}&\mapsto R\,.
\end{aligned}
\end{equation}
Since $R$ is irreducible, the commutant of $\mathfrak{g}$ in $\mathfrak{su}(\dim(R) /2)$ is trivial by Schur's lemma. 

Let us now determine whether there is a commutant of $\mathfrak{g}$ in $\mathfrak{so}(\dim(R))$, the symmetry algebra acting on the quarks. The Lie algebra $\mathfrak{su}(\dim(R) /2)$ embeds into the $\mathfrak{so}(\dim(R))$ symmetry algebra of the quarks as 
\begin{equation}
\hat t=
\begin{pmatrix}
\re(t)& \im(t)\\
-\im(t)&\re(t)
\end{pmatrix}\subset \mathfrak{so}(\dim(R))\,,
\end{equation}
where $\re(t)^T=-\re(t)$, $\im(t)^T=\im(t)$, with $T$ denoting the transpose, and $\tr(t)=0$.
Since 
\begin{equation}
 U=
\begin{pmatrix}
0&1\\
-1&0
\end{pmatrix}\subset \mathfrak{so}(\dim(R))\,
\end{equation}
commutes with $\hat t$ and $U\not\subset \mathfrak{su}(\dim(R)/2)$, a chiral quark in a complex representation has a chiral $U(1)$ global symmetry. This also follows from the following sequence of embeddings:
\begin{equation}
\mathfrak{so}(\dim(R))\supset \mathfrak{su}(\dim(R)/2)\oplus \mathfrak{u}(1)\supset \mathfrak{g}\oplus \mathfrak{u}(1)\,.
\end{equation}

 \medskip
 
 \noindent 
$\bullet$ A chiral quark in a \underline{pseudoreal representation} has a chiral $Sp(1)\simeq SU(2)$ global symmetry. 
 
 \medskip
 
A pseudoreal representation $R$ of a Lie algebra $\mathfrak{g}$ is described by traceless, antihermitian $(\dim(R)/2)\times (\dim(R)/2)$ matrices $t$ obeying
\begin{equation}
-t^T=J t J^{-1}\,,
\label{eq:pseudoreal}
\end{equation}
where $J$ is the canonical antisymmetric matrix 
\begin{equation}
 J=
\begin{pmatrix}
0&1\\
-1&0
\end{pmatrix}\,.
\end{equation}
These are precisely the generators of the $\mathfrak{sp}(\dim(R)/4 )$ Lie algebra in the fundamental representation. Therefore the pair $(\mathfrak{g}, R)$ defines the following Lie algebra embedding and branching 
\begin{equation}
\begin{aligned}
\mathfrak{sp}(\dim(R)/4)&\supset \mathfrak{g}\\
\text{fundamental}&\mapsto R\,.
\end{aligned}
\end{equation}
Since $R$ is irreducible, the commutant of $\mathfrak{g}$ in $\mathfrak{sp}(\dim(R)/4)$ is trivial by Schur's lemma. 

Let us now determine whether there is a commutant of $\mathfrak{g}$ in $\mathfrak{so}(\dim(R))$, the symmetry algebra acting on the quarks. 
Since $J= i\sigma_2\otimes 1$, the $\mathfrak{sp}(\dim(R) /4)$ matrices $t$, which obey~\eqref{eq:pseudoreal}, can be written as 
\begin{equation}
t=\sum_{M=1}^4 t_M \otimes q_M\,,
\end{equation}
where $t_M$ are real matrices obeying $t_a^T=t_a$ for $a=1,2,3$ and $t_4^T=-t_4$.\footnote{There are three symmetric matrices and one antisymmetric, each being $n\times n$ with $n=\dim(R)/4$. Thus, there are $\frac32n(n+1)+\frac12n(n-1)=n(2n+1)$ degrees of freedom, which is precisely the dimension of the algebra $\mathfrak{sp}(n)\equiv\mathfrak{sp}(\dim(R)/4)$.} Here we denote $q_M=(i\vec\sigma,1)$, with $\vec\sigma$ the Pauli matrices, a two-dimensional complex-valued representation of the quaternions. 

The embedding of $\mathfrak{sp}(\dim(R) /4)$ into the $\mathfrak{so}(\dim(R))$ symmetry algebra of the quarks is 
\begin{equation}
\hat t=
\sum_{M=1}^4 t_M \otimes \hat \sigma_M\subset \mathfrak{so}(\dim(R))\,,
\end{equation}
where $\hat\sigma_M=(\sigma_1\otimes i \sigma_2, i\sigma_2\otimes 1,\sigma_3\otimes i \sigma_2, 1\otimes 1)$ is a four-dimensional real-valued representation of the quaternions. Since the matrices $U_a\subset \mathfrak{so}(\dim(R))$
\begin{equation}
\begin{aligned}
U_1&=1\otimes i \sigma_2\otimes\sigma_1\\
U_2&=1\otimes 1 \otimes i \sigma_2\\
U_3&=1\otimes i \sigma_2\otimes\sigma_3
\end{aligned}
\end{equation}
commute with $\hat t$, generate an $\mathfrak{sp}(1)$ algebra (namely, $[\tau_a,\tau_b]=i\epsilon_{abc}\tau_c$ with $\tau_a=\frac{1}{2i}U_a$) and $U_a\not\subset \mathfrak{sp}(\dim(R)/4)$, a chiral quark in a pseudoreal representation has a chiral $\mathfrak{sp}(1)$ global symmetry. This also follows from the following sequence of embeddings:
\begin{equation}
\mathfrak{so}(\dim(R))\supset \mathfrak{sp}(\dim(R)/4)\oplus \mathfrak{sp}(1)\supset \mathfrak{g}\oplus \mathfrak{sp}(1)\,.
\end{equation}

\medskip
\noindent
$\bullet$ A chiral quark in a \underline{real representation} has no continuous global symmetry. 
\medskip
 
%
 
A real irreducible representation $R$ of a Lie algebra $\mathfrak{g}$ is described by traceless, antihermitian $(\dim(R))\times (\dim(R))$ matrices $t$ obeying
\begin{equation}
-t^T= t\,.
\end{equation}
These are precisely the generators of the $\mathfrak{so}(\dim(R))$ Lie algebra in the fundamental representation. Therefore the pair $(\mathfrak{g}, R)$ defines the following Lie algebra embedding and branching 
\begin{equation}
\begin{aligned}
\mathfrak{so}(\dim(R))&\supset \mathfrak{g}\\
\text{fundamental}&\mapsto R\,.
\end{aligned}
\end{equation}
Since $R$ is irreducible, the commutant of $\mathfrak{g}$ in $\mathfrak{so}(\dim(R))$ is trivial by Schur's lemma. Therefore, a chiral quark in a real representation has no continuous global symmetry. 

\medskip

Let us now consider the case where the representation $R$ is reducible. Since we are seeking QCD theories that are gapped, which means that they cannot have any continuous flavor symmetries, we take $R$ to be the direct sum of irreducible, real representations $R_\alpha$ with multiplicity $M_\alpha$
\begin{equation}
R=\bigoplus_\alpha M_\alpha \cdot R_\alpha\,,\qquad~\text{with}\qquad M_\alpha\in\{0,1,2,\dots \}\,.
\end{equation}
If $M_\alpha >1$, then there is an $\mathfrak{so}(M_\alpha)$ chiral flavor symmetry acting on the quarks. Indeed, the representation matrix for the reducible representation $M_\alpha \cdot R_\alpha$ can be written as 
\begin{equation}
t=1\otimes t_\alpha\,,
\end{equation}
where $t_\alpha$ is a representation of $R_\alpha$. The matrix $U\subset \mathfrak{so}(M_\alpha\cdot\dim(R_\alpha))$
\begin{equation}
U=O\otimes 1\qquad \text{with}\qquad O^T=-O
\end{equation}
is a representation of $\mathfrak{so}(M_\alpha)$ and commutes with $t$. It therefore generates an $\mathfrak{so}(M_\alpha)$ flavor symmetry. This is also a consequence of the sequence of embeddings:
\begin{equation}
 \mathfrak{so}(M_\alpha\cdot\dim(R_\alpha))\supset \mathfrak{so}(\dim(R_\alpha))\oplus \mathfrak{so}(M_\alpha)\supset \mathfrak{g}\oplus \mathfrak{so}(M_\alpha)\,.
\end{equation}
Finally, by virtue of Schur's lemma, the direct sum of distinct irreducible real representations $\oplus_\alpha R_\alpha$ does not have a continuous flavor symmetry, as the commutant of $\mathfrak{so}(\sum_\alpha \dim(R_\alpha))\supset \mathfrak{g}$ is trivial.

Let us now consider QCD with a reductive gauge group $G=K\times U(1)^n$, where $K$ is semisimple. It suffices to consider the left chiral fermions, as an identical discussion holds for the right chiral ones. When the gauge group has $U(1)$ factors, a classical $U(1)_F$ chiral flavor symmetry may be broken by the Adler-Bell-Jackiw (ABJ) anomaly, that is, by a mixed a $U(1)$-$U(1)_F$ anomaly. A classical semisimple symmetry always remains unbroken. Therefore, for the purposes of classifying QCD theories with $K\times U(1)^n$ gauge group and no flavor symmetries, consider QCD with chiral quarks transforming under $K\times U(1)^n$ as 
\begin{equation}
\bigoplus_{I=1}^{N_F} (R_I,\vec q_I)\,,
\label{repsreduct}
\end{equation}
where $R_I$ is an irreducible representation of $K$, $q_I$ is an integral $n$-component charge vector under $U(1)^n$ and all pairs $(R_I,\vec q_I)$ are distinct (any other quark content leads to a flavor symmetry). The classical chiral flavor symmetry is
\begin{equation}
\mathfrak h_\text{classical}=\bigoplus_{I=1}^{N_F} \mathfrak u(1)_I\,.
\end{equation}
We want to determine under what conditions this symmetry is completely broken by the ABJ anomaly. Upon defining the integral matrix $Q$ whose columns are 
\begin{equation}
Q=(\dim(R_1)\vec q_1,\dim(R_2)\vec q_2,\dots,\dim(R_{N_F})\vec q_{N_F})
\end{equation}
we arrive at the following result 
\begin{mdframed}
\begin{lemma}\label{lm:necessary_gap}\normalfont 
A necessary condition for $(G;R_\ell,R_r)$ with $G=K\times U(1)^n$ to be gapped is that the representations $R_\ell$ and $R_r$ of $G$ are
of the irreducible form~\eqref{repsreduct} and the charge matrices $Q_\ell,Q_r$ have trivial kernel. A QCD theory with a quark content that is not of this type is necessarily gapless. 
\end{lemma}
\end{mdframed}

The proof is straightforward. An arbitrary chiral $\mathfrak u(1)\subseteq \mathfrak h_\text{classical}$ flavor symmetry is specified by an integer vector $\vec n=(n_1,n_2,\dots,n_{N_F})$ such that, given an angle $\alpha\in \mathfrak u(1)$, the $I$-th quark is rotated by an angle $\alpha n_I$. The mixed ABJ anomaly between this $\mathfrak u(1)$ flavor symmetry and the $U(1)^n$ gauge group is $Q\vec{n}$. Therefore, no flavor symmetries remain if and only if $Q$ has empty kernel (so that there are no nontrivial solutions to $Q\vec{n}=0$). See also appendix~\ref{sec:abelian_G} for a more in-depth discussion of QCD theories with $U(1)^n$ gauge group.

Our findings thus far are summarized in the two lemmas of this subsection. This is as far as one can get using 't Hooft anomaly considerations. Answering whether a QCD theory $(G;R_\ell,R_r)$ obeying the conditions in the lemmas is gapped or gapless requires studying the dynamics. The analysis thus far does not say, for example, whether the vector-like QCD theory with $G=SU(2)$ and quarks in the isospin $j\in \mathbb Z$ representation (which is real) is gapped or gapless. We will provide a complete answer to these questions in the rest of the paper.

Before closing this section let us make one final remark. In this section we have capitalized on the symmetries of $2d$ QFTs as much as we could. In a nutshell, we showed that, if $\mathfrak h$ denotes the chiral symmetry algebra of a $2d$ system, then the system is automatically gapless as soon as $\mathfrak h$ is non-trivial. There is a nice physical interpretation of this result. In a unitary CFT, a chiral symmetry $\mathfrak h$ is always enhanced to $\mathfrak h_k$ affine algebra, for a suitable level $k$. Therefore, if $\mathfrak h$ is non-zero, the system contains $\mathfrak h_k$ massless currents which automatically make the system gapless: the infrared contains, at the very least, an $\mathfrak h_k$ WZW CFT subsector. Therefore, a necessary condition for being gapped is that the chiral symmetry is trivial, $\mathfrak h\equiv0$. We will see the $\mathfrak h_k$ currents reappear explicitly in the Hamiltonian of QCD in the following section, and we will study them in more detail when we look at the infrared of QCD in section~\ref{sec:Infrareddynamnics}.

\section{Mass spectrum and QCD Hamiltonians}\label{sec:DLCQ}

In this section we analyze the mass spectrum of QCD by studying the quantization of the QCD Hamiltonians. The main result is a derivation of the necessary and sufficient conditions for a QCD theory to be gapped. Along the way, we prove that a QCD theory with a continuous global symmetry has massless particles in the spectrum, reproducing the result derived in section~\ref{sec:symmanom} by symmetry and 't Hooft anomaly arguments. We analyze the lightcone and temporal Hamiltonians, both yielding the same conditions for QCD to be gapped.

 \subsection{Lightcone Hamiltonian}

Our aim in this section is to study the mass spectrum of QCD by quantizing the theory in the lightcone frame~\cite{THOOFT1974461,PhysRevD.32.1993,PhysRevD.32.2001}. We review here the most salient features and formulas (see~\cite{Brodsky:1997de,Dempsey:2021xpf} for reviews and recent work). The basic idea is to use the a lightcone coordinate, say $x^+=\frac{1}{\sqrt2}(x^0+x^1)$, as the time variable. This quantization defines the Hilbert space and the Cauchy data of the theory on a constant $x^+$ surface, and the conjugate lightcone Hamiltonian $P^-$ evolves states in $x^+$. When we choose $x^+$ to play the role of time, the lightcone coordinate $x^-$ plays the role of a spatial coordinate. The momentum $P^+$ conjugate to $x^-$ commutes with the lightcone Hamiltonian 
 $P^-$, i.e.,~$[P^+,P^-]=0$. Therefore, the mass spectrum of QCD can be obtained by simultaneously diagonalizing the operators $P^+$ and $P^-$, since 
\begin{equation}
M^2=2P^+P^-\,.
\label{masshell}
\end{equation}
Positive semidefiniteness of $M^2$ and of the lightcone Hamiltonian $P^-$ implies that all states in the Hilbert space have $P^+\geq 0$. This combined with the fact that interactions preserve $P^+$ implies that the vacuum state in lightcone quantization is trivial, that is, the vacuum has no particles in it, and the nonperturbative vacuum coincides with the Fock vacuum. This makes lightcone quantization well adapted to study the meson and hadron spectrum of QCD. 
 
Since $P^-$ evolves states along $x^+$, left-moving massless particles are not visible in the $P^-$ Hamiltonian. This implies that the spectrum of the Hamiltonian $P^-$ correctly accounts for all massive and right-moving massless particles, but does not detect left-moving massless particles. This shortcoming can be overcome by quantizing QCD using instead $x^-=\frac{1}{\sqrt2}(x^0-x^1)$ as the lightcone time. In this frame, the lightcone Hamiltonian is instead $\hat P^+$ and its spectrum contains all the massive and left-moving massless particles. Therefore, by diagonalizing the QCD lightcone Hamiltonians $P^-$ and $\hat P^+$ in the quantizations where $x^+$ and $x^-$ is time respectively, all the massless particles of QCD are accounted for.

Let us proceed with the lightcone quantization of QCD with left and right chiral quarks in representations $R_\ell$ and $R_r$ of the gauge group $G$. 
We start with the QCD Lagrangian 
\begin{equation}
{\mathcal L}_{\text{QCD}}=
\frac{1}{g^2} F^a_{+-}F^a_{+-} +i \psi^\dagger_\ell(\partial_--i A^a_- t^a_\ell)\psi_\ell+i \psi^\dagger_r(\partial_+-i A^a_+ t^a_{r})\psi_r\,.
\end{equation}
We fix the gauge $A^a_-=0$, in which all states have positive norm and there are no ghosts. In the lightcone gauge with $x^+$ being time, the left-chiral quarks $\psi_\ell$ and $A^a_+$ are not dynamical. They can be integrated out to yield
\begin{equation}
{\mathcal L}_{\text{QCD}}=
i \psi^\dagger_r\partial_+\psi_r -g^2 J^a_r\frac{1}{\partial^2_-} J^a_r\,,
\end{equation}
where 
\begin{equation}
J^a_r= \frac{1}{2}{:} \psi^\dagger_r t^a_r \psi_r {:}
\end{equation}
generates the right-moving quark current for $\mathfrak g$, the Lie algebra of $G$, which obeys $\partial_-J^a_r=0$. The Noether charges for the lightcone Hamiltonian and momentum are
\begin{equation}
\begin{aligned}
P^-&=-g^2 \int \mathrm dx^-\, {:}J^a_r\frac{1}{\partial^2_-} J^a_r{:} \label{LCone}\\[+2pt]
P^+&=i\int \mathrm dx^-\, {:} \psi^\dagger_r\partial_-\psi_r{:}\,.
\end{aligned}
\end{equation}
Choosing $x^-$ as lightcone time instead results in the associated lightcone Hamiltonian and momentum
\begin{equation}
\begin{aligned}
\hat P^+&=-g^2 \int \mathrm dx^+\,{:}J^a_\ell\frac{1}{\partial^2_+} J^a_\ell{:}\label{LCtwo}\\[+2pt]
\hat P^-&=i\int \mathrm dx^+\, {:}\psi^\dagger_\ell\partial_+\psi_\ell{:}\,,
\end{aligned}
\end{equation}
where now 
\begin{equation}
J^a_\ell=\frac12{:} \psi^\dagger_\ell t_\ell^a \psi_\ell {:}
\end{equation}
generates the left-moving quark current for $G$, with $\partial_+J^a_\ell=0$.

We are interested in determining when a QCD theory is gapped, and when it is gapless. This requires determining when the lightcone Hamiltonians~\eqref{LCone} and~\eqref{LCtwo} have zero eigenvalues. We will answer this question by studying the eigenvalues of the operators 
\begin{equation}
\begin{aligned}
H&=-g^2 \int \mathrm dx\, {:}J^a\frac{1}{\partial^2_x} J^a{:}\\[+2pt]
\hat P&=i\int \mathrm dx\, {:} \psi^\dagger\partial_x \psi{:}\,,
\end{aligned}
\label{LCthree}
\end{equation}
where $x=x^-$ or $x=x^+$ and $J^a=J^a_r$ or $J^a=J^a_\ell$ depending on whether $x^+$ or $x^-$ is the lightcone time. Canonical quantization yields the following equal-time commutation relations
\begin{equation}
\{ \psi^\dagger_i (x), \psi^j (y) \}=\delta^j_i\delta(x-y)\,,
\label{ETC}
\end{equation}
where the Latin indices are the representation labels for the representation $R$ of $\mathfrak{g}$ of the relevant chiral quarks. The quark field expansion in Fourier modes is
\begin{equation}
\psi^i(x)=\frac{1}{\sqrt{2\pi}}\int_0^\infty \mathrm dk\left(a^i(k)e^{-ikx}+b_i^\dagger(k) e^{ikx}\right)\,,
\end{equation}
where $k=k_-$ or $k=k_+$ depending on whether $x^+$ or $x^-$ is taken as the lightcone time. Note that in the lightcone frame the Fourier modes carry nonnegative longitudinal momentum. Using~\eqref{ETC} we get the anticommutation relations 
\begin{equation}
\{ a^i(k),a_j^\dagger(k') \}=\delta^i_j\delta(k-k')\,,\qquad \{ b^i(k) ,b_j^\dagger(k')\}=\delta^i_j\delta(k-k')\,,
\end{equation}
with $a_i^\dagger(k)=b_i^\dagger(k)$ if the quark field is a Majorana fermion. These operators define the Fock vacuum, and in fact the nonperturbative vacuum $|0 \rangle$ of lightcone QCD
\begin{equation}
a^i(k)|0\rangle=0\,, \qquad b^i(k)|0\rangle=0 \qquad \forall~k\,.
\end{equation}
Normal ordering in~\eqref{LCthree} implies that the vacuum has zero lightcone energy and momentum
\begin{equation}
H|0\rangle=0\,,\qquad P|0\rangle=0\,.
\end{equation}
 
The goal is to diagonalize the lightcone Hamiltonian(s) $H$ on the Hilbert space $\mathcal H$ created by the quarks
\begin{equation}
|\Psi^{i_1i_2\ldots i_L}\rangle\equiv a_{i_1}^\dagger(k_1)a_{i_2}^\dagger(k_2)\ldots a_{i_L}^\dagger(k_L)|0\rangle\,.
\end{equation}
The physical states of QCD must be gauge invariant, which implies that all physical states must be invariant under the action of $\mathfrak{g}$
\begin{equation}
\int \mathrm dx\, J^a(x)|\Psi^{i_1i_2\ldots i_L}\rangle=0\,.
\end{equation}
While $H$ mixes states in the Hilbert space $\mathcal H$, the longitudinal momentum operator $P$ is diagonal, it is the sum of the longitudinal momentum of each parton.

QCD has massless particles if and only if there exist states $|\Psi^{i_1i_2\ldots i_L}\rangle\in\mathcal H$, other than the vacuum state, that are gauge invariant and have zero lightcone energy,
\begin{equation}
\int \mathrm dx\, {:} J^a\frac{1}{\partial^2} J^a{:} |\Psi^{i_1i_2\ldots i_L}\rangle=0\,.
\label{zeroen}
\end{equation}
The currents $J^a= \frac{1}{2}{:} \psi^\dagger t^a \psi{:}$ constructed out of the chiral quarks $\psi$ transforming in a representation $R$ of $\mathfrak{g}$ generate an affine chiral current algebra $\mathfrak g_{I(R)}$ 
\begin{equation}
J^a(x) J^b(0)\sim \frac{I(R)\delta^{ab}}{x^2}+\frac{i f^{ab}{}_{c}J^c(0)}{x}\,,
\label{OPEcurrent}
\end{equation} 
where the level is the Dynkin index $I(R)$ of the representation $R$. The OPE~\eqref{OPEcurrent} implies, upon putting the longitudinal coordinate on the circle, that the Fourier modes obey
\begin{equation}
[J_n^a,J_m^b]=i f^{ab}{}_{c}J_{n+m}^c+I(R) \delta^{ab} n \delta_{m+n,0}\,.
\end{equation}
The zero energy state condition~\eqref{zeroen} takes the form 
\begin{equation}
\sum_{n=1}^\infty\frac{1}{n^2}J_{-n}^a J_{n}^a |\Psi^{i_1i_2\ldots i_L}\rangle=0\,.
\end{equation}
Since the Hamiltonian $H$ in~\eqref{LCthree} is a positive semidefinite operator, the necessary and sufficient conditions for a state $|\Psi^{i_1i_2\ldots i_L}\rangle$ to have zero energy are:
\begin{enumerate}
\item $|\Psi^{i_1i_2\ldots i_L}\rangle$ is a primary state of the current algebra $\mathfrak g_{I(R)}$, that is $J_n^a|\Psi^{i_1i_2\ldots i_L}\rangle=0$ ~~$\forall~n\geq 1$.
\item $|\Psi^{i_1i_2\ldots i_L}\rangle$ transforms in the trivial representation of $\mathfrak{g}$.
\end{enumerate}
We now proceed to study under what conditions these states exist. 

The quark Hilbert space $\mathcal H$ decomposes into modules of the $\mathfrak{so}(\dim(R))_1$ current algebra at level one~\cite{Witten:1983ar} (see also~\cite{Ji:2019ugf}), labeled by representations of $\mathfrak{so}(\dim(R))$. These are labeled by $({\bf 0},{\bf v},{\bf s},{\bf c})$ when $\dim(R)$ is even and by $({\bf 0},{\bf v},{\bf s})$ when it is odd, where ${\bf 0}, {\bf v},{\bf s},{\bf c}$ are the trivial, the vector and spinor representation(s) of $\mathfrak{so}(\dim(R))$. The precise relation between the modules of $\mathfrak{so}(\dim(R))_1$ current algebra and fermion Hilbert space is (e.g.~for $\dim(R)$ even)
\begin{equation} 
 \begin{aligned}
 \mathcal H_{\text{NS}}&=\mathcal H_{\bf 0}\oplus \mathcal H_{\bf v}\\
 \mathcal H_{\text{R}}&=\mathcal H_{\bf s}\oplus\mathcal H_{\bf c}\,,
 \label{decompose}
 \end{aligned}
\end{equation}
where $\mathcal H_\text{X}$ denotes the fermion Hilbert space with fermions obeying $\text X\in\{\text{NS},\text{R}\}$ (Neveu-Schwarz and Ramond) boundary conditions on the circle. 
 
Since QCD is obtained by gauging the subalgebra $\mathfrak{g}\subset\mathfrak{so}(\dim(R))$, the current algebra embeds as $ \mathfrak{g}_{I(R)}\subset\mathfrak{so}(\dim(R))_1$ into the fermion current algebra. That means that any state in $\mathcal H$ fits inside a module of the $\mathfrak{g}_{I(R)}$ current algebra. An elegant way to describe how states embed is through the branching functions $b_{\Lambda \lambda}$, which encode how the characters of the $\mathfrak{so}(\dim(R))_1$ current algebra decompose into $\mathfrak{g}_{I(R)}$ characters:
\begin{equation}
\chi_\Lambda (q)=\sum_{\lambda} b_{\Lambda \lambda}(q) \chi_\lambda (q)\,.
\end{equation}
Here $\Lambda\in\{{\bf 0},{\bf v},{\bf s},{\bf c}\}$ or $\{{\bf 0},{\bf v},{\bf s}\}$ and $\lambda$ is a highest weight vector labeling the integrable representations of $\mathfrak{g}_{I(R)}$, which obeys
\begin{equation}
\sum_{i=1}^{\operatorname{rank}(\mathfrak{g})}a^\vee_i \lambda_i\le I(R)\,,
\end{equation}
where $a^\vee_i$ are the comarks of $\mathfrak{g}$. The function $b_{\Lambda \lambda}(q)$ counts how many primary states of $\mathfrak g_{I(R)}$ with highest weight $\lambda$ appear in the decomposition of the module of $\mathfrak{so}(\dim(R))_1$ with highest weight $\Lambda$. They also capture at which level the primaries of $\mathfrak{g}_{I(R)}$ appear in the $\mathfrak{so}(\dim(R))_1$ modules. 
 
The branching function $b_{\Lambda \lambda}(q)$ has a module interpretation. $b_{\Lambda \lambda}(q)$ is a character of the chiral algebra $\mathcal A$, where $\mathcal A$ is the commutant chiral algebra of $\mathfrak{g}_{I(R)}$ inside $\mathfrak{so}(\dim(R))_1$. $\mathcal A$ necessarily contains the Virasoro algebra with central charge the difference of the central charges of the two current algebras $c_\mathcal A=c(\mathfrak{so}(\dim R)_1)- c(\mathfrak g_{I(R)})=\frac12\dim R-\frac{I(R)\dim(\mathfrak{g})}{I(R)+h}$, where $h$ is the dual Coxeter number of $\mathfrak{g}$ (see table~\ref{tab:comarksone}). Also, if $\mathfrak{h}$ is the commutant of $\mathfrak{g}$ inside $\mathfrak{so}(\dim R)$, then ${\mathcal A}$ contains a current algebra $\mathfrak{h}_{k}$. This current algebra has the interpretation as the flavor symmetry current algebra in QCD.

We have established that a QCD theory is gapless if and only if the fermion Hilbert space contains a nontrivial primary state of $\mathfrak{g}_{I(R)}$ labeled by the trivial representation of $\mathfrak{g}$. We are therefore interested in the functions $b_{\Lambda \hat {\bf 0}}(q)$, where we denote the trivial representation of $\mathfrak{g}$ by $\hat {\bf 0}$. In order to determine whether a QCD theory has a massless state it will suffice to look at the branching function for the integrable representation of $\mathfrak{so}(\dim(R))_1$ labeled by the trivial representation of $\mathfrak{so}(\dim(R))$ into the trivial representation of $\mathfrak{g}_{I(R)}$. Whether the theory has massless states or not is encoded in the properties of the function $b_{{\bf 0}\hat {\bf 0}}(q)$. 

Given that $b_{{\bf 0}\hat {\bf 0}}(q)$ is the vacuum character of the chiral algebra ${\mathcal A}$, this branching function takes the following general form\footnote{There are no $q^{n+1/2}$ terms in the expansion since $\mathcal H_{\bf 0}$ contains states created with an even number of fermions.}
\begin{equation}
b_{{\bf 0}\hat {\bf 0}}(q) =q^{-c_\mathcal A/24}\bigl(1+ a_1q+a_2q^2+\ldots \bigr)\,.
\end{equation}
The theory has massless particles if $a_l\neq 0$ for any $l$.

 We explain now the physical meaning of the coefficients $a_l$. The coefficient of the $q^0$ term being $a_0=1$ corresponds to the vacuum state $|0\rangle$, which is unique. The coefficient $a_1=\dim(\mathfrak{h})$ is the dimension of the commutant of $\mathfrak{g}$ in $\mathfrak{so}(\dim(R))$. If $\mathfrak{h}$ is nontrivial, we can build the primary, singlet states of $\mathfrak{g}_{I(R)}$ at level one by acting on the vacuum with the flavor symmetry currents 
\begin{equation}
\tilde J^\alpha=\frac{1}{2}\psi^\dagger \tilde t^\alpha\psi,\qquad \alpha=1,\ldots, \dim(\mathfrak{h})\,.
\end{equation}
 The currents $\tilde J^\alpha$ generate an $\mathfrak{h}_k$ current algebra, the level $k$ being determined by the embedding $\mathfrak{so}(\dim R)\supset \mathfrak h\oplus \mathfrak g$. Indeed, these states are annihilated by $J^a$ since $[\tilde J^\alpha, J^a ]=0$, by virtue of $\mathfrak{h}$ commuting with $\mathfrak{g}$. If such an operator $\tilde J^\alpha$ exists, then the theory is gapless. This reproduces the result we proved in section~\ref{sec:symmanom} stating that any theory with a continuous, chiral global symmetry --- which means $\mathfrak{h}$ is nontrivial --- is necessarily gapless. Therefore, a necessary condition for the QCD theory to be gapped is that the theory has no continuous, chiral flavor symmetries and therefore that $a_1=0$.

We turn our attention to the physics of the coefficient $a_2$. Recall that given a current algebra $\mathfrak g_{I(R)}$, one can construct the chiral energy momentum tensor~\cite{PhysRev.170.1659,DiFrancesco:1997nk} 
\begin{equation}
T_{{\mathfrak g}_{I(R)}}=\frac{1}{8({I(R)}+h)} {:}J^a J^a{:}\,.
\end{equation}
 The operator $T_{{\mathfrak g}_{I(R)}}$ generates a Virasoro algebra of central charge $c(\mathfrak g_{I(R)})=\frac{I(R)\dim(G)}{I(R)+h}$. The canonical level $2$ state in the Hilbert space $\mathcal H$ 
\begin{equation}
(T_{\mathfrak{so}(\dim(R))_1}-T_{{\mathfrak g}_{I(R)}})|0\rangle
\label{states}
\end{equation}
is a singlet, primary state of the $\mathfrak g_{I(R)}$ current algebra, where $T_{\mathfrak{so}(\dim(R))_1}$ is the energy momentum tensor of the $\mathfrak{so}(\dim(R))_1$ current algebra generated by the quarks in the ultraviolet. This is a consequence of the OPEs 
\begin{equation}
\begin{aligned}
T_{\mathfrak{g}_I(R)}(x)J^a(0)&\sim \frac{J^a(0)}{x^2}+\frac{\partial J^a(0)}{x}\\
T_{\mathfrak{so}(\dim(R))_1}(x)J^a(0)&\sim \frac{J^a(0)}{x^2}+\frac{\partial J^a(0)}{x}\,,
\end{aligned}
\end{equation}
so that
\begin{equation}
[J^a,T_{\mathfrak{so}(\dim(R))_1}-T_{{\mathfrak g}_{I(R)}}]=0\,.
\end{equation}
This proves that the state~\eqref{states} is a gauge invariant, primary state of $\mathfrak g_{I(R)}$. Therefore as long as $T_{\mathfrak{so}(\dim(R))_1}-T_{{\mathfrak g}_{I(R)}} \neq 0$, so that $a_2\neq 0$, the theory has massless states and the spectrum is gapless.\footnote{In the presence of a global symmetry $\mathfrak{h}$ there are additional level $2$ primary, singlet states of $\mathfrak g_{I(R)}$ 
\begin{equation}
\tilde J_{-2}^\alpha |0\rangle\,, \qquad \tilde J_{-1}^\alpha\tilde J_{-1}^\beta |0\rangle\,.
\end{equation}
Generically, there are $\dim(\mathfrak{h})(\dim(\mathfrak{h})+3)/2$ such states constructed using the flavor affine algebra $\mathfrak h_k$ currents (for very small values of $k$ one may need to subtract some null states).}

This implies that a necessary (and as we will show also sufficient) condition for a QCD theory to be gapped is that the following operator equation holds
\begin{equation}
T_{\mathfrak{so}(\dim(R))_1}-T_{{\mathfrak g}_{I(R)}}=0\,. 
\label{EMzero}
\end{equation} 
 This equation is very constraining, it implies that $a_l=0$ for all $l$. Indeed, when~\eqref{EMzero} is obeyed then $b_{\Lambda \lambda}(q)$ is as a character of the Virasoro algebra with $c_\mathcal A=0$, which has a unique, trivial unitary representation. Therefore, when equation~\eqref{EMzero} holds then\footnote{This follows by comparing the scaling dimensions of operators.}
 \begin{equation}
 b_{\Lambda \hat {\bf 0}}(q)=\delta_{\Lambda {\bf 0}} \,,
 \end{equation}
and the only singlet primary state of ${\mathfrak g}_{I(R)}$ is the vacuum state: the theory has no massless particles. 

By demanding that there are no left-moving or right-moving massless particles, we arrive at the following lemma:

\begin{mdframed}
\begin{lemma}\normalfont QCD theory $(G;R_\ell,R_r)$ is gapped if and only if {\it both} operator equations hold
\begin{equation}
\begin{aligned}
T{}_{\mathfrak{so}(\dim R_\ell)_1}-T{}_{{\mathfrak g}_{I(R_\ell)}}&=0\,,\\[+4pt]
\overline T_{\mathfrak{so}(\dim R_r)_1}-\overline T_{{\mathfrak g}_{I(R_r)}}&=0\,. 
\end{aligned}
\label{lemma4}
\end{equation} 
\end{lemma}
\end{mdframed}
This was derived by looking at the lightcone Hamiltonians with $x^+$ and $x^-$ being time.

We will come back to these equations momentarily. Before we do that we shall also derive these equations in a different quantization scheme in order to gain more insight into the mechanism behind the gap.

\subsection{Temporal gauge Hamiltonian}\label{sec:temporal_gauge}

In this section we rederive the necessary and sufficient conditions~\eqref{lemma4} for a QCD theory $(G;R_\ell,R_r)$ to be gapped by studying the canonical Hamiltonian where time is the time-like coordinate $x^0=t$. The lightcone coordinates are well-suited to algebraic considerations because the two chiralities are mostly decoupled. By contrast, a time-like coordinate requires more work but it also leads to a more transparent understanding of the spectrum, because the Hamiltonian takes the traditional form, which evolves states in physical time. For previous work on the temporal Hamiltonian of QCD see~\cite{Zeppenfeld:1983xg, Kutasov:1994xq,Langmann:1994vb}.

We follow the same conventions as in \S\ref{sec:QCD2}, which we present here for convenience
\begin{equation}
 x^{\pm} = \frac{1}{\sqrt{2}}(x^0\pm x^1),\quad A^a_{\pm} = \frac{1}{\sqrt{2}}(A^a_0\pm A^a_1)\,.
\end{equation}
We start with the Lagrangian 
\begin{equation}
 \mathcal{L} = i\psi^\dagger_{i\,\ell} \partial_- \psi_{i\,\ell} + i\psi^\dagger_{i\,r} \partial_+ \psi_{i\,r} +A^a_-J^a_\ell+A^a_+ J^a_r-\frac{1}{2g^{2}}\text{tr}(F_{\mu\nu}F^{\mu \nu})\,,
 \label{lagoon}
\end{equation}
where $J^a_{\ell,r}=\frac{1}{2}\psi^\dagger_{i\,\ell,r}t^a_{ij} \psi_{j\,\ell,r}$.
Canonical quantization leads to the following equal-time commutation relations
\begin{equation}
\begin{aligned}\label{commutations}
 \{\psi^\dagger_{i\,\ell}(x),\psi_{j\,\ell}(y)\} &= \delta_{ij}\delta(x-y)\,, \\[+2pt]
 \{\psi^\dagger_{i\, r}(x),\psi_{j\,r}(y)\} &= \delta_{ij}\delta(x-y)\,, \\[+2pt]
 [\Pi^a(x), A_1^b(y)] &= i\delta^{ab}\delta(x-y)\,\,,
\end{aligned}
\end{equation}
where $\Pi^a =\frac{\delta\mathcal{L}}{\delta \dot{A}^a_0}= \frac{1}{ g^2} F^a_{10}=\frac{1}{ g^2} E^a(x)$. Consequently, classically the currents obey the commutation relations
\begin{equation}
 [J^a_{\ell,r}(x), J^b_{\ell,r}(y) ]= if^{abc}J^c_{\ell,r}(x)\delta(x-y) \,.
\label{classcomm}
\end{equation}

The Hamiltonian density is
\begin{equation}\label{Hamiltonianzero}
\begin{aligned}
\mathscr H &= g^2 E^a(x)^2 +A_0^a G^a+\frac{1}{\sqrt{2}}A_1^a(x)(J^a_\ell - J^a_r)(x)\\
 & +\frac{1}{\sqrt{2}}i\psi^\dagger_{i\,\ell} \partial_x \psi_{i\,\ell}(x) - \frac{1}{\sqrt{2}}i\psi^\dagger_{i\,r} \partial_x \psi_{i\,r}(x) \,,\\
\end{aligned}
\end{equation}
and 
\begin{equation}
G^a(x)=J^a_\ell(x)+ J^a_r(x)-(\partial_x \Pi^a +if^{abc} A_1^b \Pi^c)(x)
\end{equation}
is the Gauss' law operator, which obeys the following commutation relation with the currents
\begin{equation}
[G^a(x), J^b_{\ell,r}(y) ]= if^{abc}J^c_{\ell,r}(x)\delta(x-y) \,.
\label{gausscomm}
\end{equation}
The Gauss law operator commutes with the Hamiltonian, that is $[G^a(x),H]=0$, where $H=\int dx\, \mathscr H$.

 The phase space of QCD has a primary constraint, namely the momentum conjugate to $A_0$ vanishes
\begin{equation}
\Pi^a=\frac{\delta \mathcal{L}}{\delta \dot{A}^a_0}=0\,.
\end{equation}
Demanding stability of this constraint leads to the secondary constraint 
\begin{equation}
\frac{d\Pi^a}{dt}=[H,\Pi^a]=G^a=0\,,
\end{equation}
Since $\frac{dG^a}{dt}=[H,G^a]=0$ on the constraint surface, there are no further constraints. Hamilton's equations derived from~\eqref{Hamiltonianzero} reproduce the equations of motion obtained by varying the Lagrangian~\eqref{lagoon}. We work in the gauge with $A^a_0=0$ so that $A^a
_+ = A^a_1\equiv A^a$.

In the quantum theory, fields are promoted to operators and composites need to be renormalized due to quantum fluctuations at arbitrary short distances. The quark currents in the quantum theory must be normal ordered $\widehat{J}^a_{\ell,r} =\, \frac{1}{2}{:}\psi^\dagger_{i\,\ell,r}t^a_{ij} \psi_{j\,\ell,r}{:}$. In the quantum theory, the commutation relations become
\begin{equation}
\begin{aligned}\label{qcommutator}
 [\widehat{J}^a_{\ell,r}(x), \widehat{J}^b_{\ell,r}(y) ]&= if^{abc}\widehat{J}^c_{\ell,r}(x)\delta(x-y) \pm iI(R) \delta^{ab}\partial_x\delta(x-y)\,, \\[+2pt]
 [\widehat{\Pi}^a(x), \widehat{A}^b(y)] &= i\delta^{ab}\delta(x-y)\,,
\end{aligned}
\end{equation}
where $I(R)\equiv I(R_\ell)=I(R_r)$ by virtue of gauge anomaly cancellation, and $\pm$ corresponds to $\ell$ and $r$ respectively. The operators $\widehat{J}^a_{\ell,r}$ generate the current algebra $\mathfrak{g}_{I(R_{\ell,r})}$. Quantization leads to the Schwinger term in the current commutators~\eqref{qcommutator} (cf.~with~\eqref{classcomm}), which will have important implications.

In order to determine the conditions for the spectrum of Hamiltonian to be gapped, we first define the fermion operators in~\eqref{Hamiltonianzero} in normal ordered form
\begin{equation}
\begin{aligned}
 \frac{1}{\sqrt{2}}i\psi^\dagger_{i\,\ell} \partial_x \psi_{i\,\ell}(x) &= \frac{1}{\sqrt{2}}{:}i\psi^\dagger_{i\,\ell} \partial_x \psi_{i\,\ell}(x){:} -\frac{1}{\sqrt{2}}i\lim_{\epsilon \to 0}\langle \psi^\dagger_{i\,\ell}(x+\epsilon)\partial_{x}\psi_{i\,\ell}(x-\epsilon) \rangle \\
 -\frac{1}{\sqrt{2}}i\psi^\dagger_{i\,r} \partial_{x} \psi_{i\,r}(x) &= -\frac{1}{\sqrt{2}}{:}i\psi^\dagger_{i\,r} \partial_{x} \psi_{i\,r}(x){:} +\frac{1}{\sqrt{2}}i\lim_{\epsilon \to 0}\langle \psi^\dagger_{i\,r}(x+\epsilon)\partial_{x}\psi_{i\,r}(x-\epsilon) \rangle\,.
\end{aligned}
\end{equation}
where $\langle \psi^\dagger_{i \ell ,\,r }(x) \psi_{j \ell, \, r}(y) \rangle \sim \frac{i \delta_{ij}}{x-y}$.
We then express $\pm \frac{1}{\sqrt{2}}{:}i\psi^\dagger_{\ell,r} \partial_x \psi_{\ell,r}{:}$ in terms of the Sugawara tensor for $\mathfrak{so}(\dim R_{\ell,r})_1$ via
\begin{equation}
\begin{aligned}
 \frac{1}{2}T{}_{\mathfrak{so}(\dim R_\ell)_1}&=\frac{i}{8(I(R)+h)}{:}\widehat{J}^a_{\ell} \widehat{J}^a_{\ell}(x){:} = \frac{1}{\sqrt{2}} {:}i\psi^\dagger_{i\,\ell} \partial_x \psi_{i\,\ell}(x) {:} \\[+2pt]
 \frac{1}{2} \overline T_{\mathfrak{so}(\dim R_r)_1}& =\frac{i}{8(I(R)+h)}{:}\widehat{J}^a_{r} \widehat{J}^a_{r}(x){:} = -\frac{1}{\sqrt{2}} {:}i\psi^\dagger_{i\,r} \partial_x \psi_{i\,r}(x) {:}\,.
\end{aligned}
\end{equation}
The crucial insight (see also~\cite{Kutasov:1994xq,Langmann:1994vb}) is that we can split the energy momentum tensor into a piece that couples to the gauge fields and a piece that is decoupled
\begin{equation}\label{splittt}
\begin{aligned}
T{}_{\mathfrak{so}(\dim R_\ell)_1}&= T{}_{\mathfrak{g}_{I(R_\ell)}}+\bigl(T{}_{\mathfrak{so}(\dim R_\ell)_1}- T{}_{\mathfrak{g}_{I(R_\ell)}}\bigr) \\ \overline T_{\mathfrak{so}(\dim R_r)_1}&= \overline T_{\mathfrak{g}_{I(R_r)}}+\bigl( \overline T_{\mathfrak{so}(\dim R_r)_1}- \overline T_{\mathfrak{g}_{I(R_r)}}\bigr)\,.
\end{aligned}
\end{equation}

The quantized Hamiltonian must commute with the quantum Gauss' law operator 
\begin{equation}
\widehat G^a(x)=\widehat J^a_\ell(x)+ \widehat J^a_r(x)-(\partial_x \widehat \Pi^a +if^{abc} \widehat A^b\, \widehat \Pi^c)(x)\,.
\end{equation}
It is given by (see appendix~\ref{temporalgaugeH} for details)
\begin{equation}\label{subhamiltonian}
\begin{aligned}
 \widehat{\mathscr H} &= \frac{1}{2}\bigl(T_{\mathfrak{g}_{I(R_\ell)}}+T_{\mathfrak{g}_{I(R_r)}}\bigr) + \frac{1}{\sqrt{2}}\widehat{A}^a(x)(\widehat{J}^a_\ell - \widehat{J}^a_r)(x) + \frac{1}{\sqrt{2}} I(R)\widehat{A}^a(x)^2 +g^2 \widehat{E}^a(x)^2 \\
 &\hspace{5mm}+ \frac{1}{\sqrt{2}}i\lim_{ \epsilon \to 0}\bigl(\langle \psi^\dagger_{i\,\ell}(x+\epsilon)\partial_x\psi_{i\,\ell}(x-\epsilon)\rangle-\langle \psi^\dagger_{i\,r}(x+\epsilon)\partial_x\psi_{i\,r}(x-\epsilon)\rangle\bigr) \\
 &\hspace{5mm}+\frac{1}{2}\bigl(T_{\mathfrak{so}(\dim R_\ell)_1}- T_{\mathfrak{g}_{I(R_\ell)}} \bigr)+\frac{1}{2}\bigl( \overline T_{\mathfrak{so}(\dim R_r)_1}- \overline T_{\mathfrak{g}_{I(R_r)}} \bigr)\,.
\end{aligned}
\end{equation}
Let us discuss some of the most salient features of this Hamiltonian. Expressing the Hamiltonian in terms of the energy momentum tensor and splitting it as in~\eqref{splittt} shows that there is a decoupled CFT with energy momentum tensors $T{}_{\mathfrak{so}(\dim R_\ell)_1}- T{}_{\mathfrak{g}_{I(R_\ell)}}$ and $\overline T_{\mathfrak{so}(\dim R_r)_1}- \overline T_{\mathfrak{g}_{I(R_r)}}$. The last line in~\eqref{subhamiltonian} describes a gapless sector. The term $I(R)\widehat{A}^a(x)^2$ in~\eqref{subhamiltonian}, which is present due to the Schwinger term in~\eqref{qcommutator}, gaps out the gauge fields, and strongly suggests that the first two lines in~\eqref{subhamiltonian} describe a gapped Hamiltonian.\footnote{It would be interesting to give a rigorous proof that it is gapped.} This analysis makes manifest that the massless degrees of freedom decouple in the ultraviolet and go along for the ride during the renormalization group flow (see~\cite{Kutasov:1994xq}). In conclusion, the QCD theory $(G;R_\ell,R_r)$ is gapped if and only if the decoupled CFT is trivial, that is if 
\begin{equation}
\begin{aligned}
T{}_{\mathfrak{so}(\dim R_\ell)_1}- T{}_{{\mathfrak g}_{I(R_\ell)}}&=0\\
\overline T_{\mathfrak{so}(\dim R_r)_1}- \overline T_{{\mathfrak g}_{I(R_r)}}&=0
\end{aligned}
\end{equation} 
We have thus recovered the conditions~\eqref{lemma4} we had derived in the previous section using the lightcone Hamiltonians.

\subsection{Classification of gapped theories}\label{sec:class_gap_QCD}
 
We now return to our main task: deducing whether a given QCD theory is gapped or not. We showed, both looking at lightcone quantization and standard canonical quantization, that a theory labelled by $(G;R_\ell,R_r)$ is gapped if and only if the two operator equations hold:
\begin{equation}\label{eq:condition_gap_QCD}
\begin{aligned}
T{}_{\mathfrak{so}(\dim R_\ell)_1}- T{}_{{\mathfrak g}_{I(R_\ell)}}&=0\,,\\
\overline T_{\mathfrak{so}(\dim R_r)_1}- \overline T_{{\mathfrak g}_{I(R_r)}}&=0\,.
\end{aligned}
\end{equation} 
 
The key point is that the equations~\eqref{eq:condition_gap_QCD} can in fact be solved. It suffices to consider one chirality first, and then combine solutions that merge left and right chiral sectors. In~\cite{GODDARD1985226,Goddard:1985jp} it was shown that the energy-momentum tensor of the affine algebra $\mathfrak g_{I(R)}$ coincides with that of a free fermion theory $\mathfrak{so}(\dim(R))_1$ if and only if the matrices $t^a_{ij}$ that generate the representation $R$ satisfy the Jacobi-like identity (see appendix \ref{app:match_T} for derivation)
\begin{equation}
t^a_{ij}t^a_{k\ell}+t^a_{ik}t^a_{\ell j}+t^a_{i\ell}t^a_{jk}=0\,.
\end{equation}
In turn, this identity is satisfied if and only if there exists some Lie algebra $\hat{\mathfrak g}$ that contains $\mathfrak g$ such that the homogeneous space\footnote{The global structure of $\hat G,G$ is arbitrary, the condition $t^a_{ij}t^a_{k\ell}+t^a_{ik}t^a_{\ell j}+t^a_{i\ell}t^a_{jk}=0$ is purely algebraic and insensitive to the choice of $\hat G,G$ for a given $\hat{\mathfrak g},\mathfrak g$. At the algebraic level, the condition can be recast as the existence of an algebra $\hat{\mathfrak g}=\mathfrak g+\mathfrak p$ such that $[p^i,p^j]\in\mathfrak g$ for $p^i,p^j\in\mathfrak p$, and $[p^i,g^a]=t^a_{ij}p^j$ for $g^a\in\mathfrak g$.} $\hat G/G$ is symmetric, and the fermions transform with respect to $\mathfrak g$ in the same way as the generators of $\hat G/G$. The condition $t^a_{ij}t^a_{k\ell}+t^a_{ik}t^a_{\ell j}+t^a_{i\ell}t^a_{jk}=0$ is nothing but the Bianchi identity for the Riemann tensor of $\hat G/G$. The symmetric spaces have been fully classified~\cite{BSMF_1926__54__214_0}, and we can read off the list of gapped QCD theories from this classification: for each symmetric space of the form $\hat G/G$ where the symmetric space generators transform according to a representation $R$ of $G$, there is a gapped QCD theory with gauge group $G$ and quarks in the representation $R$, and vice versa.

A different perspective yields the same answer. The equality of the energy-momentum tensors of $\mathfrak{so}(\dim(R))_1$ and $\mathfrak g_{I(R)}$ implies, by definition, that the affine algebra $\mathfrak g_{I(R)}$ embeds conformally into the affine algebra $\mathfrak{so}(\dim(R))_1$. Conformal embeddings have been fully classified~\cite{PhysRevD.34.3092,ALEXANDERBAIS1987561,ARCURI1987327}, and we can read off the list of gapped QCD theories from this classification: for each conformal embedding of an algebra $\mathfrak g_k$ into $\mathfrak{so}(n)_1$ via a representation $R$ of $\mathfrak g$, there is a gapped QCD theory with gauge group $G$ and quarks in the representation $R$, and vice versa.

Either point of view yields table~\ref{tab:classification_gapped}. This table contains the list of ``minimal'' gapped QCD theories. Naturally, one can also take the tensor product of two gapped theories to obtain another gapped theory. This operation corresponds to reducible symmetric spaces, or non-maximal conformal embeddings; the most general symmetric space is a product of irreducible ones, and the most general conformal embedding is a sequence of maximal ones.

\begin{table}[!h]
\begin{equation*}
\arraycolsep=8pt
\begin{array}{c|c|c|c}
\mathfrak g&R & \text{IR TQFT} & \hat{\mathfrak g} \\ \hline
\mathfrak g&\text{Adj} & SO(\dim G)_1/G_h & \mathfrak g+ \mathfrak g \\
\mathfrak{so}(N)&\ydiagram2 & SO(\frac12(N+2)(N-1))_1/Spin(N)_{N+2} & \mathfrak{su}(N)\\
\mathfrak{sp}(N)&\ydiagram{1,1} & SO((2N+1)(N-1))_1/Sp(N)_{N-1} & \mathfrak{su}(2N)\\
\mathfrak s(\mathfrak u(M)+\mathfrak u(N))&\hphantom{_q}(\ydiagram1,\ydiagram1)_q&U(MN)_1/S(U(M)_N\times U(N)_M)&\mathfrak{su}(M+N)\\
\mathfrak{so}(M)+ \mathfrak{so}(N)&(\ydiagram1,\ydiagram1) & SO(MN)_1/(Spin(M)_N\times Spin(N)_M) & \mathfrak{so}(M+N)\\
\mathfrak u(N)&\hphantom{_q}\ydiagram{1,1}_q & U(\frac12 N(N-1))_1/U(N)_{N-2,2(N-1)q^2} & \mathfrak{so}(2N)\\
\mathfrak u(N)&\hphantom{_q}\ydiagram2_q & U(\frac12 N(N+1))_1/U(N)_{N+2,2(N+1)q^2} & \mathfrak{sp}(N)\\
\mathfrak{sp}(M)+ \mathfrak{sp}(N)&(\ydiagram1,\ydiagram1) & SO(4MN)_1/(Sp(M)_N\times Sp(N)_M) & \mathfrak{sp}(M+N)\\
\mathfrak{sp}(4)&\boldsymbol{42} & SO(42)_1/Sp(4)_7 & E_6\\
\mathfrak{su}(2)+ \mathfrak{su}(6)&(\boldsymbol{2},\boldsymbol{20}) & SO(40)_1/(SU(2)_{10}\times SU(6)_6) & E_6\\
\mathfrak{so}(10)+ \mathfrak u(1)&\boldsymbol{16}_q&U(16)_1/(Spin(10)_{4}\times U(1)_{16q^2})&E_6\\
F_4&\boldsymbol{26} & SO(26)_1/F_{4,3} & E_6\\
\mathfrak{su}(8)&\boldsymbol{70} & SO(70)_1/SU(8)_{10} & E_7\\
\mathfrak{su}(2)+ \mathfrak{so}(12)&(\boldsymbol{2},\boldsymbol{32}) & SO(64)_1/(SU(2)_{16}\times Spin(12)_8) & E_7\\
E_6+ \mathfrak u(1)&\boldsymbol{27}_q&U(27)_1/(E_{6,6}\times U(1)_{27q^2})&E_7\\
\mathfrak{so}(16)&\boldsymbol{128} & SO(128)_1/Spin(16)_{16} & E_8\\
\mathfrak{su}(2)+ E_7&(\boldsymbol{2},\boldsymbol{56}) & SO(112)_1/(SU(2)_{28}\times E_{7,12}) & E_8\\
\mathfrak{su}(2)+ \mathfrak{sp}(3)&(\boldsymbol{2},\boldsymbol{14}) & SO(28)_1/(SU(2)_{7}\times Sp(3)_{5}) & F_4\\
\mathfrak{so}(9)&\boldsymbol{16} & SO(16)_1/Spin(9)_2 & F_4\\
\mathfrak{so}(4)&(\boldsymbol{2},\boldsymbol{4}) & SO(8)_1/Spin(4)_{10,2} & G_2
\end{array}
\end{equation*}
\caption{List of irreducible gapped theories. The \textbf{first column} denotes the gauge algebra. Any global choice of $G$ for a given $\mathfrak g$ leads to a gapped theory. The \textbf{second column} denotes the representation of the quarks, either in the Young diagram notation or directly in terms of its dimension. $\ydiagram1,\ydiagram2,\ydiagram{1,1}$ denote the fundamental, symmetric, and anti-symmetric representations, respectively (with appropriate reality conditions, e.g.~Majorana if real, and with traces removed, if possible). $q$ denotes the charge of the fermions under $\mathfrak u(1)$ factors, if any; this charge can be chosen arbitrarily. The \textbf{third column} denotes the TQFT that describes the space of vacua of these gapped theories (see section~\ref{sec:Infrareddynamnics}); here we choose the simply-connected form $G_\text{sc}$ for concreteness. The \textbf{fourth column} denotes the Lie algebra $\hat{\mathfrak g}$ that makes $\hat G/G$ a symmetric spaces.}
\label{tab:classification_gapped}
\end{table} 

In QCD, this stacking operation gives rise to theories with decoupled gapped sectors so they are also gapped in a trivial way -- and so they are of little interest by themselves. The exception is when the different theories contain abelian $\mathfrak u(1)$ factors in their gauge group, in which case we can couple the minimal theories through these factors, which generates another gapped theory which is \emph{not} just the product of decoupled theories. With this in mind, the most general gapped QCD theory is either a theory in table~\ref{tab:classification_gapped}, or a product of such theories provided they contain a $\mathfrak u(1)$ gauge group, in which case the matrix of charges for these $\mathfrak u(1)$ factors must be non-singular. Any other gapped theory is a trivial product of these two options.

For illustration purposes, consider the gapped theory $\mathfrak u(N)+\ydiagram1_q$, where $\ydiagram 1_q$ denotes the fundamental representation and $q\in\mathbb Z$ is an arbitrary integer that specifies the charge of the quark under the trace part $\mathfrak u(1)\subset \mathfrak u(N)$. Stacking a family of these gapped theories, and coupling the abelian factors via an arbitrary matrix of charges, one obtains the gapped theory
\begin{equation}\label{eq:reducible_u}
\prod_{i=1}^m \mathfrak u(N_i),\qquad R=\bigoplus_{i=1}^m (\boldsymbol1,\dots,\boldsymbol1,\ydiagram1,\boldsymbol1,\dots,\boldsymbol1)_{\vec q_i}\,,
\end{equation}
where $\vec q_i$ is a vector of charges that specifies how the $i$-th fermion couples to $\mathfrak u(1)^m$. The special case where $N_i=1$ for all $i$ corresponds to abelian QCD, i.e., QED with $m$ photons and $m$ fermions, which we analyze in more detail in appendix~\ref{sec:abelian_G}.

The theories in table~\ref{tab:classification_gapped} are written in terms of non-chiral data. As a matter of fact, one can also modify these theories to obtain gapped theories that are chiral. The idea is that, if the theory with vector-like matter $(R_\ell,R_r)=(R,R)$ is gapped, then the theory with chiral matter $(R_\ell,R_r)=(\sigma_\ell\cdot R,\sigma_r\cdot R)$ is also gapped, where $\sigma_\ell,\sigma_r$ denote outer automorphisms of $\mathfrak g$ (see table~\ref{tab:comarksone}). For example, for simply-laced groups $\sigma\cdot R$ may denote the conjugate representation $\bar R$, while for $\mathfrak g=\mathfrak{so}(8)$, $\sigma\cdot R$ may denote any representation related by triality. As $\dim(\sigma\cdot R)=\dim(R)$ and $I(\sigma\cdot R)=I(R)$, the chiral theory with $(R_\ell,R_r)=(\sigma_\ell\cdot R,\sigma_r\cdot R)$ is also gapped. In the case of theories that contain abelian gauge groups, the statement becomes that one can use different charges for the two chiralities, $(q_\ell,q_r)$, provided they satisfy the gauge anomaly cancellation condition~\eqref{eq:abelian_gauge_anomaly_cancel}.

For example, given the gapped theory in~\eqref{eq:reducible_u}, one can generate other gapped theories by replacing some of the fundamentals by anti-fundamentals (for one chirality only, or for both), and also by assigning generically different $U(1)$ charges to the two chiralities $\vec q_i\mapsto(\vec q_{\ell,i},\vec q_{r,i}$).

Table~\ref{tab:classification_gapped}, together with the two operations we just described (stacking gapped theories and coupling them together via their abelian factors, and acting with outer automorphisms on the representations), give the extensive list of gapped QCD theories. See tables~\ref{tab:classification_gapped_algebra} and \ref{tab:classification_gapped_chiral} for summary of gapped vector-like and chiral theories respectively. Any other theory is either a trivial product of gapped theories, or is gapless.

\section{Infrared dynamics of $\boldsymbol{2d}$ QCD}
\label{sec:Infrareddynamnics}

Having classified all QCD theories that are gapped, and consequently, those that are gapless, it remains an interesting open question to determine the effective field theory describing the low energy dynamics. The most natural proposal is that the low energy theory is a gauged WZW coset model with chiral algebra (see~\cite{Kutasov:1994xq} and more recently~\cite{Isachenkov:2014zua,Komargodski:2020mxz})
\begin{mdframed}
\begin{equation}\label{eq:conj_IR_coset}
\frac{SO(\dim R_\ell)_1}{G_{I(R_\ell)}}\times\frac{SO(\dim R_r)_1}{G_{I(R_r)}}\,.
\end{equation}
\end{mdframed}
In order to simplify notation we will focus on the chiral half, with the understanding that the full theory is constructed by putting together the left and right sectors.

The idea behind~\eqref{eq:conj_IR_coset} is that QCD can be thought of as $\dim(R)$ free fermions, which can be described as the fermionic WZW theory $SO(\dim(R))_1$, where a symmetry $G\subset SO(\dim(R))$ has been gauged, and one has added a kinetic term for the gluons. The coupling constant $g$ is dimensionful so it grows in the infrared and it is self-consistent to assume that $g\to\infty$ as $E\to0$, which means that we can drop the gluon kinetic term for very low energies. All in all, it is expected that the deep infrared of QCD theories is described by the CFT coset~\eqref{eq:conj_IR_coset}, namely an $SO(\dim(R))_1$ WZW model with gauged $G_{I(R)}$ symmetry. The level of the gauge current algebra is determined by the Dynkin embedding index of $SO(\dim(R))\supset G$; this embedding is defined by the branching rule $\ydiagram1\mapsto R$, and hence the embedding index is $I(R)$.


This proposal is also suggested by our canonical analysis of section~\ref{sec:DLCQ}, where we highlighted the presence of $\mathfrak{so}(\dim(R))_1,\mathfrak g_{I(R)}$ current algebras in the Hamiltonian of QCD, and the fact that the operator $T_{\mathfrak{so}(\dim(R))_1}-T_{\mathfrak g_{I(R)}}$ naturally appears in this Hamiltonian, playing the role of the energy-momentum of a low-energy CFT that is decoupled from massive modes, which disappear in the deep infrared. 

\paragraph{Gapped spectrum.} One nice aspect of~\eqref{eq:conj_IR_coset} is that it is perfectly consistent with our classification from the previous section, because the coset~\eqref{eq:conj_IR_coset} is a full-fledged CFT if its central charge is non-zero, but describes a TQFT when its central charge vanishes. In other words, the chiral energy-momentum tensor of the coset is $T_{\mathfrak{so}(\dim(R))_1/\mathfrak g_{I(R)}}\equiv T_{\mathfrak{so}(\dim(R))_1}-T_{{\mathfrak g}_{I(R)}}$, and this is a non-trivial operator if and only if the theory is gapless. In any case, it is important to stress that the criterion for masslessness $T_{\mathfrak{so}(\dim(R))_1/\mathfrak g_{I(R)}}\neq 0$ was obtained in previous sections independently of the conjecture~\eqref{eq:conj_IR_coset}, but the two are perfectly consistent with each other, a fact that gives more evidence for the latter.

\paragraph{Continuous symmetries.} In $2d$, continuous chiral symmetries cannot appear nor disappear along a symmetric renormalization group flow. Therefore, the effective low energy description of QCD must have the exact same symmetries as the original ultraviolet theory. This is nicely reproduced by the coset, because the symmetries of both theories have the same definition: the flavor symmetry group is the commutant of $G$ inside $SO(\dim(R))$, i.e., the rotations of the chiral quarks that commute with gauge transformations.

\paragraph{'t Hooft anomalies.} Another nice property of the conjecture~\eqref{eq:conj_IR_coset} is that it automatically matches all the 't Hooft anomalies of the original QCD theory. Indeed, while the argument above does not strictly speaking prove that this coset is the low energy limit of the ultraviolet theory, it does prove that they are in the same deformation class. In other words, even though in principle the limits $E\to0$ and $g^2\to\infty$ need not be equivalent, it is still true that they are connected by a path in parameter space. Therefore, these two theories will carry the same 't Hooft anomalies for all the symmetries that are preserved along the path. This provides a strong consistency check on the proposal that the coset really is the low-energy limit of QCD.

The case of perturbative anomalies can be exhibited explicitly. The chiral flavor symmetry $H$ in the ultraviolet is generated by the free fermion currents that commute with the gauge group, that is, commutant of $G$ inside $SO(\dim(R))$. If $R$ is given by $N_F$ copies of a given irreducible representation $R_0$, that is $R=N_F\cdot R_0$, the flavor symmetry is $H(N_F)$, with $H=O,Sp,U$ for real, pseudo-real, and complex representations, respectively (in the complex case, the symmetry may be either $U(N_F)$ or $SU(N_F)$, depending on whether the diagonal $U(1)$ is broken by the ABJ anomaly or not, see section~\ref{sec:towards}).

The 't Hooft anomaly for $H$ is the Dynkin index of the representation under the flavor group~\eqref{eq:anomP}, in this case the fundamental representation. This means that the flavor symmetry carries $\dim(R_0)$ units of anomaly. This is reproduced by the coset in a straightforward manner, because one can write
\begin{equation}\label{eq:factor_WZW_coset}
\frac{SO(\dim(R))_1}{G_{I(R)}}\equiv H_{\dim(R_0)}\times \frac{SO(\dim(R))_1}{G_{I(R)}\times H_{\dim(R_0)}}\,.
\end{equation}
The factor $H_{\dim(R_0)}$ matches the ultraviolet 't Hooft anomaly, and the factor $\frac{SO(\dim(R))_1}{G_{I(R)}\times H_{\dim(R_0)}}$ has no continuous global symmetries (no commutant). We point out that this latter coset is actually well-defined, which might not be entirely obvious. One way to see this is that one could imagine gauging the diagonal symmetry $H$ in the ultraviolet (which is anomaly-free), to yield the gauge theory $G\times H+(R_0,\ydiagram1)$. The infrared coset for this theory is precisely $\frac{SO(\dim(R))_1}{G_{I(R)}\times H_{\dim(R_0)}}$.

The case of nonperturbative global anomalies is more subtle, and requires a case-by-case analysis. That being said, the argument above proves that the coset CFT will automatically match all the anomalies, perturbative and global. This has a nice bonus consequence, namely that it predicts that many well-known CFTs actually carry nonperturbative anomalies, a fact that may not have been fully appreciated in the past. For example, below we will describe many gauge theories that flow in the infrared to common CFTs such as minimal models or WZW models. These theories necessarily carry the same nonperturbative anomalies of the ultraviolet theory, and the latter are often easy to determine (because one can flow to the deep ultraviolet, where the fermions and gluons are essentially free and semiclassical considerations often suffice). Among others, this predicts global 't Hooft anomalies for discrete symmetries such time-reversal, whose presence is seldom discussed in the CFT literature.

While there is not much one can say about global anomalies in full generality, there is one feature that is actually rather universal. There are several discrete symmetries, such as discrete chiral symmetries or antiunitary time-reversal symmetry, whose anomalies have the following effect on the Hilbert space: when the number fermions in ultraviolet is odd, the Ramond Hilbert space is automatically supersymmetric~\cite{Delmastro:2021xox}. This is a nonperturbative statement that affects the whole spectrum of the theory and, in particular, the low-energy spectrum. Therefore, the effective infrared description must satisfy this property as well. This is indeed reproduced by the coset~\eqref{eq:conj_IR_coset}, because the states in the Ramond sector come from branchings from the spinor representation(s) of $SO(\dim(R))_1$; and, famously, when $\dim(R)$ is odd there is a single spinor whose Ramond-Ramond character is identically zero, a property that is inherited to the full coset. (Another diagnosis of this anomaly is that the twisted Hilbert space becomes ill-defined, which is also reproduced by the coset because the spinor character of $SO(\dim(R))_1$ carries a factor of $\sqrt2$ for odd $\dim(R)$, and hence the twisted partition function does not have an integral expansion; see~\eqref{eq:boson_fermion_spin_so} for the characters of $SO(n)_1$).

\paragraph{One-form symmetry.} QCD theories can have one-form symmetry associated to a subgroup  center of the gauge group (see table~\ref{tab:comarksone}). This symmetry is discrete,\footnote{If the gauge group is reductive, then the one-form symmetry may include $U(1)$ factors associated to the photons. These $U(1)$ groups exist only when the photons are free; otherwise the screening by quarks explicitly breaks $U(1)$ down to a discrete subgroup.} and hence by the generalized Coleman-Mermin-Wagner theorem~\cite{Gaiotto:2014kfa}, it cannot break spontaneously. Therefore, the infrared effective description must realize all the one-form symmetries of the ultraviolet theory.

In two dimensions, the effect of a one-form symmetry is to break up the theory into distinct sectors, or universes~\cite{Hellerman:2006zs}. The full Hilbert space of the theory is the direct sum of the Hilbert spaces of the different universes (see section \ref{sec:QCD2}). The total theory suffers from a mild violation of cluster decomposition, but the theory projected to a given universe is perfectly well-defined by itself, and satisfies decomposition.

Given a QFT with one-form symmetry, the emergent infrared CFT inherits it. Hence, in these CFTs  the vacuum is not unique (the coefficient of the vacuum character in the torus partition function is an integer larger than $1$). Instead, the infrared CFT is a direct sum (not a direct product) of ``conventional'' CFTs with a unique vacuum each.

In QCD, the one-form symmetry is the subgroup of the center that is not screened by the fermions, namely the kernel of the representation $R$ under which the quarks transform, $\Gamma=\ker (R)\subseteq Z(G)$. This is a symmetry for all $g^2$ and in particular it remains a symmetry in the $g^2\to\infty$ limit, and therefore the coset CFT also has a $\Gamma$ one-form symmetry. As $\Gamma$ does not act on the fermions, it does not embed into $SO(\dim(R))$, and hence in the quotient $SO(\dim(R))_1/G_{I(R)}$ we are trying to gauge a group $\Gamma$ that does not act on anything -- this is an orbifold by a symmetry that does not act faithfully (cf. with \cite{Sharpe:2019ddn,Robbins:2021ylj,Robbins:2021ibx}). This indeed leads to $|\Gamma|$ different universes, labelled by elements $\rho\in\Gamma^\vee$ (see section \ref{sec:QCD2}).

The CFT on a given universe labelled by $\rho$ corresponds to the coset $SO(\dim(R))_1/(G/\Gamma)_{I(R)}$ with a    theta term labelled by $\rho\in \Gamma^\vee$. The functional integral of the coset sums over $G/\Gamma$ bundles, which are labeled by $\Gamma$. The sum over bundles is weighted by the theta term. It is interesting to compare this perspective with the algebraic approach to cosets in the literature  \cite{Gepner:1989jq,Schellekens:1989uf,DiFrancesco:1997nk}. In the algebraic approach to cosets one organizes representations of the coset into long and short(er) orbits under the action of $\Gamma$, as $\Gamma$ permutes the coset representations. When the action of $\Gamma$ has only long orbits, the algebraic prescription is to divide the partition function by $\Gamma$, so that the vacuum character appears with multiplicity one, and only the trivial bundles contribute. This yields the partition function in one universe which, when there are no fixed points, is the same in all universes. When the model has shorter orbits, one has to deal with ``fixed point resolution", and correct by a series of prescriptions and ans\"atze for the fact that   characters enter with fractional multiplicity.  These prescriptions have a rather clear interpretation from our perspective. When the coset has no fixed points, the CFT in each universe is the same and only trivial bundles contribute. Instead, when the coset has fixed points, the CFT  in each universe is generically different. In order to identify the partition function in a given universe when there are fixed points, one must sum over nontrivial bundles, weighted by a discrete theta term, which gives a non-vanishing constant partition function \cite{Hori:1994uf}.  These contributions combine with   those of the long orbits to produce a partition function that is modular invariant in each universe. In a sense, the algebraic approach to cosets in the literature constructs the partition function in one universe, while from our  perspective one can construct more modular invariant partition functions by weighing the sum over nontrivial bundles (which are constant) by distinct discrete theta terms.


\paragraph{Central charge.} The central charge of the CFT in the deep ultraviolet is $\frac12\dim(R_{\ell/r})$, and in the deep infrared is $\frac12\dim(R_{\ell/r})-c(G_{I(R_{\ell/r})})$. Note that $c$ decreases and the dynamics are compatible with the $c$-theorem. Note also that both $c_\ell$ and $c_r$ decrease by the same amount (because $I(R_\ell)\equiv I(R_r)$, by gauge anomaly cancellation, cf.~\eqref{eq:UV_anomaly_simple}), which is a consequence of the conservation of the gravitational anomaly $c_\ell-c_r$. It might be interesting to note that gapped theories ``erase information maximally'' in the sense that they decrease the $c$ function as much as possible.

It should be pointed out that the infrared theory described by the coset~\eqref{eq:conj_IR_coset} is not expected to be robust under deformations in the ultraviolet. If we add mass terms or four-fermi terms, in general one would find that the infrared theory is deformed as well, and the coset~\eqref{eq:conj_IR_coset} flows to a different theory. This new theory has smaller (or equal) central charge. In the case of TQFTs, the central charge is already zero so deformations in the ultraviolet will map the infrared theory to a different TQFT, with generically fewer vacua. This is to be contrasted with the similar situation in $3d$: here, infrared TQFTs are actually robust under small ultraviolet deformations. The reason is that $2d$ TQFTs have local operators, while $3d$ TQFTs do not; therefore, local deformations in the ultraviolet map to non-trivial infrared operators in $2d$, but to the trivial operator in $3d$.

\paragraph{Some simple examples.} While we will work out plenty of examples in the next few subsections, we can list a couple of simple examples here, which will hopefully illustrate some of the main features.

Take the QCD theory $(SU(2); \boldsymbol{7}, \boldsymbol{7})$. The infrared dynamics is conjecturally described by the coset $SO(7)_1/SU(2)_{28}$. This CFT has central charge $c=7/10$, which agrees with the central charge of the tricritical Ising model. There are only two fermionic CFTs with this central charge: the tricritical Ising model itself (thought of as a fermionic CFT that does not in fact depend on the spin structure), or its fermionization. In other words, it is either a bosonic minimal model, promoted to fermionic in a trivial way, or it is a fermionic minimal model~\cite{Runkel:2020zgg,Hsieh:2020uwb,Kulp:2020iet}.

Here it is easy to determine which of these options is correct. In the deep ultraviolet there are $7$ free fermions, so the system carries $-1\mod8$ units of 't Hooft anomaly under the chiral $\mathbb Z_2$ symmetry. A bosonic theory cannot match this, so the second option is correct: this QCD system flows in the infrared to the fermionized tricritical Ising model.\footnote{We also show this directly in appendix~\ref{ap:minimal_model_coset}.} Note that this theory precisely matches the 't Hooft anomaly for the discrete chiral symmetry~\cite{Smith:2021luc}.


Once the correct low-energy degrees of freedom have been identified, one can ask several interesting questions. For example, one could try to determine the mapping between relevant operators in the ultraviolet to operators in the infrared. The spectrum of infrared operators, together with their quantum numbers, is well understood. The most relevant operator in the ultraviolet is the mass term, and the most relevant operator in the infrared is the $(1/10,1/10)$ operator, so it is a very natural guess that these operators are identified. Moreover, both are odd under the chiral $\mathbb Z_2$ symmetry. A similar analysis can be performed for the rest of operators. When the mapping is complete, one can study the deformed theory, where one adds suitable scalar operators to the Lagrangian; this gives us a window to the infrared of the \emph{massive} QCD theory, by turning on the deformation $(1/10,1/10)$ to the infrared CFT.

Finally, this scenario predicts that the fermionic tricritical Ising model is invariant under time-reversal, with $\mathsf T^2=(-1)^F$, and that this symmetry has a nonperturbative 't Hooft anomaly. This symmetry, and anomaly, are manifest in the ultraviolet, where it acts as $\psi(t)\mapsto \gamma^0\psi(-t)$, with 't Hooft anomaly measured by the number of fermions mod 2, in this case $7\equiv 1\mod 2$. It would be interesting to understand how this symmetry acts on the infrared CFT, and to determine its anomaly directly.

A very similar story holds for the QCD theory $(Spin(7); \boldsymbol 8,\boldsymbol 8)$. The infrared dynamics is conjecturally described by the coset $SO(8)_1/Spin(7)_1$. This CFT has central charge $c=1/2$, so it is either the bosonic Ising model (promoted to a fermionic theory in a trivial way), or the fermionized Ising model, i.e., a free Majorana fermion. As before, it is easy to determine which of these options is actually realized: there are $8$ fermions in the deep ultraviolet, so the $\mathbb Z_2$ chiral symmetry has no 't Hooft anomalies. This is only matched by the first option, namely the bosonic Ising model; hence, this is what QCD flows to in the infrared.\footnote{We also show this directly in appendix~\ref{ap:minimal_model_coset}.}

Much like above, one can try to determine how the ultraviolet operators are mapped to the infrared ones, and what happens when we deform the theory by these operators.

In these two examples we extracted the physics of the coset directly from its central charge. This was possible thanks to the fact that they are both smaller than unity: $c<1$. For generic QCD theories, the central charge is $c>1$ and its knowledge alone does not uniquely determine the CFT. In this situation, the properties of the infrared are to be extracted from the CFT $SO(n)/G_k$ by the standard coset construction. We review this construction in appendix~\ref{app:coset_cft}. Here we also revisit the $c=7/10$ and $c=1/2$ examples again, and confirm that they correspond to the fermionic tricritical Ising model and the bosonic Ising mode, respectively, by explicitly working out the branching functions of the coset.

From now on we will assume that the conjecture~\eqref{eq:conj_IR_coset} is correct. We can use it to propose explicit descriptions of the strongly coupled infrared dynamics of interesting QCD theories.

\subsection{Gapped theories}

Let us make a few remarks about QCD theories on table~\ref{tab:classification_gapped}; these theories are gapped, so their infrared involves a certain TQFT that describes their vacua.

\paragraph{Adjoint QCD.} The first interesting example is adjoint QCD, namely the gauge theory with gauge group $G$ and a fermion in the adjoint representation. This theory has received a lot of attention in the past, see~\cite{Dalley:1992yy,Kutasov:1993gq,Boorstein:1993nd,Gross:1997mx,Katz:2013qua,Dubovsky:2018dlk,Cherman:2019hbq,Komargodski:2020mxz} for a sample of papers.

The vacua of these theories are described by the topological coset
\begin{equation}
\frac{SO(\dim(\mathfrak g))_1}{G_h}\,,
\end{equation}
where $h$ is the dual Coxeter number of $\mathfrak g$ (cf.~table~\ref{tab:comarksone}). The branching functions of this coset are well understood~\cite{KAC1988156}:
\begin{equation}\label{eq:adjoint_branch_main}
\begin{aligned}
d_\text{NS-NS}&=\sum_{\lambda\in \mathcal R}\chi_\lambda\\
d_\text{NS-R}&=\sum_{\lambda\in \mathcal R}(-1)^{h_\lambda}\chi_\lambda\\
d_\text{R-NS}&=2^{r/2}\chi_\rho\\
d_\text{R-R}&=0\,,
\end{aligned}
\end{equation}
where $r=\operatorname{rank}(\mathfrak g)$, $\rho$ denotes the Weyl vector and
\begin{equation}
\mathcal R=\{\lambda\,\mid\,\exists\hat w\in\hat W\text{ such that }\lambda=h\hat\omega_0+(\hat w-1)\hat\rho\}\,.
\end{equation}

From these equations it immediately follows that $\mathcal H_\text{R}$ is always supersymmetric and has $2^r$ states. The space $\mathcal H_\text{NS}$ also has these many states and it is purely bosonic. In other words,
\begin{equation}
\text{infrared of adjoint QCD:}\qquad \begin{cases}
\mathcal H_\text{NS}\hspace{-6pt}&=\ \mathbb C^{2N|0}\\
\mathcal H_\text{R}\hspace{-6pt}&=\ \mathbb C^{N|N}
\end{cases},\qquad N:=2^{r-1}\,.
\end{equation}

Furthermore, by explicitly constructing $\mathcal R$ for the different simple algebras, one observes that half the states in $\mathcal H_\text{NS}$ are charged under $(-1)^{F_L}$, and the other half is not. In other words, half the representations in $\mathcal R$ have integral spin $h_\lambda\in\mathbb Z$, and the other half have half-integral spin $h_\lambda\in\mathbb Z+\frac12$. The only exception is $SU(2n+1)$, which has $2^{n-1}(2^n+1)$ states with integral spin and $2^{n-1}(2^n-1)$ states with half-integral spin. 

\paragraph{QCD with bifundamentals.} The next few interesting examples correspond to theories with gauge group $G\times G$ and fermions in the bifundamental representation, namely
\begin{equation}
\begin{aligned}
S(U(N)\times U(M))&+(\ydiagram1,\ydiagram1)\\
SO(N)\times SO(M)&+(\ydiagram1,\ydiagram1)\\
Sp(N)\times Sp(M)&+(\ydiagram1,\ydiagram1)\,,
\end{aligned}
\end{equation}
whose space of vacua are described by the following cosets:
\begin{equation}
\frac{U(NM)_1}{S(U(N)_M\times U(M)_N)},\qquad \frac{SO(NM)_1}{SO(N)_M\times SO(M)_N},\qquad\frac{SO(4NM)_1}{Sp(N)_M\times Sp(M)_N}\,.
\end{equation}

The branching rules of these cosets are well-known~\cite{1989741,NACULICH1990417,cmp/1104202741,cmp/1104249321}: they describe the level-rank dualities $\mathfrak g(N)_k\leftrightarrow\mathfrak g(k)_N$. The decomposition of numerator characters $d_{\pm,\pm}$ into denominator characters $\chi$ takes the following general form:
\begin{equation}\label{eq:level_rank_duality_branch}
\begin{aligned}
d_\text{NS-X}(q,g_1,g_2)&=\sum_\lambda (\pm1)^{2(h_\lambda+h_{\lambda^t})}\chi_\lambda(q,g_1)\chi_{\lambda^t}(q,g_2)\\
d_\text{R-X}(q,g_1,g_2)&=\sum_\lambda (\pm1)^{2(h_\lambda+h_{\lambda^t})}\chi_{\gamma\cdot\lambda}(q,g_1)\chi_{\lambda^t}(q,g_2)\,,
\end{aligned}
\end{equation}
where $\lambda$ denotes a primary of $\mathfrak g(N)_k$, and $\lambda^t$ the primary of $\mathfrak g(k)_N$ obtained from $\lambda$ by transposing the Young diagram. Moreover, $h_\lambda$ denotes the conformal dimension of $\lambda$, and $\gamma\in Z(G_\text{sc})$ a suitable simple current. Finally, $g_1,g_2$ denote flavor symmetry elements of $\mathfrak g(N),\mathfrak g(k)$, respectively. 

These branching functions imply that there are as many vacua as there are primaries in $\mathfrak g$, i.e., the number of vacua is (see e.g.~\cite{Delmastro:2020dkz})
\begin{equation}
\begin{aligned}
S(U(N)\times U(M))&\colon\quad\binom{N+M-1}{M}\\[2ex]
SO(N)\times SO(M)&\colon\quad \begin{cases}
\binom{n+m}{m}+\binom{n+m-1}{m-1}& (N,M)=(2n,2m+1)\\[2ex]
\frac12\binom{n+m}{m}+\binom{n+m-1}{m-1}+\frac12\binom{n+m-2}{m-2}+\frac32\binom{n+m-2}{m}& (N,M)=(2n,2m)\\[2ex]
\frac12\binom{n+m}{m}+\frac12\binom{n+m-1}{m-1}+\frac32\binom{n+m-1}{m} & (N,M)=(2n+1,2m)\\[2ex]
\binom{n+m}{m}& (N,M)=(2n+1,2m+1)
\end{cases}\\[2ex]
Sp(N)\times Sp(M)&\colon\quad\binom{N+M}{M}\,.
\end{aligned} 
\end{equation}

\paragraph{QCD with rank-2.} Another interesting example is the theories with rank-2 quarks. Here we discuss the two theories
\begin{equation}
\begin{aligned}
Spin(N)&+\ydiagram2\\
Sp(N)&+\,\ydiagram{1,1}
\end{aligned}
\end{equation}
whose vacua are described by the cosets
\begin{equation}
\frac{SO((N+2)(N-1)/2)_1}{Spin(N)_{N+2}},\qquad \frac{SO((2N+1)(N-1))_1}{Sp(N)_{N-1}}\,.
\end{equation}
We are not aware of an explicit discussion of the branchings of these cosets in the literature. That being said, the Ramond sector turns out to be particularly simple, and is reminiscent of the adjoint case~\eqref{eq:adjoint_branch_main}:
\begin{equation}\label{eq:branch_sym_asym}
\begin{aligned}
d_\text{R-R}&=0\,,\\
\text{$Spin(N)$, $N$ odd:}\qquad d_\text{R-NS}&=2^{(N-1)/4}\chi_{[3, 1,1,\dots,1,1, 3]}+2^{(N-1)/4}\chi_{[1, 3,1,1,\dots,1,1, 3]}\\
\text{$Spin(N)$, $N$ even:}\qquad d_\text{R-NS}&=2^{(N-2)/4}\chi_{[3, 1,1,\dots,1,1, 3]}+2^{(N-2)/4}\chi_{[1, 3,1,1,\dots,1,1, 3]}+\text{c.c.}\\
\text{$Sp(N)$:}\qquad d_\text{R-NS}&=2^{(N-1)/2}\chi_{[0,1,\dots,1,0]}\,.
\end{aligned}
\end{equation}
From this we automatically conclude that there are $2^{\lfloor N/2\rfloor+1}$ and $2^{N-1}$ vacua, respectively. In the Ramond sector these vacua are split half-and-half into bosons and fermions, while in the Neveu-Schwartz sector they are all bosonic. In this latter sector, half the states are charged under $(-1)^{F_L}$ and the other half is neutral (i.e., half the primaries have integral spin and the other half have half-integral spin).

The branching rules for the other gapped theories with rank-2 quarks, namely $U(N)$ plus a symmetric or anti-symmetric quark, are analyzed in~\cite{cmp/1104274518}.

\paragraph{Exceptionals.} We close this section with an example involving an exceptional Lie group, to wit
\begin{equation}
\begin{aligned}
F_4+\boldsymbol{26}\,.
\end{aligned}
\end{equation}
The vacua of this theory are described by the coset
\begin{equation}
\frac{SO(26)_1}{F_{4,3}}\,,
\end{equation}
whose branching functions are
\begin{equation}
\begin{aligned}
d_\text{NS-X}&=\chi_{\boldsymbol 1}\pm\chi_{\boldsymbol{26}}+\chi_{\boldsymbol{273}}\pm \chi_{\boldsymbol{1274}}\\
d_\text{R-NS}&=2\chi_{\boldsymbol{4096}}\\
d_\text{R-R}&=0\,.
\end{aligned}
\end{equation}
Hence, this theory has $\mathcal H_\text{NS}=\mathbb C^{4|0}$ and $\mathcal H_\text{R}=\mathbb C^{2|2}$.

\subsection{QCD with fundamental matter}

Here we describe the infrared dynamics of QCD with quarks in the fundamental representation. More precisely, we shall discuss the following theories
\begin{itemize}
\item $SU(N)+N_F\,\ydiagram1$.
\item $SO(N)+N_F\,\ydiagram1$.
\item $Sp(N)+N_F\,\ydiagram1$.
\end{itemize}
These describe the celebrated 't Hooft model~\cite{THOOFT1974461}.

The coset CFTs that describe the low energy limit of these theories are
\begin{equation}
\frac{U(NN_F)_1}{SU(N)_{N_F}},\qquad \frac{SO(NN_F)_1}{SO(N)_{N_F}},\qquad \frac{SO(4NN_F)_1}{Sp(N)_{N_F}}\,.
\end{equation}
We now claim that these CFTs are in fact the well-known WZW theories
\begin{equation}
U(N_F)_N,\qquad SO(N_F)_N,\qquad Sp(N_F)_N\,.
\end{equation}

Indeed, the characters of the coset $SO(\cdots)_1/\mathfrak g(N)_{N_F}$ are given by the coefficients of the characters of $\mathfrak g(N)_{N_F}$ in the decomposition of $SO(\cdots)_1$; but, as in~\eqref{eq:level_rank_duality_branch}, these coefficients are precisely the characters of $\mathfrak g(N_F)_N$. In other words, the equality $SO(\cdots)_1/\mathfrak g(N)_{N_F}\equiv\mathfrak g(N_F)_N$ is tantamount to the level-rank duality $\mathfrak g(N)_{N_F}\leftrightarrow\mathfrak g(N_F)_N$.

Let us make a few remarks:
\begin{itemize}
\item Note that the infrared CFT is just the WZW model for the flavor symmetry. This CFT manifestly matches the perturbative 't Hooft anomalies for the flavor symmetry in the ultraviolet. So this is the simplest scenario for the infrared dynamics, and could have been guessed independently of the general conjecture~\eqref{eq:conj_IR_coset}. These WZW models also match the nonperturbative anomalies, although in a less obvious way (see below for an explicit example).

\item The equality $SO(\cdots)_1/\mathfrak g(N)_{N_F}\equiv\mathfrak g(N_F)_N$ can also be understood as the consequence of the triviality of the coset $SO(\cdots)_1/(\mathfrak g(N)_{N_F}\times\mathfrak g(N_F)_N)$, i.e., of the fact that the gauge theory obtained from $G+N_F\,\ydiagram1$ by gauging the flavor symmetry is gapped.

\item Similar considerations hold for other gauge groups; for example, if we use $Spin(N)$ instead of $SO(N)$, the flavor symmetry is $O(N_F)$ instead of $SO(N_F)$, and the infrared CFT is a WZW model with target space $O(N_F)$. This is again a consequence of the level-rank duality $Spin(N)_{N_F}\leftrightarrow O(N_F)_N$~\cite{Cordova:2017vab}. Similarly, one could use $U(N)$ instead of $SU(N)$, in which case the infrared CFT is $SU(N_F)_N$, again by level-rank duality.

\item This predicts for example that $SO(N_F)_N$ has an 't Hooft anomaly for time-reversal, measured by $NN_F\mod2$.

\end{itemize}

An interesting special case is $SU(N)+\ydiagram1$, i.e., a single copy of the fundamental representation $N_F=1$. The infrared coset in this case is $U(1)_N$. For $N=3$ this coset is actually a supersymmetric minimal model, $U(1)_3=\mathcal M^{\mathcal N=2}_1$, and therefore $SU(3)+\boldsymbol 3$ has emergent supersymmetry in the infrared. This is a consequence of the fact that $\wedge^3\ydiagram 1$ contains a gauge singlet, if and only if $N=3$~\cite{GODDARD1985226}.

In the ultraviolet of $SU(N)+\ydiagram1$ there is a manifest $\mathbb Z_2$ chiral symmetry that acts as $(-1)^{F_L}:\psi\mapsto\gamma^3\psi$, and whose 't Hooft anomaly is the number of Majorana fermions, $2N\mod8$. This anomaly must be reproduced by the infrared degrees of freedom, i.e., by $U(1)_N$. We check this as follows.

The equality $U(N)_1/SU(N)_1=U(1)_N$ is due to the character decomposition
\begin{equation}
\begin{aligned}
d_\text{NS-X}(q,g,\theta)&=\sum_{\ell=0}^{N-1}(\pm1)^\ell \chi^\text{NS-X}_\ell(q,\theta)\chi_{\gamma^\ell\cdot\boldsymbol0}(q,g)\\
d_\text{R-X}(q,g,\theta)&=\sum_{\ell=0}^{N-1}(\pm1)^{\ell+1} \chi^\text{R-X}_{\ell+\lfloor N/2\rfloor}(q,\theta)\chi_{\gamma^\ell\cdot\boldsymbol0}(q,g)\,,
\end{aligned}
\end{equation}
where $\gamma$ is the generator of the $Z(SU(N))=\mathbb Z_N$ center symmetry and $\boldsymbol0$ is the vacuum character. Also, $\chi_\ell(q,\theta)$ denote the regular (bosonic) characters of $U(1)_N$ if $N$ is even, and the super-characters if $N$ is odd; and $\theta$ denotes a $U(1)$ flavor fugacity. Finally, $\chi_\lambda(q,g)$ denotes an $SU(N)_1$ character with $g\in SU(N)$ flavor fugacity.

These branching relations imply that the characters of the coset CFT are
\begin{equation}
\begin{aligned}
b^\text{NS-NS}_\ell&=\chi^\text{NS-NS}_\ell\\
b^\text{NS-R}_\ell&=(-1)^\ell\chi^\text{NS-R}_\ell\\
b^\text{R-NS}_\ell&=\chi^\text{R-NS}_{\ell+\lfloor N/2\rfloor}\\
b^\text{R-R}_\ell&=(-1)^\ell\chi^\text{R-R}_{\ell+\lfloor N/2\rfloor}\,.
\end{aligned}
\end{equation}
Hence, the partition function twisted by the $(-1)^{F_L}$ symmetry is
\begin{equation}
\begin{aligned}
\tr_{\mathcal H_\text{NS}}\bigl((-1)^{F_L}q^{L_0-1/24}\bar q^{\bar L_0-1/24}\bigr)&=\sum_{\ell=0}^{N-1}\bar b^\text{NS-R}_\ell b^\text{NS-NS}_\ell\\
&=\sum_{\ell=0}^{N-1}(-1)^\ell\bar\chi^\text{NS-R}_\ell \chi^\text{NS-NS}_\ell\,.
\end{aligned}
\end{equation}
The 't Hooft anomaly under $(-1)^{F_L}$ is measured by the phase acquired by this partition function under an $ST^2S^{-1}$ modular transformation. The idea is that $S$ moves the operator $(-1)^{F_L}$ from the spatial cycle into the temporal cycle, so it allows us to access the spin of the operators in the twisted sector. In a non-anomalous fermionic theory, the spin should be half-integral; hence, $T^2$ measures precisely the extent to which this condition fails. If we use the modular matrices of $U(1)_N$ (see e.g.~\cite{Okuda:2020fyl,Delmastro:2021xox}), we obtain $ST^2S^{-1}=e^{2\pi i \frac{2N}{8}}$, precisely as in the ultraviolet.


\subsection{WZW models}

We noticed in the previous section that the infrared CFT that describes $G+N_F\,\ydiagram1$ is \emph{just} the WZW model for the chiral flavor symmetry. An interesting question one could ask is how general this situation is, i.e., for which QCD theories are the infrared degrees of freedom just the affinization of the ultraviolet currents. As stated in~\eqref{eq:factor_WZW_coset}
\begin{equation}
\frac{SO(\dim(R))_1}{G_{I(R)}}\equiv H_{\dim(R_0)}\times \frac{SO(\dim(R))_1}{G_{I(R)}\times H_{\dim(R_0)}}\,,
\end{equation}
one can always factor out these currents from the infrared coset, and the question becomes for which theories is the remaining sector a trivial CFT. Note that this extra part is precisely the infrared coset of the QCD theory where one gauges the (diagonal) flavor symmetry, $G\times H+(R_0,\ydiagram1)$. From this perspective, the answer is straightforward: the theory $G+R$ flows in the infrared to $H_{\dim(R_0)}$ (plus possibly a trivial CFT, i.e., a TQFT) if and only if the theory $G\times H+(R_0,\ydiagram1)$ is gapped. But now we can utilize our classification of gapped theories (cf.~table~\ref{tab:classification_gapped}) to give the list we are after. This way one obtains table~\ref{tab:classification_WZW}.

\begin{table}[h!]
\begin{equation*}
\begin{array}{c|c|c}
G&R & \text{IR WZW} \\ \hline
G\in \text{table~\ref{tab:classification_gapped}}&R & \varnothing \\
SU(N)&N_F\,\ydiagram1&U(N_F)_N\\
U(N)&N_F\,\ydiagram1&SU(N_F)_N\\
SO(N)&N_F\,\ydiagram1 & SO(N_F)_N\\
Sp(N)&N_F\,\ydiagram1 & Sp(N_F)_N\\
SU(N)&\ydiagram{1,1} & U(1)_{\frac12N(N-1)}\\
SU(N)&\ydiagram2 & U(1)_{\frac12N(N+1)}\\
Spin(10)&\boldsymbol{16}&U(1)_{16}\\
E_6&\boldsymbol{27}&U(1)_{27}\\
SU(2)&\boldsymbol{2} &SU(2)_1\\
SU(2)&\boldsymbol{4} &SU(2)_2\\
Sp(3)&\boldsymbol{14} & SU(2)_7\\
SU(6)&\boldsymbol{20} & SU(2)_{10}\\
Spin(12)&\boldsymbol{32} & SU(2)_{16}\\
E_7&\boldsymbol{56} & SU(2)_{28}
\end{array}
\end{equation*}
\caption{Classification of QCD theories that realize in the infrared a pure WZW model for the ultraviolet flavor symmetry (modulo a TQFT). Any theory not on this table will flow in the infrared to $H$ current algebra plus a non-trivial CFT (which has no continuous flavor symmetry). The first line ``$G\in \text{table~\ref{tab:classification_gapped}}$'' refers to the fact that gapped theories themselves satisfy this criterion, in the sense that both their flavor symmetry and their infrared CFT are trivial.}
\label{tab:classification_WZW}
\end{table}

\subsection{Minimal models}

It is interesting to note that (both SUSY and non-SUSY) minimal models~\cite{10.1007/3-540-09238-2_102,BELAVIN1984333,FRIEDAN198537,Petersen:1985bs,Goddard:1984vk,Boucher:1986bh,DiVecchia:1986fwg}, a celebrated family of $2d$ CFTs, also appear in the infrared of QCD gauge theories. We already noticed three instances of this phenomenon so far, where we found QCD theories that flow to Ising, tri-critical Ising, and a compact boson at the SUSY radius $R^2=3$ in the infrared (the latter being the first $\mathcal N=2$ minimal model). Here we describe some families of QCD theories that realize all the minimal models. The examples are by no means exhaustive: there are many other QCD theories that also flow to minimal models in the infrared. In order to simplify the discussion, in this and subsequent examples we shall make no distinction between the bosonized/fermionized versions of a given CFT, and we will not be careful with certain discrete quotients of the gauge group (so for example $SO(4)=SU(2)\times SU(2)$).

\paragraph{Virasoro minimal models.} Consider the following QCD theories:
\begin{equation}
\begin{aligned}
SU(2)^3\times SO(k)&+(\boldsymbol2,\boldsymbol1,\boldsymbol 2,\boldsymbol1)+(\boldsymbol2,\boldsymbol2,\boldsymbol1,\ydiagram1)\\
SU(2)^2\times Sp(k)&+(\boldsymbol2,\boldsymbol2,\boldsymbol1)+(\boldsymbol1,\boldsymbol2,\ydiagram1)\,.
\end{aligned}
\end{equation}
Their infrared is
\begin{equation}
\frac{SO(4k+4)_1}{SU(2)_{k+1}\times SU(2)_k\times SU(2)_1\times SO(k)_4},\qquad \frac{SO(4k+4)_1}{SU(2)_1\times SU(2)_{k+1}\times Sp(k)_1}\,,
\end{equation}
and we claim that these are both coset realizations of $\mathcal M_k$. Indeed, they can both be written as
\begin{equation}
\mathcal M_k\equiv \frac{SU(2)_k\times SU(2)_1}{SU(2)_{k+1}}
\end{equation}
thanks to the level-rank dualities $SO(nm)_1/SO(n)_m=SO(m)_n$ and $SO(4nm)_1/Sp(n)_m=Sp(m)_n$.

\paragraph{$\boldsymbol{\mathcal N=1}$ minimal model.} Consider the QCD theories
\begin{equation}
\begin{aligned}
SU(2)^2\times Sp(2)\times SO(k)+(\boldsymbol2,\boldsymbol1,\boldsymbol4,\boldsymbol1)+(\boldsymbol2,\boldsymbol2,\boldsymbol1,\ydiagram1)\\
SU(2)^2\times SO(2)\times Sp(k)+(\boldsymbol2,\boldsymbol1,0,\ydiagram1)+(\boldsymbol2,\boldsymbol2,1,\boldsymbol1)\,.
\end{aligned}
\end{equation}
Their infrared is
\begin{equation}
\frac{SO(4k+8)_1}{SU(2)_{k+2}\times SU(2)_k\times Sp(2)_1\times SO(k)_4},\quad \frac{SO(4k+8)_1}{SU(2)_{k+2}\times SU(2)_2\times SO(2)_4\times Sp(k)_1}
\end{equation}
and we claim that these are both coset realizations of $\mathcal M^{\mathcal N=1}_k$. Indeed, they can both be written as
\begin{equation}
\mathcal M_k^{\mathcal N=1}=\frac{SU(2)_k\times SU(2)_2}{SU(2)_{k+2}}
\end{equation}
for the same reason as for $\mathcal M_k$.


%

\subsection{Diagonal coset}

Here we discuss a class of QCD theories whose infrared leads to the so-called diagonal cosets $(\mathfrak g_k\times\mathfrak g_{k'})/\mathfrak g_{k+k'}$, whose structure is better understood than that of generic cosets~\cite{1989IJMPA...4..897C,FUCHS1996371}. In particular, consider the following linear quivers:
\begin{equation}
\begin{aligned}
S(U(N)\times U(M)\times U(L))+(\ydiagram 1,\ydiagram 1,\boldsymbol1)+(\boldsymbol1,\ydiagram 1,\ydiagram 1)\\
SO(N)\times SO(M)\times SO(L)+(\ydiagram 1,\ydiagram 1,\boldsymbol1)+(\boldsymbol1,\ydiagram 1,\ydiagram 1)\\
Sp(N)\times Sp(M)\times Sp(L)+(\ydiagram 1,\ydiagram 1,\boldsymbol1)+(\boldsymbol1,\ydiagram 1,\ydiagram 1)\,.
\end{aligned}
\end{equation}
Their infrared theories are
\begin{equation}
\begin{aligned}
&\frac{U(NM+LM)_1}{S(U(N)_M\times U(M)_{N+L}\times U(L)_M)},\\
&\hspace{4cm}\frac{SO(NM+LM)_1}{SO(N)_M\times SO(M)_{N+L}\times SO(L)_M},\\
&\hspace{8cm} \frac{SO(4NM+4LM)_1}{Sp(N)_M\times Sp(M)_{N+L}\times Sp(L)_M}\,,
\end{aligned}
\end{equation}
which, thanks to level-rank duality, become the following diagonal cosets
\begin{equation}
\frac{SU(M)_N\times SU(M)_L}{SU(M)_{N+L}},\qquad \frac{SO(M)_N\times SO(M)_L}{ SO(M)_{N+L}},\qquad \frac{Sp(M)_N\times Sp(M)_L}{Sp(M)_{N+L}}\,.
\end{equation}

\subsection{Kazama-Suzuki}

Here we describe QCD theories that acquire an emergent $\mathcal N=2$ supersymmetry in the infrared. In particular, they become Kazama-Suzuki models~\cite{KAZAMA1989112}.

\begin{figure}[!h]
\centering
\begin{tikzpicture}

\fill (0,0) circle (2pt);

\fill (1.5,0) circle (2pt);
\fill (2.5,0) circle (2pt);
\draw[thick] (1.5,0) -- (2.5,0);

\fill (4,-.3) circle (2pt);
\fill (4.8,-.3) circle (2pt);
\fill (4.4,.4) circle (2pt);
\draw[thick] (4,-.3) -- (4.8,-.3) -- (4.4,.4) -- cycle;

\fill (6.3,-.4) circle (2pt);
\fill (7.1,-.4) circle (2pt);
\fill (6.3,.4) circle (2pt);
\fill (7.1,.4) circle (2pt);
\draw[thick] (6.3,-.4) -- (7.1,-.4) -- (7.1,.4) -- (6.3,.4) -- cycle;
\draw[thick] (6.3,-.4) -- (7.1,.4);
\draw[thick] (7.1,-.4) -- (6.3,.4);

\fill (8.4,.1) circle (2pt);
\fill (8.58541, -0.470634) circle (2pt);
\fill (9.18541, -0.470634) circle (2pt);
\fill (9.37082, 0.1) circle (2pt);
\fill (8.88541, 0.452671) circle (2pt);

\draw[thick] (8.4,.1) -- (8.58541, -0.470634) -- (9.18541, -0.470634) -- (9.37082, 0.1) -- (8.88541, 0.452671) -- cycle;
\draw[thick] (8.4,.1) -- (9.18541, -0.470634) -- (8.88541, 0.452671) -- (8.58541, -0.470634) -- (9.37082, 0.1) -- cycle;

\end{tikzpicture}
\caption{The first few complete graphs $K_1,K_2,\dots,K_5$. If we associate to each node a gauge group $U(n_i),SO(n_i),Sp(n_i)$ (with the global $U(1)$ modded out in the unitary case), and to each edge a bifundamental quark, then the quiver gauge theory associated to $K_n$ has a Kazama-Suzuki model as its effective low energy description.}
\label{fig:complete_graph}
\end{figure}
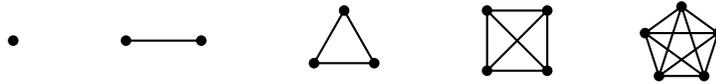

Consider the QCD quiver associated to an arbitrary complete graph $K_n$, where each node represents a gauge group $G_i$, and each edge a bifundamental quark (see figure~\ref{fig:complete_graph}). We take the gauge groups to be any of
\begin{equation}
\begin{aligned}
G&=S(U(N_1)\times U(N_2)\times\cdots \times U(N_n))\\
G&=SO(N_1)\times SO(N_2)\times\cdots \times SO(N_n)\\
G&=Sp(N_1)\times Sp(N_2)\times\cdots \times Sp(N_n)\,.
\end{aligned}
\end{equation}

Note that for $n=1,2$ these are pure Yang-Mills and $G\times G+(\ydiagram1,\ydiagram1)$, i.e., they are both gapped theories (the latter are entries on table~\ref{tab:classification_gapped}). We claim that for $n\ge3$, the theory has emergent $\mathcal N=2$ supersymmetry in the infrared, they are Kazama-Suzuki models. Indeed, their infrared cosets are
\begin{equation}
\frac{U\bigl(\sum_{i> j} N_iN_j\bigr)_1}{S\bigl(\prod_i U(N_i)_{\sum_{j\neq i}N_j}\bigr)},\qquad \frac{SO\bigl(\sum_{i> j} N_iN_j\bigr)_1}{\prod_i SO(N_i)_{\sum_{j\neq i}N_j}},\qquad \frac{SO\bigl(4\sum_{i> j} N_iN_j\bigr)_1}{\prod_i Sp(N_i)_{\sum_{j\neq i}N_j}}\,,
\end{equation}
which are the cosets that describe Kazama-Suzuki models associated to the embeddings
\begin{equation}
\begin{aligned}
&SU\bigl(\sum_{i\neq\star}N_i\bigr)\supset S(\prod_{i\neq\star}U(N_i)\bigr),\\
&\hspace{4cm}SO\bigl(\sum_{i\neq\star}N_i\bigr)\supset \prod_{i\neq\star}SO(N_i),\\
&\hspace{8cm} Sp\bigl(\sum_{i\neq\star}N_i\bigr)\supset \prod_{i\neq\star}Sp(N_i)
\end{aligned}
\end{equation}
at level $N_\star$, where $\star$ is an arbitrary node of $K_n$.

\section*{Acknowledgments}

We would like to thank Joaquim Gomis, Kentaro Hori, Theo Johnson-Freyd, Zohar Komargodski, Adam Schwimmer, and Ryan Thorngren for useful discussions. We would also like to thank the organizers of Strings 2021, where this material was first presented; and the participants of the conference for comments. Research at Perimeter Institute is supported in part by the Government of Canada through the Department of Innovation, Science and Economic Development Canada and by the Province of Ontario through the Ministry of Colleges and Universities. Any opinions, findings, and conclusions or recommendations expressed in this material are those of the authors and do not necessarily reflect the views of the funding agencies. 
\vfill\eject

\vfill\eject

\appendix
 
\section{Conventions and Background}
\label{app:conven}
 
We work in $1+1$ dimensional Minkowski spacetime with metric $\eta=\diag(-1,+1)$ and $2\times2$ gamma matrices $\gamma^\mu$, $\mu=0,1$. The Hodge dual of a one-form is $(\star j)_\mu=\epsilon_{\mu\nu}j^\nu$ where we take $\epsilon^{tx}=-\epsilon_{tx}=+1$. In particular, $\star\mathrm dt=\mathrm dx,\ \star\mathrm dx=\mathrm dt$. In null coordinates $x^\pm$ the metric is $\eta_{++}=\eta_{--}=0$, $\eta_{+-}=\eta_{-+}=1$, and the star is $\star\mathrm dx^\pm=\pm\mathrm dx^\pm$.

The minimal spinor is Majorana-Weyl, namely one can impose the simultaneous conditions $\gamma^3\psi=\pm\psi$ and $\psi^*=C\psi$, where $\gamma^3=\gamma^0\gamma^1$ is the chirality matrix and $C$ is the charge-conjugation matrix, defined by $(\gamma^\mu)^*= C\gamma^\mu C^{-1}$. It is convenient to choose the Majorana basis $\gamma^\mu=(i \sigma^2,\sigma^1)$ where $\gamma^3=\sigma^3$ and $C=1$. In this basis, Majorana fermions are real $\psi^*=\psi$, and chiral fermions are either $\psi\propto \begin{pmatrix}1\\0\end{pmatrix}$ or $\psi\propto \begin{pmatrix}0\\1\end{pmatrix}$.
We take $\psi=2^{-1/4}\begin{pmatrix}\psi_\ell\\\psi_r\end{pmatrix}$ and $x^\pm=\frac{1}{\sqrt2}(x^0\pm x^1)$. The fermion kinetic term is
\begin{equation}
-i\bar\psi\gamma^\mu\partial_\mu\psi=i\psi_\ell^*\partial_-\psi_\ell+i\psi_r^*\partial_+\psi_r\,.
\end{equation}
For Grassmann odd $a,b$ we use $(ab)^*=b^*a^*$.

\paragraph{Mass terms.} For massless fermions the two chiralities are decoupled. These couple through mass terms
\begin{equation}\label{eq:regular_mass}
\begin{aligned}
i\bar\psi \psi&=\frac{i}{\sqrt2}(\psi_\ell^*\psi_r+\psi_\ell\psi_r^*)\\
i\bar \psi \gamma^3 \psi&=\frac{-i}{\sqrt2}(\psi_\ell^*\psi_r-\psi_\ell\psi_r^*)\,.
\end{aligned}
\end{equation}
both of which are hermitian. Although less obvious, one can also use bilinears of the form
\begin{equation}\label{eq:exceptional_mass}
\begin{aligned}
\re(\bar\psi^* \psi)&=\frac{1}{\sqrt2}(\psi_\ell\psi_r-\psi_\ell^*\psi_r^*)\\
\im(\bar\psi^* \psi)&=\frac{i}{\sqrt2}(\psi_\ell\psi_r+\psi_\ell^*\psi_r^*)\,.
\end{aligned}
\end{equation}
both of which are hermitian and Lorentz scalars (recall that $\psi_\ell\mapsto e^{\eta/2}\psi_\ell$ and $\psi_r\mapsto e^{-\eta/2}\psi_r$ under a boost with rapidity $\eta\in\mathbb R$). 

If the fermion is real, $\psi_{\ell,r}^*=\psi_{\ell,r}$, then the bilinears $i\bar \psi \gamma^3 \psi$ and $\re(\bar\psi^* \psi)$ both become zero, due to fermi statistics, and the other two bilinears $i\bar\psi \psi$ and $\im(\bar\psi^* \psi)$ become identical.

\paragraph{Symmetries.} Let us list some of the manifest symmetries of QCD theories. The center one-form symmetry is straightforward: it is given by the subgroup of the center that is not screened by the fermions, namely $\ker(R)\subseteq Z(G)$, the kernel of the representation $R$. As a matter of principle, it is possible that there are other one-form symmetries that are not associated to the center of the gauge group, although exhibiting these is a much more complicated task.\footnote{For example it has recently been appreciated~\cite{Nguyen:2021naa} that pure Yang-Mills has a very large space of non-invertible one-form symmetries, valued in a maximal torus of $G$. It is not clear which if these, if any, survive the introduction of matter, or whether there are other one-form symmetries -- invertible or otherwise -- beyond these, although at face value it seems unlikely.}


Zero-form symmetries are abundant too. For example, if we have $N_F$ massless chiral fermions in a given representation $R$, one has the following continuous flavor symmetries acting on them:
\begin{equation}\label{eq:flav_chiral_symm}
\begin{array}{ll}
R~\text{complex:}&~ U(N_F)\\
R~\text{real:}&~ O(N_F)\\
R~\text{pseudoreal:}&~ Sp(N_F)\,.
\end{array}
\end{equation}
When $G$ has no $U(1)$ factors, so that the generators are traceless, the classical axial symmetries are unbroken (unlike in $3+1d$). If there are abelian factors then $U(1)$ flavor symmetries are generically broken into discrete subgroups. An important discrete chiral subgroup, which is very often present even if $U(1)$ is broken, is the chiral fermion parity $\mathbb Z^L_2$, which negates left movers and fixes right movers, to wit, $\mathbb Z_2^L\colon\psi\mapsto\gamma^3\psi$. This subgroup has a very well understood group of 't Hooft anomalies, valued in $\mathbb Z_8$ and responsible for many interesting properties of QCD theories.

In presence of mass terms, the continuous chiral symmetries~\eqref{eq:flav_chiral_symm} descend to their diagonal vector-like subgroups, or even smaller subgroups if the different flavors have different masses.

Finally, there are two other discrete symmetries that are quite useful: charge conjugation $\mathbb Z^\mathsf C_2$, which sends a representation $R$ to its conjugate $\bar R$, and time-reversal $\mathbb Z_2^\mathsf T$, which is anti-linear and satisfies $\mathsf T^2=(-1)^F$. Note that charge conjugation exists only if the group admits complex representations, for otherwise the operation is a gauge transformation and thus not a symmetry. On the other hand, time-reversal is a symmetry only if the theory is vector-like $R_\ell=R_r$, because $\gamma^0$ interchanges the two chiralities. The combination $\mathsf{CT}$ is a symmetry only if $R_\ell=\bar R_r$. Charge conjugation and time-reversal act on the gauge fields as
\begin{equation}
\begin{aligned}
\mathsf C\colon &\begin{cases} \psi(x,t)\mapsto \psi^*(x,t)\\
A_\mu(x,t)\mapsto -A_\mu^t(x,t)\\
\end{cases}\\
\mathsf T\colon &\begin{cases} \psi(x,t)\mapsto\gamma^0\psi(x,-t)\\
A_0(x,t)\mapsto +A^t_0(x,-t)\\
A_1(x,t)\mapsto -A^t_1(x,-t)
 \,,
 \end{cases}
\end{aligned}
\end{equation}
where $t$ denotes transposition.

The three discrete $\mathbb Z_2$ transformations, chiral fermion parity, charge conjugation, and time-reversal, act as follows on the fermion mass terms:
\begin{equation}
\begin{array}{ccccc}
&i\bar\psi \psi & i\bar \psi \gamma^3 \psi&\re(\bar\psi^* \psi)&\im(\bar\psi^* \psi)\\
\mathbb Z_2^L&-&-&-&-\\
\mathbb Z_2^\mathsf C&+&-&-&+\\
\mathbb Z_2^\mathsf T&-&+&-&-\\
\end{array}
\end{equation}
Note that all the mass terms are odd under the chiral symmetry, which suggests that this symmetry might be anomalous. On the other hand, there is at least one mass term that is even under the other two symmetries, so they are non-anomalous. The exception is when the fermions are real, because in that case the mass terms $i\bar \psi \gamma^3 \psi$ and $\re(\bar\psi^* \psi)$ vanish due to fermi statistics. In this situation, the remaining mass terms $i\bar\psi \psi$ and $\im(\bar\psi^* \psi)$ (which are in fact equal) are odd under time-reversal, which suggests that such symmetry might be anomalous too (and, if so, the anomaly will be at most a mod 2 effect; this is confirmed by $\Omega_3^{\text{pin}^+}=\mathbb Z_2$).

\subsection{Matching Energy Momentum Tensors}\label{app:match_T}

We have the canonical energy momentum tensor built from the fermion fields which takes the form 
\begin{equation}\label{canonical}
 T(z) = -\frac{1}{2}\sum_{i} \normord{\psi^i \partial \psi^i}(z)\,.
\end{equation}
The normal ordering is defined as the constant part of the OPE of $AB$. The notation above for two operators is defined by 
\begin{equation}\label{pairing}
 \normord{AB}(z) = \frac{1}{2 \pi i}\oint_z \frac{dx}{x-z} A(x) B(z)\,,
\end{equation}
and extracts the constant part of the OPE of $A$ and $B$ via a contour integral. We can define the current $J^a(z) = \frac{1}{2} \normord{\psi^i t^a_{ij}\psi^j}(z)$, for $t^a_{ij}$ the generators of the $\mathfrak{g}$ Lie algebra. We write the energy momentum tensor given by the Sugawara construction as 
\begin{equation}\label{Sugawara}
\begin{aligned}
 \mathcal{T}(z) &= \gamma \sum_{a} \normord{\normord{\psi^i t^a_{ij}\psi^j} \normord{\psi^k t^a_{kl}\psi^l}}(z)\\
 &=\gamma \sum_a t^a_{ij} t^a_{kl} \normord{\normord{\psi^i \psi^j}\normord{\psi^k\psi^l}}(z)\,.
\end{aligned}
\end{equation}
We first consider the term $\normord{\normord{\psi^i \psi^j}\normord{\psi^k\psi^l}}$. By the rearrangement lemma~\cite[Appendix 6.C]{DiFrancesco:1997nk} we have 
\begin{equation}\label{rearrangement}
\begin{aligned}
 \normord{\normord{\psi^i \psi^j}\normord{\psi^k\psi^l}}(z) =& \normord{\psi^i\normord{\psi^j\normord{\psi^k\psi^l}}}(z)+\normord{[\normord{\psi^i \psi^j},\normord{\psi^k \psi^l}]}(z)\\
 &+\normord{\normord{[\normord{\psi^k\psi^l},\psi^i]}\psi^j}(z)+\normord{\psi^i\normord{[\normord{\psi^k\psi^l},\psi^j]}}(z)\,,
\end{aligned}
\end{equation}
When we substitute these results into \eqref{Sugawara} we get
\begin{equation}\label{compare}
\begin{aligned}
 \gamma \sum_a t^a_{ij} t^a_{kl} &\normord{\normord{\psi^i \psi^j}\normord{\psi^k\psi^l}}(z) = \gamma \sum_a t^a_{ij} t^a_{kl} \normord{\psi^i\normord{\psi^j\normord{\psi^k\psi^l}}}(z)  \\
 &+\gamma \sum_a t^a_{ij} t^a_{kl}[-\delta^{jk}\normord{\psi^l\partial_z\psi^i}+\delta^{jl}\normord{\psi^k\partial_z\psi^{i}}-\delta^{il}\normord{\psi^k\partial_z \psi^j}+\delta^{ik}\normord{\psi^l \partial_z\psi^j}]\\
 &= \gamma \sum_a t^a_{ij} t^a_{kl} \normord{\psi^i\normord{\psi^j\normord{\psi^k\psi^l}}}(z) +\gamma \sum_a[-4t^a_{ij} t^a_{jl}\,\normord{\psi^l \partial_z \psi^i}]\,.
\end{aligned}
\end{equation}
The second term above has the form of \eqref{canonical}, where we use the fact that 
\begin{equation}
 \sum_a t^a_{ij} t^a_{jl} = 2 I(R) \frac{\dim(G)}{\dim(R)}\,\delta^{il}\,,\quad \gamma = \frac{1}{8(I(R)+h)}\,,
\end{equation}
and take $G$ and $R$ from our list of gapped theories.
A necessary condition for equality of $T(z)$ and $\mathcal{T}(z)$ is if~$\gamma \sum_a t^a_{ij} t^a_{kl} \normord{\psi^i\normord{\psi^j\normord{\psi^k\psi^l}}}(z)=0$\,. By definition, this term is 
\begin{equation}\label{contourdef}
 \frac{\gamma}{(2\pi i)^4} \sum_a\oint \frac{dw}{w-z}\,\frac{dx}{x-z} \,\frac{dy}{y-z} t^a_{ij} t^a_{kl}\, \psi^i(w)\psi^j(x)\psi^k(y) \psi^l(z),
\end{equation}
and since $\psi^i$ is Grassmann, this vanishes if the total antisymmetrization of $\sum_a t^a_{i j } t^a_{k l}$ vanishes. This gives us the condition 
\begin{equation}\label{bianchilike}
 \sum_a t^a_{i j } t^a_{k l}+t^a_{i k } t^a_{l j}+ t^a_{i l } t^a_{j k}=0\,.
\end{equation}

If the group $G= \tilde{G}\times U(1)$ which has a $U(1)$ factor, we can write $t^a_{ij}$ as a decomposition under $\tilde{G}$ and $U(1)$, where it has charge $q$ under the $U(1)$, i.e., the current is $J(z)=\frac{1}{2}\normord{\psi^i q \,\psi^i}$. As an example we take $U(N)$ with a fermion in the antisymmetric representation of $SU(N)$ and charge $q$ under $U(1)$. The $SU(N)$ part of the Sugawara tensor reads 
\begin{equation}\label{su(N)Tensor}
\begin{aligned}
 \mathcal{T}_{SU(N)}(z) &= \frac{1}{8(2N-2)}\Bigl[-4\frac{2(N^2-N-2)}{N} \Bigr]\normord{\psi^i \partial_z \psi^i}  \\
 &= -\frac{1}{2}\normord{\psi^i \partial_z \psi^i} +\frac{1}{N(N-1)}\normord{\psi^i \partial_z \psi^i}\,.
\end{aligned}
\end{equation}
The $U(1)$ part of the Sugawara tensor reads 
\begin{equation}\label{u1tensor}
 \mathcal{T}_{U(1)}(z) =\frac{1}{4N(N-1)}\normord{\normord{\psi^i \psi^i}\normord{\psi^j\psi^j}} = \frac{-1}{N(N-1)}\normord{\psi^i \partial_z \psi^i}\,,
\end{equation}
and by summing~\eqref{su(N)Tensor} and~\eqref{u1tensor} we reproduce~\eqref{canonical}. 

Now suppose we are working with an Abelian theory and we consider $n$ complex fermions, with $q_{Ia}$ charge matrix, i.e. $U(1)^n_{q^t q}$. The canonical energy momentum is given by 
\begin{equation}
 T(z) = \frac{1}{2}(\normord{\partial_z \psi^{I \dagger}\psi^I}-\normord{\psi^{I\dagger}\partial_z\psi^I})(z).
\end{equation}
With the charge matrix we can define the current $J_a = \psi^{I\dagger} q_{Ia}\psi^I$, which satisfies the OPE
\begin{equation}
 J_a(z) J_b(0) \sim \frac{k_{ab}}{z^2}.
\end{equation}
From this we build the Sugawara tensor 
\begin{equation}\label{sugawaraflavor}
 \mathcal{T}(z) = \frac{1}{2}\sum_{a,b}(k_{ab})^{-1} \normord{\normord{\psi^{I\dagger} q_{ I a }\psi^I}\normord{\psi^{J \dagger }q_{J b}\, \psi^J}}(z)\,,
\end{equation}
where $k_{ab} = (q^t q)_{ab} = q_{I a}\, q_{I b}$, so $(k_{ab})^{-1} = (q_{I a}\, q_{I b})^{-1}$, and we can define the current as $J_a = \normord{\psi^{I\dagger} q_{Ia}\psi^I}$\,. Again by the rearrangement lemma, we have 
\begin{equation}
\begin{aligned}
 q_{I a}\, q_{J b}\,\normord{\normord{\psi^{I\dagger} \psi^I}\normord{\psi^{J \dagger }\, \psi^J}}(z) &= q_{ I a }\, q_{J b} \normord{\psi^{I\dagger}\normord{\psi^I\normord{\psi^{J\dagger} \psi^J }}}(z) \\
 &+q_{ I a}\, q_{J b}(- \delta^{IJ} \psi^{J} \normord{\partial_z \psi^{I \dagger }}-\delta^{IJ}\normord{\psi^{J \dagger}\partial_z \psi^I})(z)  \\
 &=q_{I a}\, q_{I b}\,(- \normord{\psi^{I} \partial_z \psi^{I \dagger }}-\normord{\psi^{I \dagger}\partial_z \psi^I})(z)\,.
\end{aligned}
\end{equation}
After substituting into \eqref{sugawaraflavor} we get 
\begin{equation}
\begin{aligned}
 \mathcal{T}(z) &= \frac{1}{2}\sum_{a,b}(q^{-1}_{Ib} q^{-1}_{Ia})[q_{ I a }\, q_{J b} \normord{\psi^{I\dagger}\normord{\psi^I\normord{\psi^{J\dagger} \psi^J }}}(z) \\
 &\quad+q_{ I a}\, q_{J b}(- \delta^{IJ} \normord{\psi^{J } \partial_z \psi^{I \dagger }}-\delta^{IJ}\normord{\psi^{J \dagger}\partial_z \psi^I})(z)] \\
 &= \frac{1}{2}[\normord{\psi^{I\dagger}\normord{\psi^I\normord{\psi^{J\dagger} \psi^J }}}(z)+(- \delta^{IJ} \normord{\psi^{J } \partial_z \psi^{I \dagger }}-\delta^{IJ}\normord{\psi^{J \dagger}\partial_z \psi^I})(z)] \\
 &= \frac{1}{2}(\normord{\partial_z \psi^{I \dagger}\psi^I}-\normord{\psi^{I\dagger}\partial_z\psi^I})(z)\\
 &= T(z)\, ,
\end{aligned}
\end{equation}
which is the canonical energy momentum tensor. We have used the fact that $\sum_b (q_{Ib})^{-1}\,q_{Jb} = \delta_{IJ}$, and the first term in the second equality vanishes by applying~\eqref{contourdef} and evaluating the contour integrals.

\subsection{Temporal Gauge Hamiltonian commutation}\label{temporalgaugeH}
The quantized Hamiltonian is given by integrating the Hamiltonian action in \eqref{subhamiltonian}, where we take the left and right handed fermions $\psi_\ell(x)$ and $\psi_r(x)$ to be operators on a circle where $0\leq x \leq 2\pi$:
\begin{equation}\label{inthamiltonian}
\begin{aligned}
\widehat{H} = \int_0^{2\pi} &\Bigl[\frac{i}{8(I(R)+h)}\left({:}\widehat{J}^a_{\ell} \widehat{J}^a_{\ell}(x){:}+{:}\widehat{J}^a_{r} \widehat{J}^a_{r}(x){:} \right) \\
 &\hspace{5mm}+\frac{1}{\sqrt{2}}i\lim_{ \epsilon \to 0}\left(\langle \psi^\dagger_{i\,\ell}(x+\epsilon)\partial_x\psi_{i\,\ell}(x-\epsilon)\rangle-\langle \psi^\dagger_{i\,r}(x+\epsilon)\partial_x\psi_{i\,r}(x-\epsilon)\rangle\right) \\
 &\hspace{15mm}+ \frac{1}{\sqrt{2}}\widehat{A}^a(x)(\widehat{J}^a_\ell - \widehat{J}^a_r)(x) + \frac{1}{\sqrt{2}} I(R)\widehat{A}^a(x)^2\Bigr]dx \\
 & \hspace{-10mm}+g^2\int_{0}^{2\pi} (\widehat{E}^a(x))^2 dx.
\end{aligned}
\end{equation}
We use the expression for $\widehat{J}^a_{\ell\,,r}$ as specifically given in section \ref{sec:temporal_gauge}.

For the commutator $[\widehat{G}^a, \widehat{H}]=0$, the pieces proportional to $g^0$ and $g^2$ vanish separately. The commutator of $\widehat{G}^a$ with ${:}\widehat{J}^a_{\ell,r} \widehat{J}^a_{\ell,r}(x){:}$ vanishes~\cite{Antoniadis:1985eq}, and we proceed to use the commutation relations in~\eqref{qcommutator} to show that the other terms in~\eqref{inthamiltonian} vanish when commuted with $\widehat{G}^a$. Consider first the terms:
\begin{align}\label{currentl}
[\widehat{J}^a_\ell(x),\int_0^{2\pi}\!\!\! \widehat{A}^b \widehat{J}^b_{\ell}(y) dy]&=\int_{0}^{2\pi}\!\!\! \widehat{A}^b(y)\left( if^{abc}J^c_{\ell}(x)\delta(x-y) +i I(R) \delta^{ab}\partial_x\delta(x-y)\right)dy\\\label{currentr}
-[\widehat{J}^a_r(x),\int_0^{2\pi}\!\!\! \widehat{A}^b \widehat{J}^b_{r}(y) dy]&=-\int_{0}^{2\pi}\!\!\! \widehat{A}^b(y)\left( if^{abc}J^c_{r}(x)\delta(x-y) -i I(R) \delta^{ab}\partial_x\delta(x-y)\right)dy\,.
\end{align}
We then look at the terms:
\begin{align}
-[D_x\widehat{\Pi}^a(x),\int_0^{2\pi}\widehat{A}^d (\widehat{J}^d_{\ell}-\widehat{J}^d_r)(y) dy]&=\notag\\
 &\hspace{-3cm} -\int_{0}^{2\pi} i\delta^{ab}\partial_x\delta(x-y)\left(\widehat{J}^b_\ell-\widehat{J}^b_r \right)(y) dy \label{AJ} \\
 &\hspace{-3cm}-\int_{0}^{2\pi} if^{abc}\widehat{A}^b(x)\delta^{cd}\left(\widehat{J}^d_\ell-\widehat{J}^d_r \right)(y)\delta(x-y)dy\,,\notag\\
-I(R)[D_x\widehat{\Pi}^a(x),\int_0^{2\pi} \widehat{A}^d(y)^2 dy]&=\notag\\
 &\hspace{-3cm}-2iI(R)\int_{0}^{2\pi}\widehat{A}^d(y) \delta^{ad}\partial_x \delta(x-y) dy\label{A2}\\
 &\hspace{-3cm}-2iI(R)\int_{0}^{2\pi}\!\!\!if^{abc}\widehat{A}^d(y)\widehat{A}^b(x)\delta^{cd}\delta(x-y) dy\,.\notag
\end{align}
The second term in~\eqref{currentl} and~\eqref{currentr} cancel with the first term in~\eqref{A2}. The first term in~\eqref{currentl} and~\eqref{currentr} cancel with the second term in~\eqref{AJ}; the last term in~\eqref{A2} vanishes by antisymmetry. We are thus left to negotiate the term 
\begin{equation}\label{deriv}
-\int_{0}^{2\pi} i\delta^{ab}\partial_x\delta(x-y)\left(\widehat{J}^b_\ell-\widehat{J}^b_r \right)(y) dy
\end{equation}
in~\eqref{AJ}. For this we consider 
\begin{equation}\label{remainder}
[\widehat{J}^a_{\ell,r}(x), \pm \int_{0}^{2\pi}i\lim_{\epsilon \to 0}\langle \psi^\dagger_{k\,\ell,r}(y+\epsilon)\partial_y\psi_{k\,\ell,r}(y-\epsilon)\rangle dy ]
\end{equation}
 where we first compute the commutator by treating the second term as an operator, and then using the propagator while taking the $\epsilon \to 0$ limit. This is the same procedure used to prove that $[\widehat{J}^a_{\ell\,,r}(x), \widehat{J}^b_{\ell,r}(y)]$ contains a Schwinger term. Working with just the left handed part of~\eqref{remainder} we get 
\begin{equation}\label{cancel}
 \begin{aligned}
 &[\widehat{J}^a_{\ell}(x), \int_{0}^{2\pi}i\lim_{\epsilon \to 0}\langle \psi^\dagger_{k\,\ell}(y+\epsilon)\partial_y\psi_{k\,\ell}(y-\epsilon) \rangle dy ] \\
 &=[\tfrac{1}{2}\psi^\dagger_{i\, \ell} t^a_{ij} \psi_{j\,\ell}(x), \int_{0}^{2\pi}i\lim_{\epsilon \to 0}\langle \psi^\dagger_{k\,\ell}(y+\epsilon)\partial_y\psi_{k\,\ell}(y-\epsilon) \rangle dy ] \\
 &=\frac{i}{2}\lim_{\epsilon \to 0}\Bigl\{ \int_{0}^{2\pi} \delta(x-y-\epsilon)\partial_y(\psi^{\dagger}_{i\,\ell}(y+\epsilon) t^{a}_{ij} \psi_{j\,\ell}(y-\epsilon)) dy \\
 & \hspace{15mm}+ \int_{0}^{2\pi} \partial_y\delta(x-y+\epsilon) \psi_{j\, \ell}(y+\epsilon)t^a_{ij} \psi^\dagger_{i\,\ell}(y-\epsilon) dy
 \Bigl\} \\
 &= \frac{i}{2}\lim_{\epsilon \to 0} \Bigl\{-\int_{0}^{2\pi} \partial_y\delta(x-y-\epsilon) \psi^\dagger_{i\, \ell}(y+\epsilon)t^a_{ij} \psi_{j\,\ell}(y-\epsilon)dy \\
 &\hspace{15mm}+ \int_{0}^{2\pi} \partial_y\delta(x-y+\epsilon) \psi_{j\, \ell}(y+\epsilon)t^a_{ij} \psi^\dagger_{i\,\ell}(y-\epsilon) dy \Bigr\} \\
 &= i \Bigl\{ -\int_{0}^{2\pi} \partial_y \delta(x-y) {:}\psi^\dagger_{i\,\ell}(y)t^a_{ij}\psi_{j\,\ell}(y){:}dy\Bigr\} \\
 &= i\int_{0}^{2\pi} \partial_x \delta(x-y) \widehat{J}^a(y)dy\,,
 \end{aligned}
\end{equation}
which cancels the first term in \eqref{deriv}, and we can do an analogous computation for $\widehat{J}^a_{r}$. The first equality we use the fact that $\widehat{J}^a(x)= \frac{1}{2}\psi^\dagger_{i\, \ell} t^a_{ij} \psi_{j\,\ell}(x)+\text{(singular term)}$, where the singular term vanishes in the commutator. To go from the third equal sign to the fourth equal sign we replace the fermions by the normal ordered version where~\cite{Affleck:1985wb}
 \begin{equation}
 \psi^{\dagger}_{i\,\ell}(y+\epsilon) t^{a}_{ij} \psi_{j\,\ell}(y-\epsilon) = {:}\psi^{\dagger}_{i\,\ell}(y+\epsilon) t^{a}_{ij} \psi_{j\,\ell}(y-\epsilon){:}+\lim_{\tilde{\epsilon} \to 0}\langle \psi^{\dagger}_{i\,\ell}(x+\epsilon+\tilde{\epsilon}) t^{a}_{ij} \psi_{j\,\ell}(x-\epsilon-\tilde{\epsilon})\rangle\,,
 \end{equation}
 and the second term vanishes under the derivative. For the term proportional to $g^2$ we consider 
 \begin{equation}
 [D_x \widehat{\Pi}^a(x) , \int_{0}^{2\pi} \widehat{E}^d(y)^2 dy] = -2i\int_{0}^{2\pi}if^{abc} \widehat{E}^b(x) \widehat{E}^c(y) \delta(x-y) dy\,,
 \end{equation}
 which vanishes due to the antisymmetry of $f^{abc}$.

\section{Infrared coset CFTs}\label{app:coset_cft}

In this appendix we review the formalism of coset CFTs~\cite{DiFrancesco:1997nk}, our primary goal being to understand the CFTs that appear in the deep infrared of QCD~\eqref{eq:conj_IR_coset}.

\paragraph{Chiral characters.} One of the most important concepts in RCFT is that of a chiral character. These consist of a finite family of functions $\{\chi_\lambda(q)\}$ of the complex structure $q=e^{2\pi i\tau}$, holomorphic in the upper half plane, and labelled by the primaries of the theory $\lambda$ (the representations of the chiral algebra). Given these characters, the torus partition function of the theory takes the form
\begin{equation}
Z(q)=\sum_{\lambda,\bar\lambda}M_{\bar\lambda,\lambda}\,\bar\chi_{\bar\lambda}(\bar q)\chi_{\lambda}(q)\equiv \bar\chi^\dagger M\chi\,.
\end{equation}
Here $M$, the so-called mass matrix, specifies how the left-moving sectors are paired up with right-moving ones. The possible choices for $M$ are constrained by the requirement of $Z$ being a modular-invariant function of $q$. This is archived by a key property of the characters, to wit, their covariance under modular transformations. Under a generic such transformation, the characters mix with each other in a well-defined fashion, and the role of $M$ is to ensure that the sesquilinear form $Z=\chi^\dagger M\chi$ is a scalar under these transformations. As a result, while $Z(q)$ is a well-defined function on $\mathbb H/SL(2,\mathbb Z)$, the tuple $\chi$ is best thought of as a non-trivial section thereon.

When $\chi_\lambda(q)$ admits a Hilbert space interpretation, it is defined as
\begin{equation}\label{eq:chiral_char_def}
\chi_\lambda(q):=\tr_{\mathcal H_\lambda}(q^{L_0-c/24})\,,
\end{equation}
such that
\begin{equation}
Z(q)=\tr_{\mathcal H}(q^{L_0-c/24}\bar q^{\bar L_0-c/24}),\qquad \mathcal H=\bigoplus_{\lambda,\bar\lambda}M_{\bar\lambda,\lambda}\mathcal H_{\bar \lambda}\otimes\mathcal H_\lambda\,.
\end{equation}
Here $\mathcal H$ is the full Hilbert space of the theory, and $\mathcal H_\lambda$ is the representation space (module) for $\lambda$. The mass matrix $M$ dictates how these chiral modules combine into $\mathcal H$. From now on, and in order to simplify the notation and presentation, we always have in mind the diagonal theory $M_{\bar\lambda,\lambda}=\delta_{\bar\lambda,\lambda}$. Non-diagonal theories can often be thought of as the diagonal theory of a larger algebra via (potentially non-abelian) anyon condensation.

\paragraph{Fermionic CFTs.} A CFT is said to be \emph{fermionic} if, on top of the dependence on the conformal structure of spacetime, it also depends on the choice of spin structure thereof. In other words, a fermionic CFT depends on the boundary conditions for fermionic fields. The CFTs that appear at RG fixed points of QCD theories are naturally fermionic, because the microscopic theory contains quarks. Hence our main interest is in fermionic CFTs.

In the case of the torus there are four spin structures, corresponding to either periodic or anti-periodic boundary conditions around the two non-trivial cycles. We also refer to these boundary conditions as Ramond and Neveu-Schwartz, respectively, and we use the notation $+=\text R$, $-=\text{NS}$ interchangeably.

In a fermionic CFT, the characters acquire a dependence on the spin structure: they become super-characters. Consequently, we denote them as $\chi_\lambda^{\pm,\pm}$ where
\begin{equation}
\begin{aligned}
\chi_\lambda^\text{NS-NS}(q)&:=\tr_{\mathcal H_{\text{NS};\lambda}}(q^{L_0-c/24})\\
\chi_\lambda^\text{NS-R}(q)&:=\tr_{\mathcal H_{\text{NS};\lambda}}((-1)^{F_L}q^{L_0-c/24})\\
\chi_\lambda^\text{R-NS}(q)&:=\tr_{\mathcal H_{\text{R};\lambda}}(q^{L_0-c/24})\\
\chi_\lambda^\text{R-R}(q)&:=\tr_{\mathcal H_{\text{R};\lambda}}((-1)^{F_L}q^{L_0-c/24})\,.
\end{aligned}
\end{equation}
Here, $\mathcal H_{\pm;\lambda}$ denotes the module of $\lambda$ with fermion boundary conditions $\pm$, while $(-1)^{F_L}$ denotes the chiral fermion parity operator, which assigns $+1$ to bosonic left-movers and $-1$ to fermionic left-movers, while it acts trivially on the right-moving modes.

The partition function of a fermionic CFT is obtained by combining the two chiral halves in a modular covariant way:
\begin{equation}
Z_{\pm,\pm}(q)=\sum_{\bar\lambda,\lambda}M_{\bar \lambda,\lambda}^{\pm}\bar\chi^{\pm,\pm}_{\bar \lambda}(\bar q)\chi^{\pm,\pm}_{ \lambda}(q)\,,
\end{equation}
which computes
\begin{equation}
Z_{\pm,\pm}(q)=\tr_{\mathcal H_\pm}((\mp1)^Fq^{L_0-c/24}\bar q^{\bar L_0-c/24}),\qquad \mathcal H_\pm=\bigoplus_{\lambda,\bar\lambda}M^\pm_{\bar\lambda,\lambda}\mathcal H_{\pm;\bar \lambda}\otimes\mathcal H_{\pm;\lambda}\,,
\end{equation}
where $(-1)^F\equiv (-1)^{F_L}(-1)^{F_R}$ is the total fermion parity.

Here the mass matrices $M^{\pm}$, which dictate how the two chiral halves $\mathcal H_{\pm;\bar\lambda},\mathcal H_{\pm;\lambda}$ combine into the full Hilbert space $\mathcal H_\pm$, are chosen so as to ensure that $Z_{\pm,\pm}$ transforms appropriately under modular transformations. Unlike in the case of bosonic CFTs, $Z_{\pm,\pm}$ is not in general invariant under $SL(2,\mathbb Z)$. Indeed, modular transformations generically map the different spin structures into each other, which induces a reshuffling of the partition functions $Z_{\pm,\pm}$. Specifically, under the standard generators of $SL(2,\mathbb Z)=\langle S,T\rangle$, the partition functions transform as
\begin{equation}\label{eq:modularity_fCFT}
\begin{aligned}
S\tikz[baseline=24pt]{\draw[->,thick,>=stealth] ([shift=(30:.5cm)]2,1) arc (30:330:.5cm);}
\quad Z_\text{NS-NS}\quad 
\overset{T}{\tikz[baseline=-3pt]{\draw[<->,thick,>=stealth] (0,0) -- (1,0);}}
\quad &Z_\text{NS-R}\quad 
\overset{S}{\tikz[baseline=-3pt]{\draw[<->,thick,>=stealth] (0,0) -- (1,0);}}
\quad Z_\text{R-NS}\quad 
\tikz[baseline=24pt]{\draw[->,thick,>=stealth] ([shift=(150:.5cm)]2,1) arc (150:-150:.5cm);}T\\
S\tikz[baseline=24pt]{\draw[->,thick,>=stealth] ([shift=(30:.5cm)]2,1) arc (30:330:.5cm);}
\quad &Z_\text{R-R}\quad 
\tikz[baseline=24pt]{\draw[->,thick,>=stealth] ([shift=(150:.5cm)]2,1) arc (150:-150:.5cm);}T
\end{aligned}
\end{equation}
The choices for the mass matrices $M^{\pm}$ are constrained by the requirement of $Z_{\pm,\pm}$ being a modular-covariant function of $q$. As usual, we will always have in mind the diagonal theory $M^{\pm}_{\bar \lambda,\lambda}=\delta_{\bar \lambda,\lambda}$.

In order to simplify the notation, we shall frequently leave the dependence on the spin structure $\pm,\pm$ implicit. 

\paragraph{Flavor-twisted characters.} If the chiral algebra has some flavor symmetry $U$, then it is often useful to introduce flavor-twisted characters (i.e., we turn on fugacities for the Cartan generators; these are roots of $\mathfrak u$). This allows us to organize the modules $\mathcal H_\lambda$ into irreducible representations of $U$ so as to have a more transparent understanding of the structure of states therein. To this end, we can define extended characters as
\begin{equation}
\chi_\lambda(q,g):=\tr_{\mathcal H_\lambda}(q^{L_0-c/24}\rho(g))\,,
\end{equation}
where $g\in U$ is a symmetry group element and $\rho\colon g\to\mathcal H_\lambda$ is its representation on the Hilbert space. The character is a class function, so its dependence on $g$ is only through its conjugacy class.

Regular characters $\chi_\lambda(q)$ are obtained from the extended ones $\chi_\lambda(q,g)$ by setting $g=1$. The former only keep track of the conformal weights of the states in $\mathcal H_\lambda$, while the latter also keeps track of their quantum numbers under $U$.

\paragraph{Coset CFTs.} Whenever the chiral algebra has a subalgebra, one can expand the characters of the former in terms of those of the latter,
\begin{equation}
\chi_\lambda(q)=\sum_{\Lambda}b^\Lambda_\lambda(q)\chi_\Lambda(q)\,,
\end{equation}
where $\chi_\lambda$ are the characters of the original chiral algebra, and $\chi_\Lambda$ those of the subalgebra.

The key point of this construction is that, if $\chi_\lambda$ and $\chi_\Lambda$ are both modular covariant, then so are the coefficients $b_\lambda^\Lambda$. This means that one can think of these coefficients as the characters of a new theory, which we call the \emph{coset CFT}; this is the celebrated GKO coset construction~\cite{Goddard:1986ee}. If at least one of $\chi_\lambda,\chi_\Lambda$ is a super-character, then so is $b_\lambda^\Lambda$, and hence the coset is a fermionic CFT.

This new theory, the coset CFT, has characters $b_\lambda^\Lambda(q)$, and therefore its partition function takes the form
\begin{equation}
Z^\text{coset}(q)=\sum_b \bar b(\bar q)b(q)\,,
\end{equation}
where we restrict to diagonal partition functions for simplicity. If $b$ is a super-character, the expression above defines the partition function of a fermionic CFT, while if it is a regular character, it defines the partition function of a bosonic CFT.

It should be remarked that, in order to actually calculate the coefficients $b_\lambda^\Lambda$, it is often unavoidable to turn on flavor fugacities, for otherwise the computation becomes impracticable. One is therefore forced to look at the extended characters $\chi_\lambda(q,g)$, $\chi_\Lambda(q,g')$, where $g$ is a symmetry group element of the original algebra, and $g'$ its restriction to the subalgebra. Specifically, if the original algebra has flavor symmetry $U$ and its subalgebra has flavor symmetry $U'\subseteq U$, then the character decomposition can be extended to
\begin{equation}
\chi_\lambda(q,g)=\sum_{\Lambda}b^\Lambda_\lambda(q,g'')\chi_\Lambda(q,g')\,,
\end{equation}
where $g\in U$, $g'\in U'$, and $g''$ is a group element of the flavor symmetry of the coset, namely the commutant of $U'$ inside $U$.

\paragraph{WZW CFTs.} The discussion so far has been rather abstract and general. To be concrete, let us discuss Wess--Zumino--Witten (WZW) theories, which we review next. WZW theories are labelled by a compact Lie group $G$, which we take to be simple and connected, and a ``level'' $k$, an integer that specifies the central extension for the loop algebra of $G$. We denote the corresponding model by $G_k$. When $G$ is simply-connected (which we henceforth assume, unless specified otherwise) the chiral algebra is a Ka\v{c}-Moody algebra:
\begin{equation}\label{eq:KM_algebra}
[J^a_n,J^b_m]=f^{ab}{}_{c\,}J^c_{n+m}+kn\delta_{ab}\delta_{n+m}\,,
\end{equation}
while if $G$ is not simply-connected, then the chiral algebra is Ka\v{c}-Moody extended by the simple currents that generate $\pi_1(G)$. WZW theories with $\pi_1(G)=0$ are always bosonic, while those with $\pi_1(G)\neq0$ are fermionic if any of the currents that generate $\pi_1(G)$ is a fermion (it has half-integral conformal weight).

The representations of the chiral algebra are required to be unitary with respect to $(J^a_n)^\dagger:=J^a_{-n}$.

The enveloping algebra of~\eqref{eq:KM_algebra} contains the Virasoro algebra via the Sugawara construction:
\begin{equation}
L_n=\frac{1}{2(k+h)}\sum_{a,m}{:}J^a_mJ^a_{n-m}{:}\,,
\end{equation}
such that
\begin{equation}
\begin{aligned}
[L_n,L_m]&=(m-n)L_{n+m}+\frac{c(G_k)}{12}n(n^2-1)\delta_{n+m},\qquad c(G_k):=\frac{k}{k+h}\dim(\mathfrak g)\\
[L_n,J^a_m]&=-mJ^a_{n+m}\,.
\end{aligned}
\end{equation}
Note that this last expression indicates that $J_{-n}$ carries $n$ units of $L_0$ eigenvalue.

The primaries of the theory are labeled by the integrable representations of $\mathfrak g_k$, to wit, the highest-weight representations $\lambda\in\operatorname{Rep}(\mathfrak g)$ that satisfy
\begin{equation}
(\theta,\lambda)\le k\,,
\end{equation}
with $\theta$ the highest root of $\mathfrak g$ (the highest weight of the adjoint representation) and $(\cdot,\cdot)$ its Killing form, normalized to $(\theta,\theta)=2$. The module $\mathcal H_\lambda$ is constructed as follows: at lowest grade, one begins with the vacuum states $|\lambda\rangle$, which live in the finite dimensional representation $\lambda$ generated by $J^a_0$. On top of these vacua one constructs the excited states. For example, at grade-one one has the states $J_{-1}^a|\lambda\rangle$, which live inside the representation $\theta\otimes \lambda$. At grade-two one has the states $J_{-2}^a|\lambda\rangle$ and $J_{-1}^aJ_{-1}^b|\lambda\rangle$, so the states live inside $\theta\otimes \lambda+\theta^2_\text{sym}\otimes \lambda$. Etc. Importantly, not all of these states are physical: we must project out the null states, i.e., the states whose norm vanishes. Continuing this way one obtains the character
\begin{equation}
\chi_\lambda(q,g)=q^{h_\lambda-c/24}(\chi_\lambda(g)+\chi_{R_1}(g)q+\chi_{R_2}(g)q^2+\chi_{R_3}(g)q^3+\cdots)\,,
\end{equation}
where $R_n$ is the space of physical states at grade $n$, and $\chi_R(g):=\tr_R(g)$ the finite character of $R$.\footnote{By definition, $\chi_R(g):=\sum_{\lambda\in\Omega(R)}z^\lambda$, where $\Omega(R)$ is the space of weights of the representation $R$, with multiplicities.} A more convenient way to obtain the characters is the so-called Weyl-Ka\v{c} formula~\cite{KAC1974},
\begin{equation}\label{eq:weyl_kac}
\chi_\lambda(q,g)=\frac{\hat\chi_{\lambda+\rho}(q,g)}{\hat\chi_\rho(q,g)},\qquad \hat\chi_\lambda(q,g):=\sum_{\substack{w\in W\\\alpha^\vee\in Q^\vee}}\det(w)
\,z^{k \alpha^\vee+w\lambda}q^{|k\alpha^\vee+w\lambda|^2/2k}\,,
\end{equation}
where $W$ denotes the Weyl group of $\mathfrak g$, $Q^\vee$ its coroot lattice, $\rho$ its Weyl vector, and $z$ is the value of $g\in G$ when conjugated to any maximal torus.

\paragraph{$\boldsymbol{Spin(n)_1}$ CFT.} A particularly important family of WZW models is $Spin(n)_1$. For $n$ odd this theory has three primaries $0,v,\sigma$ (the scalar, vector, and spinor representation, respectively), and for $n$ even it has four, $0,v,s,c$ (where $s,c$ denote the two spinors). The corresponding characters read
\begingroup
\allowdisplaybreaks
\begin{align}
\chi_0(q,g)&=q^{-n/48}\biggl[\bullet+\ydiagram{1,1}\,q+\biggl(\bullet+\ydiagram{1,1}+\ydiagram2+\ydiagram{1,1,1,1}\,\biggr)q^2\notag\\
&\hspace{2cm}+\biggl(\bullet+3\,\ydiagram{1,1}+\ydiagram2+\ydiagram{1,1,1,1}+\ydiagram{2,1,1}+\ydiagram{1,1,1,1,1,1}\,\biggr)q^3+\cdots\biggr]\notag\\
\chi_v(q,g)&=q^{-n/48+1/2}\,\ydiagram1\,\biggl[\bullet+\bullet\,q+\biggl(\bullet+\ydiagram{1,1}\,\biggr)q^2+\biggl(\bullet+ 2\,\ydiagram{1,1}+\ydiagram{1,1,1,1}\, \biggr)q^3+\cdots\biggr]\notag\\
&\hspace{2cm}+q^{-n/48+3/2}\biggl[\ydiagram{1,1,1}+\ydiagram{1,1,1,1,1}\,q+\ydiagram{1,1,1,1,1,1,1}\,q^2+\cdots\biggr]\notag\\
&\equiv q^{-n/48+1/2}\biggl[\ydiagram{1}+\biggl(\ydiagram1+\ydiagram{1,1,1}\,\biggr)q+\biggl(2\,\ydiagram1+\ydiagram{2,1}+\ydiagram{1,1,1}+\ydiagram{1,1,1,1,1}\,\biggr)q^2\label{eq:char_spin_n}\\
&\hspace{2cm}+\biggl(3\,\ydiagram1+ 2\, \ydiagram{2,1} +3\,\ydiagram{1,1,1} +\ydiagram{2,1,1,1} +\ydiagram{1,1,1,1,1}+\ydiagram{1,1,1,1,1,1,1}\,\biggr)q^3+\cdots\biggr]\notag\\
\chi_\sigma(q,g)&=q^{n/24}\,\sigma\,\biggl[\bullet+\ydiagram{1,1}\,q^2+\biggl(\bullet+\ydiagram{1,1}+\ydiagram2\biggr)q^3+\biggl(\bullet+2\,\ydiagram{1,1}+\ydiagram2+\ydiagram{1,1,1,1}\,\biggr)q^4+\cdots\biggr]\notag\\
&\hspace{1cm}+q^{n/24+1}\,\bar\sigma\,\biggl[ \ydiagram1+\ydiagram1\,q+\biggl(\ydiagram1+\ydiagram{1,1,1}\,\biggr)q^2+\biggl(2\, \ydiagram1+\ydiagram{1,1,1}+\ydiagram{2,1}\biggr)q^3+\cdots\biggr]\notag\\
&\equiv q^{n/24}\biggl[\sigma+\bigl(\sigma+\overline{\dot{\ydiagram1}}\bigr)q+\bigl(2 \sigma+\dot{\ydiagram{1,1}}+2\,\overline{\dot{\ydiagram1}}\bigr)q^2\notag\\
&\hspace{2cm}+\bigl(4 \sigma+\overline{\dot{\ydiagram{1,1,1}}}+2\,\dot{\ydiagram{1,1}}+4\,\overline{\dot{\ydiagram1}}+\dot{\ydiagram2}\bigr)q^3+\cdots\biggr]\,,\notag
\end{align}
\endgroup
where $\sigma$ denotes the spinor of $Spin(n)$ when $n$ is odd, and any of the two spinors when $n$ is even (in which case $\bar \sigma$ denotes the conjugate spinor, obtained by permuting the last two Dynkin labels). Here, a Young diagram stands for the finite character $\chi_R(g)$ associated to the representation $R$ of $\mathfrak{so}(n)$, and a dot $\dot R$ is a short-hand notation for the representation whose highest weight is given by $\lambda_{\dot R}:=\lambda_R+\omega_{\lfloor n/2\rfloor}$ (so for example $\dot{\ydiagram1}=(1,0,\dots,0,1)$). 

These expressions make it manifest how the different states of $Spin(n)_1$ appear at each level. For example, in $\chi_0$ at grade $n=2$ the states live inside
\begin{equation}
J^a_{-2}|0\rangle+J^a_{-1}J_{-1}^b|0\rangle\subseteq \ydiagram{1,1}+ (\ydiagram{1,1}\otimes\ydiagram{1,1})_\text{sym}=\biggl(\bullet+\ydiagram{1,1}+\ydiagram2+\ydiagram{1,1,1,1}\,\biggr)+\ydiagram{2,2}\,,
\end{equation}
and one can check that the representations in parentheses are physical and the isolated one is null (its norm vanishes).

\paragraph{$\boldsymbol{SO(n)_1}$ CFT.} A related -- and also very important -- WZW model is $SO(n)_1$. This theory is obtained from $Spin(n)_1$ through fermionization. There are two key properties of $SO(n)_1$ that make it special. First, it is a \emph{fermionic} theory, meaning that its characters and partition functions depend on the choice of spin structure. Second, it is a \emph{holomorphic} theory, meaning that its partition function factorizes as $Z_{\pm,\pm}=|d_{\pm,\pm}|^2$ (as opposed to non-holomorphic theories whose partition function is a sum of such terms). In other words, $SO(n)_1$ has a \emph{unique} primary (for fixed spin structure). Following the convention of~\cite{DiFrancesco:1997nk} we denote this unique character as $d_{\pm,\pm}$.

The characters of $SO(n)_1$ can be obtained from those of $Spin(n)_1$ (cf.~equation~\eqref{eq:char_spin_n}) via the standard bosonization/fermionization dictionary:
\begin{equation}\label{eq:boson_fermion_spin_so}
\begin{aligned}
d_\text{NS-X}(q,g)&=\chi_0(q,g)\pm\chi_v(q,g)\\
\text{$n$ odd:}\quad d_\text{R-NS}(q,g)&=\sqrt2\chi_\sigma(q,g)\\
d_\text{R-R}(q,g)&=0\\
\text{$n$ even:}\quad d_\text{R-X}(q,g)&=\chi_s(q,g)\pm\chi_c(q,g)\,.
\end{aligned}
\end{equation}


\paragraph{$\boldsymbol{U(1)_k}$ CFT.} So far we have described WZW models for simple groups. The case of $U(1)$ requires a separate discussion. By $U(1)_k$, with $k$ even, we mean a free compact boson at radius $R^2=k$. This has a chiral $U(1)$ flavor symmetry; if we turn on a fugacity $z\in U(1)$ for this symmetry, the characters of this CFT are
\begin{equation}\label{eq:u1_char_bos}
\chi_\ell(q,z)=\eta(q)^{-1}\sum_{u\in\mathbb Z}q^{\frac12k(u+\ell/k)^2}z^{\sqrt ku+\ell/\sqrt k}\,,
\end{equation}
with $\ell=0,1,\dots,k-1$, and where $\eta$ is the Dedekind function.

When $k$ is odd, by $U(1)_k$ we mean the theory $U(1)_{4k}$ extended by the vertex operator of weight $k/2$, i.e., by $\ell=2k$. As this weight is half-integral, the operator is a fermion and the extension results in a fermionic CFT. Its super-characters can be obtained for example by following the rules of~\cite{Delmastro:2021xox}:
\begin{equation}\label{eq:u1_char_fer}
\begin{aligned}
\chi^{(k)}_{\text{NS-X};\ell}(q,z)&= \chi^{(4k)}_{2\ell}(q,z)\pm\chi^{(4k)}_{2\ell+2k}(q,z)\\
\chi^{(k)}_{\text{R-X};\ell}(q,z)&= \chi^{(4k)}_{2\ell+1}(q,z)\pm\chi^{(4k)}_{2\ell+1+2k}(q,z)\,.
\end{aligned}
\end{equation}
These characters have also been discussed in e.g.~\cite{Okuda:2020fyl}. The special case $k=1$ is equivalent to a free fermion theory, $U(1)_1=SO(2)_1$, as can be checked by comparing the corresponding super-characters; this is nothing but the trivial statement that one complex fermion equals two real fermions. More generally, $U(n)_1$ denotes the CFT of $n$ complex fermions, and we shall use the notation $U(n)_1=SO(2n)_1$ interchangeably. (The former is more natural when the free fermions are associated to a complex representation of the gauge group $G$).

\paragraph{WZW coset models.} We are now ready to discuss the class of models of interest, namely cosets of the form $SO(n)_1/G_k$, which appear at the deep infrared of QCD theories with gauge group $G$.

The CFT $SO(n)_1/G_k$ is obtained by embedding $G_k$ into $SO(n)_1$. As above, this embedding gives rise to a character decomposition of the form
\begin{equation}\label{eq:char_coset_SO_G}
d_{\pm,\pm}=\sum_\lambda b_\lambda^{\pm,\pm}\chi_\lambda\,,
\end{equation}
where $d_{\pm,\pm}$ are the characters of $SO(n)_1$, and $\chi_\lambda$ those of $G_k$. The denominator theory $G_k$ is allowed to be fermionic, in which case it is understood that $\chi_\lambda=\chi^{\pm,\pm}_\lambda$ is a super-character at spin structure $\pm,\pm$. In any case, whether $G_k$ is fermionic or not, the coset is a fermionic theory, because the numerator $SO(n)_1$ is fermionic. Consequently, the coefficients $b^{\pm,\pm}_\lambda$ depend on the spin structure, as indicated by the superscript. These coefficients are the super-characters of the coset CFT $SO(n)_1/G_k$, and they determine the dynamics of QCD in the infrared. In particular, at low energies the partition function of QCD becomes
\begin{equation}\label{eq:coset_partition_function}
Z_{\pm,\pm}=\sum_\lambda |b_\lambda^{\pm,\pm}|^2\,.
\end{equation}

In this sense, the whole problem of describing the strongly coupled dynamics of QCD has been reduced to the task of finding the coefficients $b_\lambda$ in~\eqref{eq:char_coset_SO_G}. While for generic cosets this is a computationally demanding task, for cosets of the form $SO(n)_1/G_k$ there is a substantial simplification: the numerator $SO(n)_1$ is in fact equivalent to $n$ free Majorana fermions.

This free fermion representation can be exploited as follows. Consider the coset $SO(n)_1/G_k$, where $n=\dim(R)$ and $k=I(R)$ with embedding $G\subseteq SO(n)$ via the representation $R$. In the Neveu-Schwartz sector the free fermions have half-integral modding $\psi_{r+1/2}$, while in the Ramond sector they have integral modding $\psi_r$. These modes are independent, so the $SO(\dim(R))_1$ partition function is just the product of the individual partition functions over all $r\in\mathbb N$. The fermions $\psi_{r+1/2},\psi_r$ all generate $R$-modules except for the Ramond zero modes $\psi_0$, which generate a spinor module. With this, the different partition functions of $SO(\dim(R))_1$ read
\begin{equation}\label{eq:char_SO_G}
\begin{aligned}
d_\text{NS-X}(q,g;R)&=q^{-\dim(R)/48}\prod_{r=0}^\infty\prod_{\lambda\in\Omega(R)}1\pm z^\lambda q^{r+1/2}\\
\text{$\dim(R)$ odd}:\quad d_\text{R-NS}(q,g;R)&=\sqrt 2\,q^{\dim(R)/24}\chi_\sigma(g)\prod_{r=1}^\infty\prod_{\lambda\in\Omega(R)}1+ z^\lambda q^r
\\
\quad d_\text{R-R}(q,g;R)&=0\\
\text{$\dim(R)$ even}:\quad d_\text{R-X}(q,g;R)&=q^{\dim(R)/24}(\chi_s(g)\pm\chi_c(g))\prod_{r=1}^\infty \prod_{\lambda\in\Omega(R)}1\pm z^\lambda q^r\,,
\end{aligned}
\end{equation}
where $g\in G$ is the restriction of any flavor $SO(\dim(R))$ symmetry to the subgroup $G$, and $z$ its value on any maximal torus.

One can use these expressions to compute the first few terms of the $q$-expansion of $d_{\pm,\pm}$. These terms are then reorganized into $G_{I(R)}$ characters, whose $q$-expansion can be obtained with e.g.~the Weyl-Ka\v{c} formula~\eqref{eq:weyl_kac}. The characters of the coset $SO(\dim(R))_1/G_{I(R)}$ are identified with the coefficients of this reorganized series. Computer software is often instrumental in these computations, for example the LieART Mathematica package~\cite{Feger:2019tvk}. The extensive tables of Lie algebras, representations, and branchings in~\cite{Yamatsu:2015npn} can also come in handy.

\paragraph{Topological cosets and conformal embeddings.} A special role is played by cosets $SO(n)_1/G_k$ where $G_k$ embeds into $SO(n)_1$ \emph{conformally}, i.e., when the central charge of $SO(n)_1/G_k$ vanishes. We argued in the main text that this happens if and only if the QCD theory with group $G$ is gapped. When this happens, the infrared theory becomes a trivial CFT. That being said, the coset is not an empty theory, even though it has no local degrees of freedom; in other words, it is a \emph{topological} QFT. The low energy dynamics of gapped theories is entirely contained in the topological degrees of freedom carried by the topological coset $SO(n)_1/G_k$.

By topological invariance, all observables of such cosets become $q$-independent, so the branching functions $b_\lambda$ are just numbers instead of functions of $q$. Note that topological invariance is just a special case of conformal invariance: TQFTs are invariant under all diffeomorphisms instead of just the conformal ones. This means, in particular, that the formula~\eqref{eq:coset_partition_function} is still valid for TQFTs. In this case, as $L_0\equiv 0$, the partition function actually computes the total number of states in the theory, which has a finite-dimensional Hilbert space:
\begin{equation}
\begin{aligned}
Z_{\pm,-}&=\tr_{\mathcal H_\pm}(1)\equiv \text{bosons plus fermions in $\mathcal H_\pm$}\\
Z_{\pm,+}&=\tr_{\mathcal H_\pm}(-1)^F\equiv \text{bosons minus fermions in $\mathcal H_\pm$}
\end{aligned}
\end{equation}
or, equivalently,
\begin{equation}
\begin{aligned}
\text{number of bosons in $\mathcal H_\pm$}=\frac12(Z_{\pm,-}+Z_{\pm,+})&\equiv \sum_\lambda \frac12(|b_\lambda^{\pm,-}|^2+|b_\lambda^{\pm,+}|^2)\\
\text{number of fermions in $\mathcal H_\pm$}=\frac12(Z_{\pm,-}-Z_{\pm,+})&\equiv \sum_\lambda \frac12(|b_\lambda^{\pm,-}|^2-|b_\lambda^{\pm,+}|^2)\,.
\end{aligned}
\end{equation}

In what follows we shall work out several examples in some detail in order to illustrate some of the previous considerations.

\subsection{Examples of topological cosets}

Here we demonstrate the coset construction for the conformal embedding $SO(8)_1\supset SU(3)_3$. As $c=8/2-3\cdot8/6=0$, the resulting theory is topological, i.e., it has a finite-dimensional Hilbert space. In order to find this Hilbert space we need to decompose the $SO(8)$ characters into $SU(3)$ characters. The former are given by~\eqref{eq:char_SO_G}:
\begin{equation}\label{eq:SO8_SU3_char}
\begin{aligned}
\chi_0(q,g)&=q^{-1/6}\biggl[\boldsymbol1+(\boldsymbol8+\boldsymbol{10}+\overline{\boldsymbol{10}})\,q+(\boldsymbol1+4\,\boldsymbol8+\boldsymbol{10}+\overline{\boldsymbol{10}}+3\,\boldsymbol{27})q^2\\
&+(2\,\boldsymbol1+ 10\,\boldsymbol8+6\,\boldsymbol{10}+6\,\overline{\boldsymbol{10}}+6\,\boldsymbol{27}+2\,\boldsymbol{35}+2\,\overline{\boldsymbol{35}}+\boldsymbol{64})q^3+\cdots\biggr]\\
\chi_v(q,g)&=q^{1/3}\biggl[\boldsymbol 8+(\boldsymbol1+2\,\boldsymbol8+\boldsymbol{10}+\overline{\boldsymbol{10}}+\boldsymbol{27})q\\
&+(2\,\boldsymbol1+6\,\boldsymbol 8+3\,\boldsymbol{10}+3\,\overline{\boldsymbol{10}}+4\,\boldsymbol{27}+\boldsymbol{35}+\overline{\boldsymbol{35}})q^2+\cdots\biggr]\,,
\end{aligned}
\end{equation}
and, by triality,
\begin{equation}
\chi_s(q,g)=\chi_c(q,g)=\chi_v(q,g)\,.
\end{equation}

We next reorganize these characters in terms of $SU(3)_3$ characters. The characters of $SU(3)_3$ are given by
\begingroup
\allowdisplaybreaks
\begin{align}
\chi_{\boldsymbol1}(q,g)&=q^{-1/6}\biggl[\boldsymbol1+\boldsymbol8q+\bigl(\boldsymbol1+2\,\boldsymbol8+\boldsymbol{27}\bigr)q^2\notag\\
&\hspace{2cm}+(2\, \boldsymbol1+ 4\, \boldsymbol8+2\, \boldsymbol{10}+ 2\, \overline{\boldsymbol{10}}+2\, \boldsymbol{27}+\boldsymbol{64})q^3+\cdots\biggr]\notag\\
\chi_{\boldsymbol3}(q,g)&=q^{1/18}\biggl[\boldsymbol3+\bigl(\boldsymbol3+\boldsymbol6+\boldsymbol{15}\bigr)q+\bigl(3\,\boldsymbol3+2\,\boldsymbol6+3\,\boldsymbol{15}+\boldsymbol{24}+\boldsymbol{42}\bigr)q^2+\cdots\biggr]\notag\\
\chi_{\boldsymbol6}(q,g)&=q^{7/18}\biggl[\boldsymbol6+\bigl(\boldsymbol3 +\boldsymbol6+ \boldsymbol{15}+\boldsymbol{24} \bigr)q\notag\\
&\hspace{2cm}+\bigl(2 \,\boldsymbol3 +4 \,\boldsymbol6+ 3 \,\boldsymbol{15}+ \boldsymbol{15}'+3 \,\boldsymbol{24}+\boldsymbol{42} \bigr)q^2+\cdots\biggr]\notag\\
\chi_{\boldsymbol8}(q,g)&=q^{1/3}\biggl[\boldsymbol8+\bigl(\boldsymbol1+ 2\,\boldsymbol8+\boldsymbol{10}+\overline{\boldsymbol{10}}+\boldsymbol{27} \bigr)q\label{eq:SU3_chars}\\
&\hspace{2cm}+\bigl(2\, \boldsymbol1+ 6\,\boldsymbol8+ 3 \,\boldsymbol{10}+3 \,\overline{\boldsymbol{10}}+ 4 \,\boldsymbol{27}+\boldsymbol{35}+ \overline{\boldsymbol{35}}\bigr)q^2+\cdots\biggr]\notag\\
\chi_{\boldsymbol{10}}(q,g)&=q^{5/6}\biggl[\boldsymbol{10}+\bigl(\boldsymbol8 +\boldsymbol{10}+\boldsymbol{27}\bigr)q\notag\\
&\hspace{2cm}+\bigl(3\,\boldsymbol8 +3 \,\boldsymbol{10}+\overline{\boldsymbol{10}}+ 2 \,\boldsymbol{27}+\boldsymbol{35}+ \overline{\boldsymbol{35}} \bigr)q^2+\cdots\biggr]\notag\\
\chi_{\boldsymbol{15}}(q,g)&=q^{13/18}\biggl[\boldsymbol{15}+\bigl(\boldsymbol3+ \boldsymbol6+ 2 \,\boldsymbol{15}+\boldsymbol{15}'+\boldsymbol{24}\bigr)q\notag\\
&\hspace{2cm}+\bigl(3 \,\boldsymbol3+ 3 \,\boldsymbol6+ 6 \,\boldsymbol{15}+2 \,\boldsymbol{15}'+ \boldsymbol{21}+ 3 \,\boldsymbol{24}+2 \,\boldsymbol{42} \bigr)q^2+\cdots\biggr]\,.\notag
\end{align}
\endgroup

By comparing~\eqref{eq:SO8_SU3_char} to~\eqref{eq:SU3_chars} it is easily checked that
\begin{equation}
\begin{aligned}
d_\text{NS-NS}&=\chi_{\boldsymbol1}+\chi_{\boldsymbol8} +\chi_{\boldsymbol{10}}+\chi_{\overline{\boldsymbol{10}}}\\
d_\text{NS-R}&=\chi_{\boldsymbol1} - \chi_{\boldsymbol8}+\chi_{\boldsymbol{10}}+\chi_{\overline{\boldsymbol{10}}}\\
d_\text{R-NS}&=2 \chi_{\boldsymbol8}\\
d_\text{R-R}&=0\,,
\end{aligned}
\end{equation}
which implies that the NS sector of $SO(8)_1/SU(3)_3$ has four bosons and no fermions, and the R sector has two and two, i.e., $\mathcal H_\text{NS}=\mathbb C^{4|0}$ and $\mathcal H_\text{R}=\mathbb C^{2|2}$.

It is interesting to note that, out of the four bosons in $\mathcal H_\text{NS}$, one of them (the one corresponding to $b_{\boldsymbol 8}$) is charged under $(-1)^{F_R}$, while the other three are neutral. In the full non-chiral theory, this state is a boson because it comes from $\bar b_{\boldsymbol 8}b_{\boldsymbol 8}$, which is charged under both $(-1)^{F_L}$ and $(-1)^{F_R}$ (and is therefore neutral under $(-1)^F=(-1)^{F_L}(-1)^{F_R}$).

In the previous example we found that $\mathcal H_\text{R}$ was supersymmetric (it contains the same number of bosons as fermions), which was a consequence of $d_\text{R-R}$ vanishing. In order to show that this is not always the case, we will describe an example where $\mathcal H_\text{R}$ is \emph{not} supersymmetric.

Consider the coset $SO(16)_1/Spin(9)_2$. Using~\eqref{eq:char_SO_G}, the characters of the numerator are
\begingroup
\allowdisplaybreaks
\begin{align}
\chi_0(q,g)&=q^{-1/3}\bigl(\boldsymbol1+(\boldsymbol{36}+\boldsymbol{84})q\notag\\
&+(\boldsymbol{9}+\boldsymbol{16}+\boldsymbol{36}+\boldsymbol{44}+\boldsymbol{84}+2 \,\boldsymbol{126}+\boldsymbol{231}+\boldsymbol{495}+\boldsymbol{924})q^2\notag\\
&+(2 \,\boldsymbol{9}+\boldsymbol{16}+5 \,\boldsymbol{36}+\boldsymbol{44}+5 \,\boldsymbol{84}+3 \,\boldsymbol{126}+3 \,\boldsymbol{231}+\boldsymbol{495}\notag\\
&+3 \,\boldsymbol{594}+\boldsymbol{910}+3 \,\boldsymbol{924}+\boldsymbol{1650}+\boldsymbol{2457}+2 \,\boldsymbol{2772})q^3+\cdots\bigr)\notag\\
\chi_v(q,g)&=q^{1/6}\bigl(\boldsymbol{16}+(\boldsymbol{16}+\boldsymbol{128}+\boldsymbol{432})q\notag\\
&+(3 \,\boldsymbol{16}+3 \,\boldsymbol{128}+3 \,\boldsymbol{432}+\boldsymbol{576}+\boldsymbol{672}+\boldsymbol{768}+\boldsymbol{2560})q^2\notag\\
&+(7 \,\boldsymbol{16}+9 \,\boldsymbol{128}+9 \,\boldsymbol{432}+4 \,\boldsymbol{576}+2 \,\boldsymbol{672}+5 \,\boldsymbol{768}\notag\\
&+\boldsymbol{1920}+4 \,\boldsymbol{2560}+\boldsymbol{4608}+\boldsymbol{4928}+2 \,\boldsymbol{5040})q^3+\cdots\bigr)\notag\\
\chi_s(q,g)&=q^{2/3}\bigl(\boldsymbol{44}+\boldsymbol{84}+(\boldsymbol{9}+\boldsymbol{36}+\boldsymbol{44}+\boldsymbol{84}+\boldsymbol{126}+\boldsymbol{231}+\boldsymbol{594}+\boldsymbol{924})q\\
&+(\boldsymbol1+2 \,\boldsymbol{9}+3 \,\boldsymbol{36}+3 \,\boldsymbol{44}+4 \,\boldsymbol{84}+3 \,\boldsymbol{126}+3 \,\boldsymbol{231}+\boldsymbol{495}\notag\\
&+3 \,\boldsymbol{594}+\boldsymbol{910}+4 \,\boldsymbol{924}+\boldsymbol{1650}+\boldsymbol{1980}+\boldsymbol{2457}+\boldsymbol{2772})q^2\notag\\
&+(2 \,\boldsymbol1+6 \,\boldsymbol{9}+8 \,\boldsymbol{36}+6 \,\boldsymbol{44}+10 \,\boldsymbol{84}+10 \,\boldsymbol{126}+2 \,\boldsymbol{156}+9 \,\boldsymbol{231}\notag\\
&+3 \,\boldsymbol{495}+11 \,\boldsymbol{594}+3 \,\boldsymbol{910}+11 \,\boldsymbol{924}+5 \,\boldsymbol{1650}+2 \,\boldsymbol{1980}\notag\\
&+3 \,\boldsymbol{2457}+\boldsymbol{2574}+5 \,\boldsymbol{2772}+3 \,\boldsymbol{3900}+2 \,\boldsymbol{4158}+\boldsymbol{9009}+\boldsymbol{15444})q^3+\cdots\bigr)\notag\\
\chi_c(q,g)&=q^{2/3}\bigl(\boldsymbol{128}+(\boldsymbol{16}+2 \,\boldsymbol{128}+\boldsymbol{432}+\boldsymbol{576}+\boldsymbol{768})q\notag\\
&+(3 \,\boldsymbol{16}+6 \,\boldsymbol{128}+4 \,\boldsymbol{432}+3 \,\boldsymbol{576}+\boldsymbol{672}+3 \,\boldsymbol{768}+2 \,\boldsymbol{2560}+\boldsymbol{5040})q^2\notag\\
&+(8 \,\boldsymbol{16}+16 \,\boldsymbol{128}+13 \,\boldsymbol{432}+9 \,\boldsymbol{576}+4 \,\boldsymbol{672}+9 \,\boldsymbol{768}+\boldsymbol{1920}\notag\\
&+8 \,\boldsymbol{2560}+2 \,\boldsymbol{4608}+\boldsymbol{4928}+5 \,\boldsymbol{5040}+\boldsymbol{9504}+\boldsymbol{12672})q^3+\cdots\bigr)\,.\notag
\end{align}
\endgroup

In order to express these in terms of the characters of the denominator, we need the $Spin(9)_2$ characters:
\begingroup
\allowdisplaybreaks
\begin{align}
\chi_{\boldsymbol1}(q,g)&=q^{-1/3}\bigl(\boldsymbol1+\boldsymbol{36}\,q+(\boldsymbol{1}+ \boldsymbol{36}+\boldsymbol{44} + \boldsymbol{126} + \boldsymbol{495} )q^2\notag\\
 &+(\boldsymbol{1}+ 4\,\boldsymbol{36}+ \boldsymbol{44}+ \boldsymbol{84} + \boldsymbol{126}+ \boldsymbol{495}+ 2\,\boldsymbol{594} + \boldsymbol{910} + \boldsymbol{2772})q^3+\cdots\bigr)\notag\\
\chi_{\boldsymbol{16}}(q,g)&=q^{1/6}\bigl(\boldsymbol{16}+(\boldsymbol{16}+ \boldsymbol{128} + \boldsymbol{432} )q\notag\\
&+(3\,\boldsymbol{16}+ 3\,\boldsymbol{128}+ 3\,\boldsymbol{432} + \boldsymbol{576}+ \boldsymbol{672} + \boldsymbol{768} + \boldsymbol{2560} )q^2\notag\\
&+(7\,\boldsymbol{16}+ 9\,\boldsymbol{128}+ 9\,\boldsymbol{432} + 4\,\boldsymbol{576}+ 2\,\boldsymbol{672} + 5\,\boldsymbol{768}\notag\\
&+ \boldsymbol{1920}+ 4\,\boldsymbol{2560}+ \boldsymbol{4608} + \boldsymbol{4928} + 2\,\boldsymbol{5040} )q^3+\cdots\bigr)\notag\\
\chi_{\boldsymbol{44}}(q,g)&=q^{2/3}\bigl(\boldsymbol{44}+(\boldsymbol{36}+ \boldsymbol{44} + \boldsymbol{594} )q\notag\\
&+(\boldsymbol{1} + 2\,\boldsymbol{36}+ 3\,\boldsymbol{44}+ \boldsymbol{126}+ \boldsymbol{495}+ 2\,\boldsymbol{594}+ \boldsymbol{910} + \boldsymbol{924}+ \boldsymbol{1980} )q^2\notag\\
&+(\boldsymbol{1}+ 6\,\boldsymbol{36} + 5\,\boldsymbol{44}+ \boldsymbol{84}+ 3\,\boldsymbol{126} + \boldsymbol{231} + 2\,\boldsymbol{495}+ 7\,\boldsymbol{594}+ 3\,\boldsymbol{910}\notag\\
&+ 2\,\boldsymbol{924}+ \boldsymbol{1980}+ 2\,\boldsymbol{2772}+ \boldsymbol{3900} + \boldsymbol{4158} + \boldsymbol{9009} )q^3+\cdots\bigr)\\
\chi_{\boldsymbol{84}}(q,g)&=q^{2/3}\bigl(\boldsymbol{84}+(\boldsymbol{9}+ \boldsymbol{84} +\boldsymbol{126}+ \boldsymbol{231} + \boldsymbol{924} )q\notag\\
&+(2\,\boldsymbol{9}+ \boldsymbol{36}+ 4\,\boldsymbol{84} +2\,\boldsymbol{126}+ 3\,\boldsymbol{231}+ \boldsymbol{594}+ 3\,\boldsymbol{924}+ \boldsymbol{1650} + \boldsymbol{2457}+ \boldsymbol{2772})q^2\notag\\
&+(\boldsymbol{1}+ 6\,\boldsymbol{9}+ 2\,\boldsymbol{36}+ \boldsymbol{44} + 9\,\boldsymbol{84} + 7\,\boldsymbol{126}+ 2\,\boldsymbol{156}+ 8\,\boldsymbol{231}+ \boldsymbol{495}+ 4\,\boldsymbol{594}+ 9\,\boldsymbol{924}\notag\\
&+ 5\,\boldsymbol{1650}+ \boldsymbol{1980}+ 3\,\boldsymbol{2457}+ \boldsymbol{2574} + 3\,\boldsymbol{2772}+ 2\,\boldsymbol{3900}+ \boldsymbol{4158} + \boldsymbol{15444})q^3+\cdots\bigr)\notag\\
\chi_{\boldsymbol{128}}(q,g)&=q^{2/3}\bigl(\boldsymbol{128}+(\boldsymbol{16}+ 2\,\boldsymbol{128}+ \boldsymbol{432}+ \boldsymbol{576} + \boldsymbol{768} )q\notag\\
&+(3\,\boldsymbol{16}+ 6\,\boldsymbol{128}+ 4\,\boldsymbol{432} + 3\,\boldsymbol{576}+ \boldsymbol{672} + 3\,\boldsymbol{768} + 2\,\boldsymbol{2560} + \boldsymbol{5040} )q^2\notag\\
&+(8\,\boldsymbol{16} + 16\,\boldsymbol{128}+ 13\,\boldsymbol{432}+ 9\,\boldsymbol{576}+ 4\,\boldsymbol{672} + 9\,\boldsymbol{768}+ \boldsymbol{1920}\notag\\
& + 8\,\boldsymbol{2560}+ 2\,\boldsymbol{4608} + \boldsymbol{4928}+ 5\,\boldsymbol{5040}+ \boldsymbol{9504}+ \boldsymbol{12672} )q^3+\cdots\bigr)\,,\notag
\end{align}
\endgroup
in terms of which one can write
\begin{equation}
\begin{aligned}
d_\text{NS-NS}(q,g)&=\chi_{\boldsymbol1}(q,g)+\chi_{\boldsymbol{16}}(q,g)+\chi_{\boldsymbol{84}}(q,g)\\
d_\text{NS-R}(q,g)&=\chi_{\boldsymbol1}(q,g)-\chi_{\boldsymbol{16}}(q,g)+\chi_{\boldsymbol{84}}(q,g)\\
d_\text{R-NS}(q,g)&=\chi_{\boldsymbol{44}}(q,g)+\chi_{\boldsymbol{84}}(q,g) +\chi_{\boldsymbol{128}}(q,g)\\
d_\text{R-R}(q,g)&=\chi_{\boldsymbol{44}}(q,g)+\chi_{\boldsymbol{84}}(q,g) -\chi_{\boldsymbol{128}}(q,g)\,.
\end{aligned}
\end{equation}

These affine branching rules imply that the Hilbert spaces of $SO(16)_1/Spin(9)_2$ are $\mathcal H_\text{NS}=\mathcal H_\text{R}=\mathbb C^{3|0}$. As promised, $\mathcal H_\text{R}$ is not supersymmetric. While the unextended character $d_\text{R-R}(q)=0$ vanishes, the extended one $d_\text{R-R}(q,g)=\chi_{\boldsymbol{84}}(q,g) + \chi_{\boldsymbol{44}}(q,g)-\chi_{\boldsymbol{128}}(q,g)$ is non-zero, and hence the coset does not have the same number of bosons and fermions.

In section~\ref{sec:Infrareddynamnics} we attributed the supersymmetry of the Ramond sector to certain 't Hooft anomalies. These anomalies are present when the number of quarks is odd. In the theory $SO(16)_1/Spin(9)_2$ the number of fermions is even, $16$, so there is no reason to expect that the Ramond sector is supersymmetric -- and indeed it is not.

\subsection{Example of a non-topological coset}\label{ap:minimal_model_coset}

Here we study the coset $Spin(7)_1/SU(2)_{28}$, which has $c=7/2-3\times28/30\equiv7/10$. This is non-zero so the coset is non-topological, i.e., it is a traditional CFT. We begin by writing the characters of the numerator $Spin(7)_1$:
\begin{equation}
\begin{aligned}
\chi_0(q,g)&=q^{-7/48}\biggl[\boldsymbol1+(\boldsymbol3+\boldsymbol7+\boldsymbol{11})q
+(2\,\boldsymbol1+\boldsymbol3+2\,\boldsymbol5+2\,\boldsymbol7+2\,\boldsymbol9+\boldsymbol{11}+2\,\boldsymbol{13})q^2\\
&+(2\,\boldsymbol1+5\,\boldsymbol3+4\,\boldsymbol5+8\,\boldsymbol7+5\,\boldsymbol9+6\,\boldsymbol{11}+4\,\boldsymbol{13}
+2\,\boldsymbol{15}+\boldsymbol{17}+\boldsymbol{19})q^3+\cdots\biggr]\\
\chi_v(q,g)&=q^{17/48}\biggl[\boldsymbol7+(\boldsymbol1+\boldsymbol5+2\,\boldsymbol7+\boldsymbol9+\boldsymbol{13})q\\
&+(\boldsymbol1+2\,\boldsymbol3+3\,\boldsymbol7+\boldsymbol{11}
+3\,\boldsymbol5+2\,\boldsymbol7+3\,\boldsymbol9+2\,\boldsymbol{11}+2\,\boldsymbol{13}+\boldsymbol{15}+\boldsymbol{17})q^2\\
&+(4\,\boldsymbol1+5\,\boldsymbol3+9 \,\boldsymbol5+12 \,\boldsymbol7+10 \,\boldsymbol9+8\,\boldsymbol{11}+7 \,\boldsymbol{13}
+2 \,\boldsymbol{15}+3 \,\boldsymbol{17}+\boldsymbol{19})q^3+\cdots\biggr]\\
\chi_\sigma(q,g)&=q^{7/24}\biggl[
\boldsymbol1+\boldsymbol7+
(\boldsymbol1+\boldsymbol3+\boldsymbol5+2\,\boldsymbol7+\boldsymbol9+\boldsymbol{11}+\boldsymbol{13})q\\
&+(2 \,\boldsymbol1+3 \,\boldsymbol3+4 \,\boldsymbol5+6 \,\boldsymbol7+4 \,\boldsymbol9+4 \,\boldsymbol{11}+3 \,\boldsymbol{13}+\boldsymbol{15}+\boldsymbol{17})q^2+\\
&+(2\,\boldsymbol1+8 \,\boldsymbol3+11 \,\boldsymbol5+16 \,\boldsymbol7+13 \,\boldsymbol9+11 \,\boldsymbol{11}\\
&\qquad+10 \,\boldsymbol{13}+5 \,\boldsymbol{15}+3 \,\boldsymbol{17}+2\,\boldsymbol{19})q^3+\cdots\biggr]\,.
\end{aligned}
\end{equation}

Similarly, the characters of the denominator $SU(2)_{28}$ are
\begin{equation}
\begin{aligned}
\chi_{\boldsymbol1}(q,g)&=q^{-7/60}(\boldsymbol1+\boldsymbol3\,q+(\boldsymbol1+\boldsymbol3+\boldsymbol5)q^2+(\boldsymbol1+3 \,\boldsymbol3+\boldsymbol5+\boldsymbol7)q^3+\cdots)\\
\chi_{\boldsymbol7}(q,g)&=q^{17/60}(\boldsymbol7+(\boldsymbol5+\boldsymbol7+\boldsymbol9)q+(\boldsymbol3+2 \,\boldsymbol5+3 \,\boldsymbol7+2 \,\boldsymbol9+\boldsymbol{11})q^2\\
&+(\boldsymbol1+2 \,\boldsymbol3+5 \,\boldsymbol5+6 \,\boldsymbol7+5 \,\boldsymbol9+2 \,\boldsymbol{11}+\boldsymbol{13})q^3+\cdots)\\
\chi_{\boldsymbol{11}}(q,g)&=q^{53/60}(\boldsymbol{11}+(\boldsymbol9+\boldsymbol{11}+\boldsymbol{13})q+(\boldsymbol7+2 \,\boldsymbol9+3 \,\boldsymbol{11}+2 \,\boldsymbol{13}+\boldsymbol{15})q^2\\
&+(\boldsymbol5+2 \,\boldsymbol7+5 \,\boldsymbol9+6 \,\boldsymbol{11}+5 \,\boldsymbol{13}+2 \,\boldsymbol{15}+\boldsymbol{17})q^3+\cdots)\\
\chi_{\boldsymbol{13}}(q,g)&=q^{77/60}(\boldsymbol{13}+(\boldsymbol{11}+\boldsymbol{13}+\boldsymbol{15})q+(\boldsymbol9+2 \,\boldsymbol{11}+3 \,\boldsymbol{13}+2 \,\boldsymbol{15}+\boldsymbol{17})q^2\\
&+(\boldsymbol7+2 \,\boldsymbol9+5 \,\boldsymbol{11}+6 \,\boldsymbol{13}+5 \,\boldsymbol{15}+2 \,\boldsymbol{17}+\boldsymbol{19})q^3+\cdots)\\
\chi_{\boldsymbol{17}}(q,g)&=q^{137/60}(\boldsymbol{17}+(\boldsymbol{15}+\boldsymbol{17}+\boldsymbol{19})q+(\boldsymbol{13}+2 \,\boldsymbol{15}+3 \,\boldsymbol{17}+2 \,\boldsymbol{19}+\boldsymbol{21})q^2\\
&+(\boldsymbol{11}+2 \,\boldsymbol{13}+5 \,\boldsymbol{15}+6 \,\boldsymbol{17}+5 \,\boldsymbol{19}+2 \,\boldsymbol{21}+\boldsymbol{23})q^3+\cdots)\\
\chi_{\boldsymbol{19}}(q,g)&=q^{173/60}(\boldsymbol{19}+(\boldsymbol{17}+\boldsymbol{19}+\boldsymbol{21})q+(\boldsymbol{15}+2 \,\boldsymbol{17}+3 \,\boldsymbol{19}+2 \,\boldsymbol{21}+\boldsymbol{23})q^2\\
&+(\boldsymbol{13}+2 \,\boldsymbol{15}+5 \,\boldsymbol{17}+6 \,\boldsymbol{19}+5 \,\boldsymbol{21}+2 \,\boldsymbol{23}+\boldsymbol{25})q^3+\cdots)\\
\chi_{\boldsymbol{23}}(q,g)&=q^{257/60}(\boldsymbol{23}+(\boldsymbol{21}+\boldsymbol{23}+\boldsymbol{25})q+(\boldsymbol{19}+2 \,\boldsymbol{21}+3 \,\boldsymbol{23}+2 \,\boldsymbol{25}+\boldsymbol{27})q^2\\
&+(\boldsymbol{17}+2 \,\boldsymbol{19}+5 \,\boldsymbol{21}+6 \,\boldsymbol{23}+5 \,\boldsymbol{25}+2 \,\boldsymbol{27}+\boldsymbol{29})q^3+\cdots)\\
\chi_{\boldsymbol{29}}(q,g)&=q^{413/60}(\boldsymbol{29}+(\boldsymbol{27}+\boldsymbol{29})q+(\boldsymbol{25}+2 \,\boldsymbol{27}+2 \,\boldsymbol{29}+\boldsymbol{31})q^2\\
&+(\boldsymbol{23}+2 \,\boldsymbol{25}+4 \,\boldsymbol{27}+4 \,\boldsymbol{29}+2 \,\boldsymbol{31})q^3+\cdots)\\
&\cdots
\end{aligned}
\end{equation}

Given these expressions one can check that the $Spin(7)_1$ characters decompose into $SU(2)_{28}$ characters as
\begin{equation}
\begin{aligned}
\chi_0(q,g)&=q^{-7/240}(\chi_{\boldsymbol1}(q,g)+\chi_{\boldsymbol{11}}(q,g)+\chi_{\boldsymbol{19}}(q,g)+\chi_{\boldsymbol{29}}(q,g))(1+q^2+q^3+2 q^4+2 q^5+4 q^6+\cdots)\\
&+q^{137/240}(\chi_{\boldsymbol 7}(q,g)+\chi_{\boldsymbol{13}}(q,g)+\chi_{\boldsymbol{17}}(q,g)+\chi_{\boldsymbol{23}}(q,g))(1+q+2 q^2+2 q^3+4 q^4+5 q^5+\cdots)\\
\chi_v(q,g)&=q^{353/240}(\chi_{\boldsymbol1}(q,g)+\chi_{\boldsymbol{11}}(q,g)+\chi_{\boldsymbol{19}}(q,g)+\chi_{\boldsymbol{29}}(q,g))(1+q+2 q^2+2 q^3+3 q^4+4 q^5+\cdots)\\
&+q^{17/240}(\chi_{\boldsymbol 7}(q,g)+\chi_{\boldsymbol{13}}(q,g)+\chi_{\boldsymbol{17}}(q,g)+\chi_{\boldsymbol{23}}(q,g))(1+q+q^2+2 q^3+3 q^4+4 q^5+\cdots)\\
\chi_\sigma(q,g)&=q^{49/120}(\chi_{\boldsymbol1}(q,g)+\chi_{\boldsymbol{11}}(q,g)+\chi_{\boldsymbol{19}}(q,g)+\chi_{\boldsymbol{29}}(q,g))(1+q+q^2+2 q^3+3 q^4+4 q^5+\cdots)\\
&+q^{1/120}(\chi_{\boldsymbol 7}(q,g)+\chi_{\boldsymbol{13}}(q,g)+\chi_{\boldsymbol{17}}(q,g)+\chi_{\boldsymbol{23}}(q,g))(1+q+2 q^2+3 q^3+4 q^4+6 q^5+\cdots)\,.
\end{aligned}
\end{equation}

According to the coset prescription we are instructed to regard the $q$-dependent coefficients as the characters of a new theory, which we denote as
\begin{equation}
\begin{aligned}
\chi_{1,1}(q)&=q^{-7/240}(1+q^2+q^3+2 q^4+2 q^5+4 q^6+4 q^7+7 q^8+\cdots)\\
\chi_{1,3}(q)&=q^{137/240}(1+q+2 q^2+2 q^3+4 q^4+5 q^5+7 q^6+9 q^7+\cdots)\\
\chi_{1,4}(q)&=q^{353/240}(1+q+2 q^2+2 q^3+3 q^4+4 q^5+6 q^6+7 q^7+\cdots)\\
\chi_{1,2}(q)&=q^{17/240}(1+q+q^2+2 q^3+3 q^4+4 q^5+6 q^6+8 q^7+\cdots)\\
\chi_{2,1}(q)&=q^{49/120}(1+q+q^2+2 q^3+3 q^4+4 q^5+6 q^6+8 q^7+\cdots)\\
\chi_{2,2}(q)&=q^{1/120}(1+q+2 q^2+3 q^3+4 q^4+6 q^5+8 q^6+11 q^7+\cdots)\,,
\end{aligned}
\end{equation}
which we recognize as the Virasoro characters of the minimal model $M(4,5)$ with central charge $c=7/10$:
\begin{equation}
\chi_{r,s}(q)=k_s(q)-k_{-s}(q),\qquad k_s(q):=\eta(q)^{-1}\sum_{n\in\mathbb Z}q^{((2 n p (p+1)+r (p+1) - s p)^2)/4 p (p+1)}\,.
\end{equation}

Fermionizing, we get the super-characters
\begin{equation}
\begin{aligned}
\chi^\text{NS-X}(q)&=\chi_{1,1}\pm\chi_{1,4}\quad \&\quad \chi_{1,3}\pm\chi_{1,2}\\
\chi^\text{R-NS}(q)&=\chi_{2,1}\quad\&\quad \chi_{2,2}\,,
\end{aligned}
\end{equation}
which, nonsurprisingly, are the characters of the fermionic $c=7/10$ minimal model. In other words, the infrared theory of $SU(2)+\boldsymbol{7}$ is the fermionic $M(4,5)$ minimal model, with coset realization $SO(7)_1/SU(2)_{28}$.

Another interesting example is $SO(8)_1/Spin(7)_1$. The numerator and denominator algebras are both of the type $\mathfrak{so}(n)_1$, so the characters are straightforward. By working out the decomposition one obtains
\begin{equation}
\begin{aligned}
d_\text{NS-X}&=\chi^{(7)}_0\chi^{(1)}_0+\chi^{(7)}_v\chi^{(1)}_v\pm\chi^{(7)}_\sigma\chi^{(1)}_\sigma\\
d_\text{R-X}&=\chi^{(7)}_0\chi^{(1)}_v+\chi^{(7)}_v\chi^{(1)}_0\pm\chi^{(7)}_\sigma\chi^{(1)}_\sigma\,,
\end{aligned}
\end{equation}
where $\chi^{(7)}$ are the characters of $Spin(7)_1$ and $\chi^{(1)}$ are the Ising characters. Therefore, the infrared chiral algebra of $Spin(8)+\boldsymbol8$ is the (bosonic) Ising CFT.

\section{Abelian theories}\label{sec:abelian_G}

Consider a QED theory with $N_c$ photons and $N_F$ non-chiral Dirac fermions:
\begin{equation}\label{eq:non_chiral_QED}
\mathcal L=\frac12\sum_{i,j=1}^{N_c}g_{ij}^{-2}\mathrm da_i\wedge\star\mathrm da_j+\frac{i}{2\pi}\sum_{i=1}^{N_c}\theta_i\mathrm da_i+\sum_{I=1}^{N_F}\bar\psi_I\!\not\!\! D_I\psi^I\,,
\end{equation}
where the gauge fields are normalized to integral periods:
\begin{equation}\label{eq:QED_normalization}
\int\frac{\mathrm da_i}{2\pi}\in \mathbb Z\,,
\end{equation}
and $D$ denotes the covariant derivative
\begin{equation}
 D_I= \partial+i\sum_{i=1}^{N_c}Q_{Ii} a{}_i\,,
\end{equation}
with $Q_{Ii}\in\mathbb Z$ the charge of the field $\psi_I$ under $U(1)_i$.

In order to simplify the notation we will often think of indexed objects $g_{ij},\theta_i,a_i,Q_{Ii}$ as arrays of suitable shape, e.g., $g$ is a square matrix of dimension $N_c\times N_c$, $\theta$ a row vector of dimension $1\times N_c$, $Q$ a rectangular matrix of dimension $N_F\times N_c$, etc. 

In~\eqref{eq:non_chiral_QED}, $g_{ij},\theta_i$ denote the coupling constants of the model. These are not all independent: for example, one can always perform linear changes of basis in photon space $a\mapsto Aa$, under which $g^{-2}\mapsto A^tg^{-2}A$, $\theta\mapsto \theta A$. Here $A\in GL(N_c,\mathbb Z)$ is an integral unimodular matrix so as to preserve the quantization condition~\eqref{eq:QED_normalization}. Under this redefinition, the matrix of charges $Q$ transforms as $Q\mapsto QA$. We will come back to this momentarily.

Classically, the model has flavor symmetry $U(1)^{N_F}$ corresponding to axial rotations of the fermions $\psi_I\mapsto e^{i \alpha_I\gamma^3}\psi_I$. This flavor symmetry may be enhanced to a non-abelian group if some of the rows of $Q$ are equal. These classical symmetries often have a mixed anomaly with the gauge group $U(1)^{N_c}$. Specifically, consider the flavor subgroup $U(1)_F\subseteq U(1)^{N_F}$ defined by a certain row vector of integers $n=(n_1,n_2,\dots,n_{N_F})$ such that, under $\alpha\in U(1)_F$, the fermions rotate with angle $\alpha_I=\alpha n_I$. Under $U(1)_F$ the left-movers have charge $+n_I$, while the right-movers have charge $-n_I$. Under $U(1)_i$, both chiralities have charge $Q_{Ii}$. Therefore, the mixed flavor-gauge anomaly is
\begin{equation}
U(1)_F\text{ - }U(1)_i\colon\qquad \sum_{I=1}^{N_F}2n_IQ_{Ii}\equiv 2(nQ)_i\,.
\end{equation}
This mixed anomaly has two (dual) interpretations: first, it stems from the fact that, under $U(1)_F$ rotations, the theta terms in~\eqref{eq:non_chiral_QED} shift, so these terms are rendered unphysical; and second, it corresponds to the fact that the current that generates $U(1)_F$ is not conserved, and hence $U(1)_F$ is not an actual symmetry. Let us analyze both these points in turn.

\paragraph{Theta terms.} The mixed anomaly between $U(1)^{N_F}$ and $U(1)^{N_c}$ can be understood as the statement that, under $U(1)^{N_F}$ rotations, the theta terms in~\eqref{eq:non_chiral_QED} shift. In particular, under the $U(1)_F\subseteq U(1)^{N_F}$ subgroup specified by the integer vector $n$, one has
\begin{equation}
U(1)_F\colon \theta\mapsto \theta+2\alpha nQ\,.
\end{equation}
This means that $\theta$ are unphysical parameters, as they can generically be rotated away. More precisely, the linear combination $\theta v$, where $v$ is a $N_c\times 1$ column vector, shifts as $\Delta(\theta v)=2 \alpha nQv$, and this is zero for all $\alpha\in U(1)_F$ if and only if $Qv\equiv 0$. In other words, the physical theta parameters of the system are in correspondence with the vectors that are annihilated by $Q$ on the right:
\begin{mdframed}
$\bullet$ In the non-chiral QED system~\eqref{eq:non_chiral_QED}, the space of physical theta parameters is given by the kernel (right-null-space) of $Q$. The linear combination $\theta v$ is physical if and only if $v\in \ker Q$.
\end{mdframed}

\paragraph{Axial currents.} The mixed anomaly between $U(1)^{N_F}$ and $U(1)^{N_c}$ can also be understood as the statement that, in the quantum theory, some of the currents that generate $U(1)^{N_F}$ are not conserved. In particular, the subgroup $U(1)_F\subseteq U(1)^{N_F}$ specified by the integer vector $n$ is generated by the current
\begin{equation}
J_n:=\sum_{I=1}^{N_F}n_IJ_I\,,
\end{equation}
where $J^\mu_I=\bar\psi_I\gamma^3\gamma^\mu\psi_I$ is the current that generates axial rotations of the $I$-th fermion. The mixed $U(1)_F$-$U(1)_i$ anomaly violates the conservation law for $J_n$ if and only if $2nQ$ is non-zero. In other words, there are as many conserved axial currents as there are row vectors that are annihilated by $Q$ on the left:
\begin{mdframed}
$\bullet$ In the non-chiral QED system~\eqref{eq:non_chiral_QED}, the algebra of axial flavor symmetries is given by the cokernel (left-null-space) of $Q$. The linear combination $n\cdot J$ is conserved if and only if $n\in\operatorname{coker}Q$.
\end{mdframed}

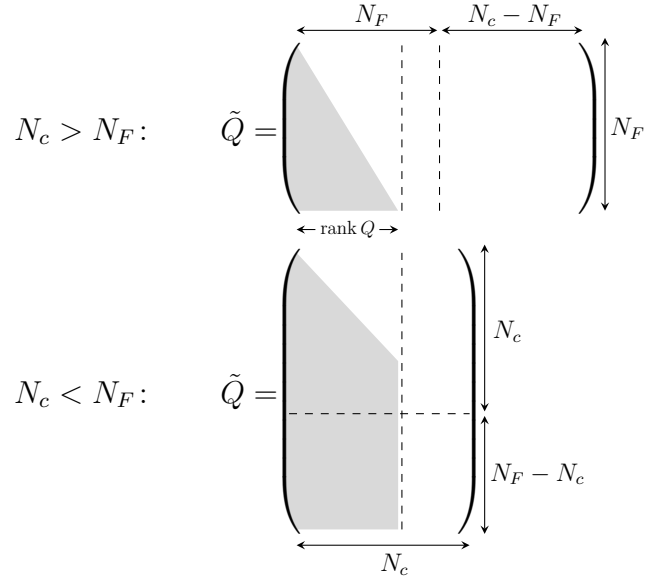
\begin{figure}[!h]
\centering
\begin{tikzpicture}[baseline=-2cm]
\node at (-1.9,0) {$N_c>N_F\colon\qquad \tilde Q=$};
\fill[gray,opacity=.3] (.1,-.82-.28) -- (1.45,-.82-.28) -- (.1,.82+.28) -- (.03,.7+.28) -- (-.01,.6+.28) -- (-.04,.4+.28) -- (-.05,.2+.28) -- (-.05,0) -- (-.05,-.2-.28) -- (-.04,-.4-.28) -- (-.01,-.6-.28) -- (.03,-.7-.28) -- cycle;
\node at (0,0) {$\left(\vphantom{\begin{matrix}\\\\\\\\\end{matrix}}\right.$};
\draw[dashed] (1.5,-.82-.28) -- (1.5,.82+.28);
\draw[dashed] (2,-.82-.28) -- (2,.82+.28);
\node at (4,0) {$\left.\vphantom{\begin{matrix}\\\\\\\\\end{matrix}}\right)$};
\draw[<->,>=stealth] (.1,1.3) -- (2-.03,1.3);
\node[scale=.8] at (1.1,1.5) {$N_F$};

\draw[<->,>=stealth] (2.03,1.3) -- (3.9,1.3);
\node[scale=.8] at (3,1.5) {$N_c-N_F$};

\draw[<->,>=stealth] (4.2,-1.1) -- (4.2,1.2);
\node[scale=.8] at (4.5,0) {$N_F$};

\draw[<-,>=stealth] (.1,-1.35) -- (.35,-1.35);
\draw[->,>=stealth] (1.2,-1.35) -- (1.45,-1.35);
\node[scale=.6] at (.78,-1.35) {$\operatorname{rank}Q$};

\begin{scope}[shift={(0,-3.5)}]
\node at (-1.9,0) {$N_c<N_F\colon\qquad \tilde Q=$};
\fill[gray,opacity=.3] (.1,-.82-.74-.28) -- (1.45,-.82-.74-.28) -- (1.45,.4) -- (.1,.82+.74+.28) -- (.03,.7+.74+.28) -- (-.01,.6+.74+.28) -- (-.04,.4+.74+.28) -- (-.05,.2+.74+.28) -- (-.05,0) -- (-.05,-.2-.74-.28) -- (-.04,-.4-.74-.28) -- (-.01,-.6-.74-.28) -- (.03,-.7-.74-.28) -- cycle;
\draw[dashed] (1.5,-.82-.74-.28) -- (1.5,.82+.74+.28);
\draw[dashed] (0,-.3) -- (2.4,-.3);
\node at (0,0) {$\left(\vphantom{\begin{matrix}\\\\\\\\\\\\\\\end{matrix}}\right.$};
\node at (2.4,0) {$\left.\vphantom{\begin{matrix}\\\\\\\\\\\\\\\end{matrix}}\right)$};
\draw[<->,>=stealth] (2.6,-.3+.03) -- (2.6,1.95);
\node[scale=.8] at (2.9,.8) {$N_c$};

\draw[<->,>=stealth] (2.6,-.3-.03) -- (2.6,-1.9);
\node[scale=.8] at (3.3,-1.1) {$N_F-N_c$};

\draw[<->,>=stealth] (.1,-2.05) -- (2.4,-2.05);
\node[scale=.8] at (1.4,-2.3) {$N_c$};

\end{scope}
\end{tikzpicture}
\caption{Hermite normal form of an $N_F\times N_c$ integral matrix $Q$. The gray region represents the non-zero entries. This decomposition is unique if we impose some further restrictions, such as positivity of pivots; this shall play no role in this work.}
\label{fig:hermite_normal}
\end{figure}

\paragraph{Hermite Normal form.} We noticed above that one can always perform changes of basis in photon space according to $a\mapsto Aa$, where $A$ is a matrix in $GL(N_c,\mathbb Z)$. This change of basis redefines the matrix of charges according to $Q\mapsto QA$. One can always use this freedom to put $Q$ in (column) Hermite normal form, namely $Q=\tilde QA$ where $\tilde Q$ is lower triangular and columns of zeros, if any, are to the far right (see figure~\ref{fig:hermite_normal}).

In this basis the matrix of charges has $\ker Q$ columns that are identically zero. This means that the corresponding photons are essentially decoupled. (They still couple topologically, via the kinetic term. This does not affect local properties like the existence of a mass gap). This explains why the physical theta terms come from $\ker Q$: a theta term is physical if and only if it multiplies a free photon. In other words, the parameter $\theta_i v_i$ is physical if and only if the photon $a_i v_i$ is decoupled from the fermions.

Putting together the last two observations we learn that, as far as classifying gapped theories is concerned, we can assume without loss of generality that there is the same number of photons than fermions, $N_F\equiv N_c$. This follows from the rank-nullity theorem, $|\ker Q|-|\operatorname{coker} Q|\equiv N_c-N_F$, which implies that if $N_c\neq N_F$, then at least one of $\ker Q,\operatorname{coker} Q$ will be non-empty. If $\ker Q$ is non-empty the system contains decoupled photons, which are gapped and hence do not affect the classification. Conversely, if $\operatorname{coker} Q$ is non-empty, the system contains continuous chiral symmetries and hence it is automatically gapless. The interesting question is, therefore, what happens if $Q$ is square and non-singular.

If $N_F\equiv N_c$ and $Q$ is full-rank, there are no rows nor columns that are zero, so there are no decoupled sectors and no continuous axial symmetries. There are no $U(1)^{N_F}$ axial symmetries because of the mixed anomaly, and no non-abelian chiral symmetries because there are no repeated rows in $Q$ (for otherwise the matrix would not be full-rank). We now claim that these conditions are not only necessary for being gapped, but also sufficient:
\begin{mdframed}
\begin{lemma}\normalfont 
The non-chiral QED system~\eqref{eq:non_chiral_QED} defined by a square matrix of charges $Q$ is gapped if and only if $Q$ is full rank.
\end{lemma}
\end{mdframed}

This claim follows from the analysis of section~\ref{sec:DLCQ}, and the fact that free fermions have chiral algebra $U(N_F)_1$ and the photons a chiral algebra $U(1)_{Q^tQ}$. The lattice generated by the compact scalars in $U(1)_{Q^tQ}$ is non-degenerate if and only if $Q$ is full-rank. Equality of the energy-momentum tensors of $U(N_F)_1$ and $U(1)_{Q^tQ}$ is nothing but the standard boson-fermion correspondence in $2d$.

One can reach this conclusion by looking directly at the central charges. The central charge of the free fermions is $N_F$ and, and that of the compact bosons is\footnote{Here $\operatorname{sign}(K)$ denotes the signature of the matrix $K$, defined as $+1$ for each positive eigenvalue, $-1$ for each negative eigenvalue, and $0$ for each zero eigenvalue. As $K=Q^tQ$ is positive semi-definite, $\operatorname{sign}(Q^tQ)\equiv\operatorname{rank}(Q^tQ)$.} $\operatorname{sign}(Q^tQ)\equiv\operatorname{rank}(Q)$, and these match if and only if $Q$ is full-rank, as required.

As an interesting remark, note that if $Q^tQ$ is not full rank, then in the gauge chiral algebra $U(1)_{Q^tQ}$ there is a factor of $U(1)_0$ for each zero eigenvalue of $Q^tQ$. This factor of $U(1)_0$ should be thought of as a free photon (a similar phenomenon was observed in~\cite{Choi:2018tuh} in a $3d$ QCD system). This is consistent with the discussion so far, in the sense that if the rank is not maximal there will be columns of zeros in $Q$, signaling decoupled photons.

\paragraph{Discrete symmetries.} If $\operatorname{rank}Q=N_c$, we saw earlier that $Q$ defines a QED theory with no continuous chiral symmetries. That being said, the system in general enjoys several discrete symmetries. Let us look at purely left-handed transformations. If we consider a $U(1)_\ell$ transformation defined by an integer vector $n$, then the theta term shifts as
\begin{equation}
U(1)_\ell\colon \theta\mapsto \theta+\alpha nQ\,,
\end{equation}
as per the flavor-gauge mixed anomaly. This shift means that $U(1)_\ell$ is not a true symmetry of the quantum system. On the other hand, if we choose $\alpha$ in such a way that $\theta$ stays invariant modulo $2\pi$, then the corresponding transformation does constitute a true symmetry of the quantum theory. This is simplest to ensure in the Hermite basis~\ref{fig:hermite_normal}. In this basis it becomes clear that there is a $\mathbb Z_{\tilde q}$ discrete symmetry for each diagonal component of $\tilde Q$, obtained by choosing $\alpha=2\pi k/\tilde q$ with $k=0,1,\dots,\tilde q-1$. Note that there is another factor of $\mathbb Z_{\tilde q}$ that acts on the right-handed fermions alone, but the two factors of $\mathbb Z_{\tilde q}$ are not two distinct symmetries, inasmuch as their simultaneous action is nothing but a gauge transformation. Hence, all in all, the flavor symmetry group of QED is
\begin{equation}
\prod_{\tilde q\in\operatorname{diag}(\tilde Q)}\mathbb Z_{\tilde q}\,.
\end{equation}
Note that if $\operatorname{rank}Q<N_c$, then some of the diagonal components of $\tilde Q$ will be zero. If for $\tilde q=0$ we agree to denote $\mathbb Z_0\equiv U(1)$, then the group above also contains the case where the system has non-trivial continuous symmetries.

For future reference we mention the fact that the order of the symmetry group is $\prod \tilde q\equiv\det(Q)$.

\paragraph{Chiral theories.} We finally make a few remarks concerning chiral theories. These are labelled by pairs of integral matrices $Q_\ell,Q_r$ which specify the charges of the left-movers and right-movers, respectively. Gauge anomaly cancellation requires~\eqref{eq:abelian_gauge_anomaly_cancel}
\begin{equation}\label{eq:chiral_QED_anomaly}
Q_\ell^tQ_\ell\equiv Q_r^tQ_r\,.
\end{equation}

The reader might find it useful to have at their disposal examples of chiral theories. A trivial class of examples is $Q_\ell=-Q_r$. A more interesting class of examples is provided by choosing any non-symmetric normal matrix $Q$, and taking $Q_\ell=Q$, $Q_r=Q^t$. More generally, it is easy to show that if $Q_\ell,Q_r$ satisfy the gauge anomaly cancellation condition~\eqref{eq:chiral_QED_anomaly}, then there exists some orthogonal matrix $\mathcal O$ such that $Q_\ell\equiv \mathcal OQ_r$. Therefore, we can generate other families of examples by fixing $Q_r$ and looking for orthogonal matrices $\mathcal O$ that make $\mathcal OQ_r$ integral.

In any case, many of the previous claims for non-chiral theories can be easily generalized to chiral theories. For example, if $N_c>N_F$, the extra photons are still decoupled. Indeed, if the matrices $Q_\ell,Q_r$ are fat (more columns than rows, see figure~\ref{fig:hermite_normal}) then they necessarily have a non-trivial kernel. The anomaly cancellation condition says that they in fact \emph{share} the kernel: the equality $Q_\ell^tQ_\ell\equiv Q_r^tQ_r$ implies that $||Q_\ell v||^2=||Q_rv||^2$ for any vector $v$, so $v$ is either annihilated by both $Q_\ell,Q_r$ or by neither. If $v$ is in their kernel, then the linear combination $v_i a_i$ is indeed a decoupled photon.

Similarly, if $N_c<N_F$, then there will necessarily be some anomalous continuous symmetry, because the matrices $Q_\ell,Q_r$ will have a non-trivial cokernel, so the associated currents will be conserved --- the mixed anomaly with the gauged $U(1)$ will vanish.

All in all, in classifying gapped theories we can assume without loss of generality that $N_c\equiv N_F$, and that $Q_\ell,Q_r$ are full rank. In this situation, the exact same argument from before proves that these are not only necessary conditions for being gapped, but also sufficient:

\begin{mdframed}
\begin{lemma}\normalfont 
A chiral QED system defined by a pair of square matrices of charges $(Q_\ell,Q_r)$, subject to the gauge anomaly cancellation condition~\eqref{eq:chiral_QED_anomaly}, is gapped if and only if $(Q_\ell,Q_r)$ are full rank. (Both matrices necessarily have the same rank, due to~\eqref{eq:chiral_QED_anomaly}). 
\end{lemma}
\end{mdframed}

As a consistency check, note that a gapped theory cannot have continuous chiral symmetries, and it is not entirely obvious from the discussion above that a model with full rank matrices $(Q_\ell,Q_r)$ has no such symmetries. It is clear that, being full rank, there are no purely left handed (nor purely right handed) symmetries; but there is no immediate reason that excludes symmetries where both chiralities transform at the same time. It is not hard to show that, as a matter of fact, no such symmetries exist either: any would-be flavor symmetry where both chiralities transform simultaneously is either broken by a mixed flavor-gauge anomaly, or a pure gauge transformation itself. Hence, chiral models with full rank $(Q_\ell,Q_r)$ have no continuous chiral symmetries, as required for a supposedly gapped theory.

The conjectural infrared TQFT has left chiral algebra $U(N_F)_1/U(1)_{Q^t_\ell Q_\ell}$, and right chiral algebra $U(N_F)_1/U(1)_{Q^t_r Q_r}$. These two algebras are isomorphic --  cf.~the anomaly cancellation condition -- via the the orthogonal matrix $\mathcal O$ discussed above, namely $\mathcal O:=Q_\ell Q_r^{-1}$.

\subsection[{$U(1)$ with $N$ charge-$q$ Dirac fermions}]{$\boldsymbol{U(1)}$ with $\boldsymbol N$ charge-$\boldsymbol q$ Dirac fermions}

Here we analyse the infrared dynamics of $U(1)$ plus $N$ copies of a charge-$q$ Dirac fermion. The claim is that the low energy theory of this system corresponds to a copy of the $SU(N)_1$ WZW model on each of the $q$ universes. (These universes are the result of the $\mathbb Z_q$ one-form symmetry). This nicely reproduces the analysis of~\cite{Misumi:2019dwq}.

According to the general conjecture~\eqref{eq:conj_IR_coset}, the infrared dynamics of the model are described by the coset
\begin{equation}
\frac{U(N)_1}{U(1)_{Nq^2}}\,.
\end{equation}
Note that if $N>1$, the central charge is non-zero, so the coset describes a gapless theory.

In order to project the theory into a specific universe we gauge the one-form symmetry $\mathbb Z_q$, to wit
\begin{equation}
\biggl(\frac{U(N)_1}{U(1)_{Nq^2}}\biggr)/\mathbb Z_q\equiv \frac{U(N)_1}{U(1)_N}\equiv SU(N)_1\,.
\end{equation}
The second equality involves the character decomposition
\begin{equation}
\begin{aligned}
d_\text{NS-X}(q,g,\theta)&=\sum_{n=0}^{N-1}(\pm1)^n\chi_n(q,\theta)\chi_{\gamma^n\cdot\boldsymbol0}(q,g)\\
d_\text{R-X}(q,g,\theta)&=\sum_{n=0}^{N-1}(\pm1)^{n+1}\chi_{n+\lfloor N/2\rfloor}(q,\theta)\chi_{\gamma^n\cdot\boldsymbol0}(q,g)\,,
\end{aligned}
\end{equation}
where $\theta$ is a flavor $U(1)$ parameter and $g$ an $SU(N)$ flavor parameter. The characters of $U(1)_N$ are denoted by $\chi_n(q,\theta)$ and those of $SU(N)_1$ by $\chi_\lambda(q,g)$. When $N$ is odd, $\chi_n$ denotes a super-character and when even, a regular character.

Note that when $N=1$ the CFT $SU(1)_1$ becomes trivial, which means that the charge-$q$ Schwinger model has a unique, trivial vacuum in each universe. In this case, the infrared coset $U(1)_1/U(1)_{q^2}$ describes a gapped theory, and the character decomposition is
\begin{equation}
\begin{aligned}
d_\text{NS-X}(q,\theta)&=\sum_{\ell=0}^{q-1}(\pm1)^\ell\chi^\text{NS-X}_{\ell q}(q,\theta)\\
d_\text{R-X}(q,\theta)&=\sum_{\ell=0}^{q-1}(\pm1)^\ell\chi^\text{R-X}_{\ell q+\lfloor q/2\rfloor}(q,\theta)\,,
\end{aligned}
\end{equation}
where $\chi_\ell$ are the characters of $U(1)_{q^2}$; these are regular (bosonic) characters when $q$ is even (cf.~\eqref{eq:u1_char_bos}), and super-characters when odd (cf.~\eqref{eq:u1_char_fer}). From this character decomposition we learn that the theory has $q$ vacua, all bosonic, in both sectors $\mathcal H_\text{NS}=\mathcal H_\text{R}=\mathbb C^{q|0}$. These $q$ vacua live in the $q$ universes, one in each.\footnote{More precisely, the states that live in a specific universe are linear combination of these $q$ states. $(-1)^{F_L}$ does not commute with the one-form symmetry, and therefore the states in a given universe do not have well-defined charge under $(-1)^{F_L}$.} Out of these, $\lceil q/2\rceil$ are neutral under $(-1)^{F_L}$, and the rest $\lfloor q/2\rfloor$ are charged.

\subsection{Vacua of gapped theories}

In the previous section we described the infrared dynamics of a gapless theory. Here we study the gapped case, namely those theories where the matrix of charges $Q$ is square and full rank. Conjecturally, the vacua of such theories are described by the coset~\eqref{eq:conj_IR_coset}
\begin{equation}
\frac{U(n)_1}{U(1)_{Q^tQ}}\,.
\end{equation}
We focus on the case where $Q$ describes an even lattice, i.e., where all the diagonal components of $Q^tQ$ are even. In this situation the CFT $U(1)_{Q^tQ}$ is bosonic, that is, its characters do not depend on the spin structure.

The characters of the numerator $U(n)_1$ are given by
\begin{equation}
\begin{aligned}
d_\text{NS-X}(q,\theta)&=q^{-n/24}\prod_{r=0}^\infty\prod_{i=1}^n(1\pm z_iq^{r+1/2})(1\pm z^{-1}_iq^{r+1/2})\\
d_\text{R-X}(q,\theta)&=q^{n/12}\bigg[\prod_{i=1}^nz_i^{1/2}\pm z_i^{-1/2}\bigg]\prod_{r=1}^\infty\prod_{i=1}^n(1\pm z_iq^r)(1\pm z^{-1}_iq^r)\,,
\end{aligned}
\end{equation}
where $\theta\in U(n)$ is a flavor parameter conjugate to $\theta\sim\diag(z_1,\dots,z_n)$.

On the other hand, the characters of the denominator are
\begin{equation}
\chi_{\ell}(q,\theta)=\frac{1}{\eta(q)^n}\sum_{u\in \mathbb Z^n+K^{-1}\ell}q^{\frac12 u^tKu}z^{Q u},\qquad K:=Q^tQ\,,
\end{equation}
where $\eta$ is the Dedekind eta function.

The vacua of the QED theory labelled by a matrix $Q$ are determined by the branching functions of $U(n)_1$ into $U(1)_{Q^tQ}$, i.e., by the decomposition of $d_{\pm,\pm}$ characters into $\chi_\ell$ characters. We propose that the branching functions of the coset $U(n)_1/U(1)_{Q^tQ}$ are given by the following:
\begin{equation}
\begin{aligned}
d_\text{NS-X}(q,\theta)&=\sum_{\ell\in\Gamma(Q)}(\pm1)^{||\ell||^2}\chi_{Q^t\ell}(q,\theta)\\
d_\text{R-X}(q,\theta)&=\sum_{\ell\in\Gamma(Q)}(\pm1)^{||\ell||^2}\chi_{Q^t(\ell+1/2)}(q,\theta)\,,
\end{aligned}
\end{equation}
where $\Gamma(Q)\cong \mathbb Z^n/\sim$ is the set of all integral vectors modulo the identification through rows of $Q$: two vectors are declared to be equivalent if they differ by some integral linear combination of the rows of $Q$. As a consistency check, note that $2h_{Q^t\ell}=||\ell||^2$ so the chiral fermion parity in the NS-sector corresponds precisely to the spin of the characters. Similarly, in the R-sector the spin is $2h_{Q^t(\ell+1/2)}=n/4\mod 1$ so the spin is in $\mathbb Z+\frac18n$, as expected from the zero-point energy of the fermions.

The branchings above predict that the QED theory with matrix of charges $Q$ has $|\Gamma(Q)|\equiv|\det(Q)|$ vacuum states. As in the Schwinger model, these can be thought of as the result of the spontaneous symmetry breaking of the axial symmetry, which also has $|\det(Q)|$ elements. Note also that $|\det(Q)|$ is the order of the one-form symmetry, which suggests that each universe has a single vacuum state.

\clearpage
\printbibliography
\end{document}